\documentclass{aa} 

\graphicspath{{./}{figures/}}
\bibpunct{(}{)}{;}{a}{}{,} % to follow the A&A style
\usepackage{graphicx, xcolor}
\usepackage{txfonts}
\usepackage{hyperref}
\usepackage{amsmath}
\usepackage{empheq}
\usepackage{mathabx}

\hypersetup{
    colorlinks=True,
    allcolors=blue
    }
\makeatletter
\renewcommand*\aa@pageof{, page \thepage{} of \pageref*{LastPage}}
\makeatother

\begin{document}

   \title{The SRG/eROSITA diffuse soft X-ray background}      % I would write it as ab object nae: Local Hot Bubble
   \subtitle{I. The local hot bubble in the western Galactic hemisphere}

   \author{Michael~C.~H.~Yeung
         \inst{1}
         \and  
         Gabriele~Ponti \inst{2}\fnmsep\inst{1}
         \and 
         Michael~J.~Freyberg\inst{1}
         \and
         Konrad~Dennerl\inst{1}
          \and         
        Teng Liu\inst{3}\fnmsep\inst{4}\fnmsep\inst{1}
        \and
         Nicola~Locatelli \inst{2}
         \and
         Martin~G.~F.~Mayer\inst{5}
         \and
         Jeremy~S.~Sanders \inst{1}
         \and
         Manami~Sasaki\inst{5}
         \and
         Andy~Strong \inst{1}
         \and
         Yi~Zhang \inst{1}
         \and
         Xueying~Zheng \inst{1}
         \and
         Efrain~Gatuzz \inst{1}}

   \institute{Max-Planck-Institut für extraterrestrische Physik, Gießenbachstraße 1, 85748 Garching, Germany\\
              \email{myeung@mpe.mpg.de}
        \and
            INAF-Osservatorio Astronomico di Brera, Via E. Bianchi 46, I-23807 Merate (LC), Italy
        \and
        Department of Astronomy, University of Science and Technology of China, Hefei 230026, China
        \and
        School of Astronomy and Space Science, University of Science and Technology of China, Hefei 230026, China
        \and        
        Dr.~Karl Remeis Observatory, Erlangen Centre for Astroparticle Physics, Friedrich-Alexander-Universit{\"a}t Erlangen-N{\"u}rnberg, Sternwartstra{\ss}e 7, 96049 Bamberg, Germany}

   \authorrunning{M.~C.~H.~Yeung et al.}

% \abstract{}{}{}{}{} 
% 5 {} token are mandatory
 
  \abstract
   {The SRG/eROSITA All-Sky Surveys (eRASSs) combine the advantages of complete sky coverage and the energy resolution provided by the charge couple device and offer the most holistic and detailed view of the diffuse soft X-ray background (SXRB) to date. The first eRASS (eRASS1) was completed at solar minimum, when solar wind charge exchange emission was minimal, providing the clearest view of the SXRB.}
   %, the LHB and SWCX emissions can be studied in unprecedented detail.}
   {We aim to extract spatial and spectral information from each constituent of the SXRB in the western Galactic hemisphere, focusing on the local hot bubble (LHB).}
   {We extracted and analysed eRASS1 spectra from almost all directions in the western Galactic hemisphere by dividing the sky into equal signal-to-noise bins. We fitted all bins with fixed spectral templates of known background constituents.}
   {We find the temperature of the LHB exhibits a north-south dichotomy at high latitudes ($|b|>30\degr$), with the south being hotter, with a mean temperature at $kT=121.8\pm0.6\,$eV and the north at $kT=100.8\pm0.5$\,eV. At low latitudes, the LHB temperature increases towards the Galactic plane, especially towards the inner Galaxy. The LHB emission measure (${\rm EM_{LHB}}$) enhances approximately towards the Galactic poles. The ${\rm EM_{LHB}}$ map shows clear anti-correlation with the local dust column density. In particular, we found tunnels of dust cavities filled with hot plasma, potentially forming a wider network of hot interstellar medium. We also constructed a three-dimensional LHB model from ${\rm EM_{LHB}}$, assuming constant density. The average thermal pressure of the LHB is $P_{\rm thermal}/k=10100^{+1200}_{-1500}\,{\rm cm^{-3}\,K}$, a lower value than typical supernova remnants and wind-blown bubbles. This could be an indication of the LHB being open towards high Galactic latitudes.}
   %The eROSITA bubbles show a shell-like temperature structure, with the shell being cooler at $\sim0.2\,$keV, enveloping a hotter interior at $\sim0.3\,$keV, under the assumption of collisional ionisation equilibrium at solar abundance. The appearance of the cooler shell might be a result of non-equilibrium ionisation of the shocked plasma.}
   {}

   \date{Received 10 June 2024; accepted 15 August 2024}
   \titlerunning{eROSITA soft X-ray background I: Local hot bubble in the western Galactic hemisphere}
    \authorrunning{M.~C.~H.~Yeung et al.}
   \maketitle
%
%-------------------------------------------------------------------
\section{Introduction} \label{sec:intro}
The diffuse soft X-ray background (SXRB) is known to be a superposition of emission components spanning from the length scale of AU to cosmic distances \citep[e.g.][]{McCammon_1990,Kuntz2000}. All-sky analyses of the SXRB have largely relied on ROSAT/PSPC broadband count rates and ratios \citep[e.g.][]{Snowden97,Snowden98,Kuntz2000}, especially for studies on the local hot bubble (LHB; e.g. \citealt{Snowden2000,Liu2017}), as background photons with energies $\lesssim0.3\,$keV (R1: $0.11$--$0.284$\,keV; R2: $0.14$--$0.284$\,keV) are readily absorbed by the wall of the local bubble and all R12 counts are effectively of local origin. The brightest soft X-ray line feature in the SXRB is the \ion{O}{VII} triplet at $0.57\,$keV. This is an important diagnostic, as it encapsulates emissions from the LHB; the background Galactic emission, including the Milky Way circum-galactic medium (MW CGM); and depending on sight lines, intervening Galactic structures. Hence, extracting information for a particular component using broadband count rates and ratios alone becomes challenging and heavily dependent on the knowledge and the accurate subtraction of the rest. Notably, the discovery of the time-variable solar wind charge exchange (SWCX) component has aggravated the issue even further \citep{Lisse_1996,Cravens1997,Dennerl_1997}.

X-ray observatories with charge couple device (CCD) cameras, such as \textit{XMM-Newton}/EPIC, \textit{Chandra}/ACIS, and \textit{Suzaku}/XIS, have demonstrated the importance of spectral resolution to decompose the SXRB into their respective components \citep[e.g.][]{Lumb02, Markevitch03,Henley08,Yoshino09}, but they are restricted to limited `pencil-beam' sight lines due to their relatively small fields of view (FoVs) and observing strategies. Only since the last decade have a substantial number of pointings been accumulated to enable a more holistic view of the SXRB \citep{Henley10,Henley12,Henley13,Nakashima18,Gupta23,Pan24}. A notable exception is Halosat, where a large portion of the sky is covered in 333 pointings by virtue of its large FoV ($\sim 10\degr$) whilst achieving CCD-type energy resolution using non-imaging silicon drift detectors \citep{Kaaret19, Kaaret20, Ringuette21, Bluem22}.

Currently, high spectral resolution observations of the SXRB remain extremely difficult to obtain without telescopes of a large grasp\footnote{Grasp is defined as the product of FoV and effective area.}. However, successive sounding rocket launches of the X-ray Quantum Calorimeter (XQC) \citep{McCammon02,Crowder12,Wulf19} have shown promise of being able to separate the SWCX contribution from the LHB emission beyond CCD energy resolution using \ion{C}{VI} Ly-$\alpha$, $\gamma$ and the fine structure lines within the \ion{O}{VII} triplet, despite their short exposure times. 

The local interstellar medium (LISM) is known to be a volume devoid of neutral gas -- also known as the local cavity (LC) or the local bubble (LB) \citep[e.g. see a review by][]{Frisch11}. This cavity is instead filled by a $\sim0.1$\,keV ($\sim 10^6$\,K) plasma \citep[e.g.][]{Snowden1990,Snowden97,McCammon02,Liu2017,GMC_shadow}. The electron density of this plasma is uncertain without the assumption of its line-of-sight density profile. However, shadowing studies of molecular clouds on the wall of the LC at different distances have consistently shown that a uniform electron density of $\sim 4\times10^{-3}\,{\rm cm^{-3}}$ is a reasonable assumption, as probed by various sight lines \citep{Snowden14,GMC_shadow}. There is a growing number of studies that support a formation scenario where dozens of supernova explosions create and sustain the LHB \citep[e.g.][]{Fuchs06,zucker22,Schulreich23}, employing a combination of star cluster traceback, numerical simulations, and matching supernova-produced radioisotopes found in Earth's crust.

Solar wind charge exchange has long been the biggest source of uncertainty revolving around the X-ray measurements of the LHB, as its spectrum resembles that of the LHB despite being non-thermal in nature \citep[see reviews by][and references therein]{Dennerl10, Kuntz19}. SWCX can be broadly separated into two categories: magnetospheric and heliospheric. The former refers to the emission from solar wind ions interacting with neutrals in Earth's exosphere and the latter with the inflowing neutral interstellar medium (ISM) into the solar system. Observations conducted by most X-ray missions are prone to both kinds of SWCX emissions. An exception is eROSITA \citep{eROSITA}, which is on board the Spectrum-Roentgen-Gamma (SRG) \citep{SRG} observatory.
SRG orbits around the Sun-Earth Lagrangian point L2. SRG/eROSITA always pointed perpendicular to the Sun-Earth line during all-sky surveys (eRASSs). Hence, it never looked through the Earth's exosphere, avoiding the magnetospheric SWCX.
No evidence of SWCX emission from the magnetotail has been found thus far (\citealt{GMC_shadow}, Dennerl et al. in prep).
Heliospheric SWCX indeed caused variations in the SXRB in eRASSs, and it correlates with the solar cycle and ecliptic latitudes \citep{Ponti2023,GMC_shadow}. Studies using other soft X-ray instruments have found the same correlation between solar cycles and heliospheric SWCX \citep{Qu22,Ueda22,Pan24}.
% Variations caused by heliospheric SWCX were indeed observed in eRASSs, and are demonstrated to be correlated strongly with the solar cycle and ecliptic latitudes \citep{Ponti2023, GMC_shadow}, consistent with the conclusion reached by studies using other soft X-ray instruments \citep{Qu22,Ueda22,Pan24}.
The first eRASS (eRASS1) was completed during solar minimum and, on average, exhibits a low amount of heliospheric SWCX \citep{Ponti2023, GMC_shadow}.
A parallel work (Dennerl et al. in prep) will rigorously present and discuss the SWCX contributions in eRASSs. In this work, we leverage this advantage and model the SXRB in eRASS1 without the SWCX contribution, which became non-negligible in later eRASSs.

The hot phase of the CGM in a spiral galaxy is believed to trace gas from both feedback processes and the shock-heated intergalactic medium out to its virial radius \citep[e.g. see][for a review]{Putman12}. The hot halo gas is also predicted to be at approximately the virial temperature (for the MW, $\sim 0.2\,$keV) and holds a large fraction of its baryons. Numerous X-ray observations have confirmed the existence of this phase via both emission \citep[e.g.][]{Yoshino09,Henley10,Henley12,Henley13,Miller_15,Nakashima18,Kaaret20,Ponti2023} and absorption studies \citep[e.g.][]{Bregman07,Yao07,Yao08,sakai12,miller13,Fang14,Fang15}.
Recently, there have been growing reports of the presence of an additional thermal component in the CGM, at $\sim 0.7\,$keV \citep[e.g.][]{Das19,Bluem22,Ponti2023}, which could alternatively be attributed to coronal emission from M-dwarfs \citep{Wulf19}.
One of the prime goals of our spectral analysis is the characterisation of the two CGM components. The results and discussion concerning the MW CGM will be presented in Paper~II (Ponti et al. in prep) of the series.

Our understanding of the cosmic X-ray background (CXB) at the low-energy end ($0.5$--$2$\,keV) has taken a giant leap forward with \textit{Chandra} and \textit{XMM-Newton} surveys, resolving $\sim 80$\% of the CXB sources \citep{DeLuca04,Luo17,Cappelluti17}. The resolved sources are mostly identified as active galactic nuclei (AGNs). Still, contributions from galaxies, galaxy groups, and clusters are expected to rise below 1\,keV, steepening the slope at the softest end \citep{Gilli}. eROSITA is an instrument specialised in the soft X-ray band. As a result, we expect to be able to measure such a steepening and subsequently apply the result to refine our measurements on the Galactic components.

In this work, we report on the spectral analysis of the eRASS1 data in the western Galactic hemisphere, being as spatially continuous as possible by extracting a large number ($\sim 2000$ bins) of high S/N spectra. This work is the first of a series focusing on the various components of the SXRB seen by eROSITA. In particular, this paper lays out the data extraction and the general methodology adopted in the spectral modelling in Sect.~\ref{sec:ana} and \ref{sec:spec_ana}, which is the backbone for the series. Then, in Sect.~\ref{sec:result}, we report on the results, which are mainly focused on, but not limited to, the LHB and the hot LISM, with intervening discussions and interpretations. The results on the CGM, the Galactic corona, and the eROSITA bubbles will mainly be presented in other works in the series (Ponti et al. in prep; Yeung et al. in prep). Hence, they are only mentioned in this work when they directly affect results on the LHB.

% The uncertainties reported in this work are 16$^{\mathrm{th}}$ and 84$^{\mathrm{th}}$ percentiles unless stated otherwise.

\section{Data selection and spectral extraction} \label{sec:ana}
We study the diffuse soft X-ray background from the eRASS1 data of processing version \texttt{c020} of the western Galactic hemisphere ($180\degr < l < 360\degr$). We cleaned the data following a few criteria. First, we discarded the data from TM5 and TM7 due to optical light-leak \citep{eROSITA}, and removed good-time-intervals (GTIs) with count rates $> 1.435$\,counts\,s$^{-1}$\,deg$^{-2}$ in the 4--9\,keV band to minimise contaminations by flares. The particle background dominates this band, and the threshold corresponds to 1.5 times the level of particle background in this band \citep{GMC_shadow}. The eROSITA-DE consortium has also released a list of temperature-sensitive or bright pixels that occasionally produce artefacts but are not officially flagged as bad pixels. They are mostly from TM4, which suffered from a major micrometeoroid hit \citep{Freyberg22}. We rejected these sensitive pixels in addition. Last but not least, we masked regions with overdense source detection \citep{DR1} and positions of known galaxy clusters with $R_{500}\gtrsim3\arcmin$ as described in \citet[][and references therein]{Bulbul24}. The overdense source detection regions could be regions within or near extended sources, such as supernova remnants or artefacts caused by bright point sources, which triggered a high density of spurious source detections.

Subsequently, we defined our spatial binning of spectral extraction using the software \texttt{contbin} \citep{contbin}, with the primary aim of dividing the western Galactic hemisphere into bins of approximately constant S/N in the diffuse soft X-ray emission, instead of imposing a regular grid system such as the \textit{skytile} system adopted by the standard products of eROSITA. For our analysis, \texttt{contbin} also has the advantage of defining bins with edges more closely following distinct features (for example, from superbubbles, supernova remnants etc.) and being computationally efficient compared to traditional Voronoi binning codes. The binning was done on the eRASS1 0.2--0.6\,keV diffuse emission count map (all detected sources masked\footnote{More precisely, the masking of `all' detected sources is done by merging the \texttt{CheeseMask} images from the standard eSASS pipeline from all the skytiles and project them into a \texttt{HEALPix} map of $N\mathrm{side}=4096$ (pixel size $\simeq51\arcsec$) using nearest-neighbour interpolation. Contribution from the masked pixels is then removed from the final diffuse emission count map after downsampling the \texttt{HEALPix} map from $N\mathrm{side}=4096$ to $N\mathrm{side}=256$.}) after subtracting the expected counts from the non-X-ray background measured from the filter-wheel-closed data \citep{GMC_shadow}, as this band contains the bulk of the emissions from the LHB that eROSITA observes. This can be written explicitly as
\begin{eqnarray}
S(\Vec{r}) &=& C(\Vec{r}) - B_{\rm nonvig}(\Vec{r}) \\
           &=& C(\Vec{r}) - E_{\rm nonvig}(\Vec{r}) \times R_{\rm FWC}(\Vec{r})\;,   
\end{eqnarray}
where $\vec{r}$, $S$, $C$ and $B_{\rm nonvig}$ denote the sky position, signal, total counts from diffuse emission, non-X-ray or non-vignetted background counts respectively. $B_{\rm nonvig}$ can be further written as a product of non-vignetted exposure time ($E_{\rm nonvig}$) and the count rate of the filter-wheel-closed background ($R_{\rm FWC}$).
We estimate the corresponding noise map $N(\Vec{r})$ for the S/N calculation using equation (4) of \citet{contbin} as adopted from \citet{Gehrels}, that is,
\begin{eqnarray}
N(\vec{r}) &=& \sqrt{g[C(\Vec{r})] + g[B_{\rm nonvig}(\vec{r})]}\;,
\end{eqnarray}
where
\begin{eqnarray}
g(c) &=& \Bigg(1+\sqrt{c+\frac{3}{4}}\Bigg)^2
\end{eqnarray}
is an estimation of the upper limit of the squared uncertainty on $c$ counts in Poissonian statistics. Before binning, the maps were projected into the zenithal equal area (ZEA) projection. Contour-binning yielded 2010 bins larger than 1\,deg$^{2}$, which we consider valid bins for spectral analysis. $1\,$deg$^2$ is approximately the eROSITA field-of-view. This selection primarily removed areas near the south ecliptic pole and the Large Magellanic Cloud where the exposure time is maximal due to the overlapping of the scanning loci in eRASSs but are not representative of the general SXRB.

\begin{figure}   % Fig.1
    \centering
    \includegraphics[width=0.49\textwidth]{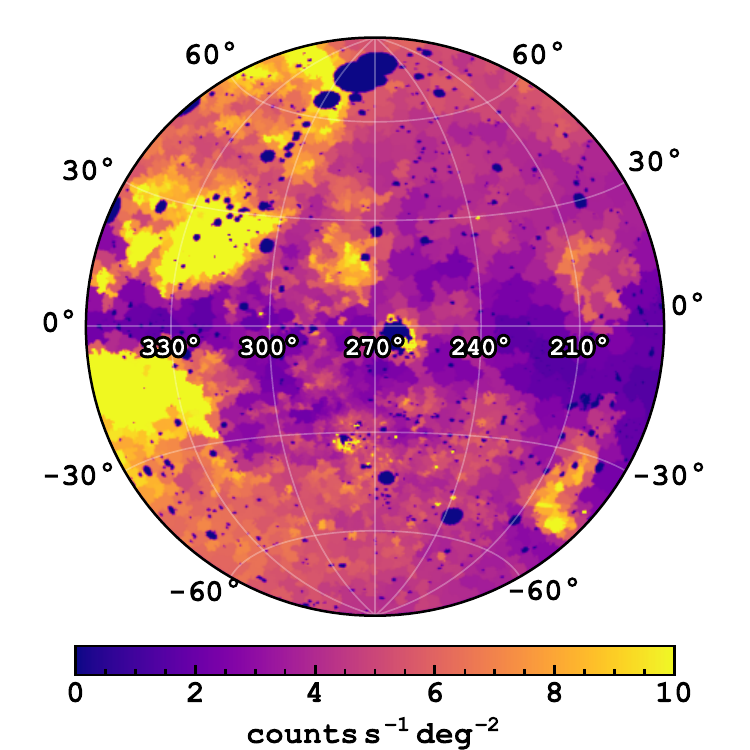}
    \caption{Contour-binned eRASS1 0.2--0.6\,keV band surface brightness map in zenithal equal-area projection. Locations of big galaxy clusters, overdense regions in source detection, and bins with sky area less than 1\,deg$^2$ were masked. Counts from the non-X-ray background and all eRASS1-catalogued sources \citep{DR1} were also removed from this image (but not in the spectra; see Sect.~\ref{sec:ana} for more details).}     %%%-MJF  really "all" detected sources or the ones above some flux limit?
    \label{fig:contbin}
\end{figure}

\begin{figure}   % Fig.2
    \centering
    \includegraphics[width=0.49\textwidth]{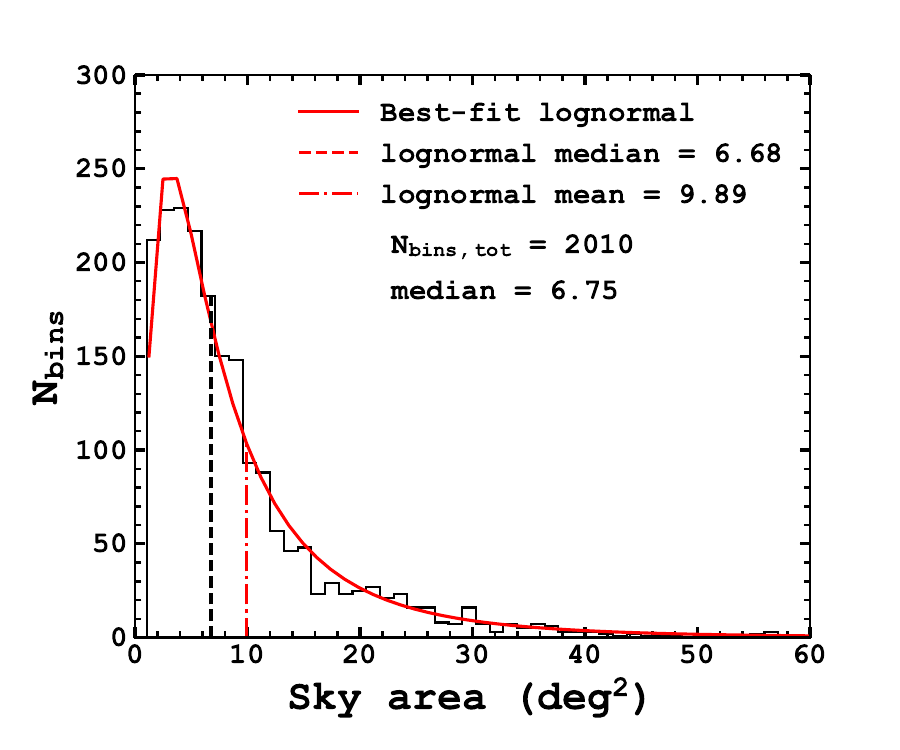}
    \caption{Sky area distribution of the contour bins. It can be approximated by a log-normal distribution, as shown with the red line.}
    \label{fig:area_dist}
\end{figure}

Fig.~{\ref{fig:contbin}} shows the contour-binned eRASS1 0.2--0.6\,keV band surface brightness map of the valid bins. Large soft X-ray emitting structures such as the eROSITA bubbles (a pair of bubbles at $l\gtrsim290\degr$ in the north and $l\gtrsim320\degr$ in the south), Antlia supernova remnant ($l,b$)$\sim$($275\degr$, $15\degr$), Monogem Ring ($l,b$)$\sim$($200\degr$, $8\degr$) and Orion-Eridanus Superbubble ($l,b$)$\sim$($205\degr$, $-30\degr$), and the Galactic disc stand out in stark relief. 
The sky area distribution of the valid contour bins is shown in Fig.~\ref{fig:area_dist}. The median bin size is $\sim7\,\mathrm{deg}^2$. All bins with an area less than 1\,deg$^2$ were removed. The distribution can be well approximated by a log-normal function, as shown by the red line.

Possible fluctuations contributed by bright X-ray sources should be removed from the analysis of diffuse emission. However, as is discussed in more detail in Sect.~\ref{sec:CXB}, removing sources alters the shape and normalisation of the cosmic X-ray background (CXB). Masking sources above a flux limit threshold substantially higher than the eRASS1 flux limit \citep[$F_{\mathrm{lim},0.5\mathrm{-}2\,\mathrm{keV}} \sim 10^{-14}\,\mathrm{erg}\,\mathrm{s}^{-1}\,\mathrm{cm}^{-2}$;][]{DR1} could avoid the problem of having an exposure-dependent CXB component in the spectra. In order words, if one were to mask all detected sources, the CXB resolve fraction would depend on the exposure depth, causing spatially correlated CXB normalisation and photon index variations. As such, we chose to mask only sources with fluxes $F_{0.5\mathrm{-}2\,\mathrm{keV}} > 10^{-12}\,\mathrm{erg}\,\mathrm{s}^{-1}\,\mathrm{cm}^{-2}$ and detection likelihood of \texttt{DET\char`_LIKE\char`_0}\,$>10$ in the eRASS1 source catalogue \citep{DR1}, using circular masks of radius four times their aperture photometry extraction radii ($4\times \texttt{APE\char`_RADIUS\char`_1}\simeq 2\arcmin$)\footnote{The masking radius of four times \texttt{APE\char`_RADIUS\char`_1} is $\simeq 2\arcmin$ for most sources since \texttt{APE\char`_RADIUS\char`_1} relates directly to the encircled energy fraction (0.75) set during the source detection pipeline, which corresponds to $\simeq0\farcm5$ for the point-spread-function of eROSITA.} during spectral extraction. This choice minimises the spectral fluctuations introduced by bright sources while maintaining a largely uniform CXB component in all spectra. This masking was applied during spectral extraction and differs from the one used to create the diffuse emission map for contour-binning.

We extracted the spectrum and its auxiliary response file for events of all valid patterns (\texttt{pat=15}) of each contour bin by providing a point source cheesemask and a contour bin mask that demarcates the bin profile to the \texttt{eSASS} task \texttt{srctool}. They serve as the basis of all analyses in this work.
%The eRASSs data are arranged in \textit{skytiles}, each uniquely defined with bounds in RA and Dec ($\sim3\degr$ in each coordinate for the majority of skytiles). For each contour bin, we merged the eventfiles of all the intersecting skytiles.

Last but not least, we also utilised publicly available ROSAT R1 and R2 diffuse background maps in our spectral fitting \citep{Snowden97}. Despite ROSAT/PSPC's poor spectral resolution, it provides a larger grasp than eROSITA at energies $\lesssim0.3$\,keV \citep[see Fig.\,10 of][]{eROSITA}. We found the addition of ROSAT data helps break the degeneracies between the LHB and the CGM components that arose in some low absorption regions (see Sect.~\ref{subsec:degen}).

The R1 and R2 band count rate and sigma maps were used to provide two more data points in each spectrum, with some manipulations. In detail, we began by binning the maps using the same set of contour bins, then we converted the map unit of $\mathrm{count}\,\mathrm{Ms}^{-1}\,\mathrm{arcmin}^{-2}$ to a flux unit of $\mathrm{erg}\,\mathrm{s}^{-1}\,\mathrm{cm}^{-2}$ by assuming a $0.1\,$keV \texttt{apec} spectral model and multiplication with the bin area. The use of a \texttt{apec} model \citep{apec} is motivated by the approximation that the contribution from the LHB dominates R1 and R2 counts. As the last step, we used the \texttt{ftflx2xsp} task in \texttt{FTOOLS} to create \texttt{Xspec}/\texttt{PyXspec}-ready spectra and diagonal response matrices to enable simultaneously fitting with the eROSITA spectra. The bin edges of the R1 and R2 maps were taken from \citep{Snowden94} where the band response drops to 10\% of the peak values, namely, $0.11$--$0.284$\,keV for R1 and $0.14$--$0.284$\,keV for R2.

\section{Spectral analysis} \label{sec:spec_ana}

Attempting to decompose the diffuse emission from half the sky both spatially and spectrally is an ambitious task. As the first study to do so, we decided to employ a more conventional approach in the analysis, whereby treating the contour bins as independent during spectral fitting. Additionally, we adopted a fixed number of spectral components to fit the spectra from all contour bins. These components include four conventional X-ray background components \citep[e.g.][]{Gupta21,Bluem22,Ponti2023,GMC_shadow}: (1) the local hot bubble (LHB), (2) the Milky Way's halo or (warm-hot) CGM, (3) the Galactic corona or the hot component of the CGM (COR) and (4) the cosmic X-ray background (CXB), and the non-X-ray background modelled by (5) the eROSITA filter-wheel-closed (FWC) background models \citep{GMC_shadow}. The SWCX emission during eRASS1 was weak, as it was the time of solar minimum, as shown by Dennerl et al. (in prep.) in a detailed eRASS1--eRASS4 half-sky analysis, \citet{Ponti2023} in the eFEDS field, and \citet{GMC_shadow} in three giant molecular cloud sight lines. Therefore, we did not include an SWCX component in our spectral fits. In addition, we included an extra thermal component for contour bins overlapping with the eROSITA bubbles (eROBub) (Sect.~\ref{subsec:erobub_method}).

\subsection{Description of model components} \label{sec:comp}
\begin{figure*}
    \centering
    \includegraphics[width=0.45\textwidth]{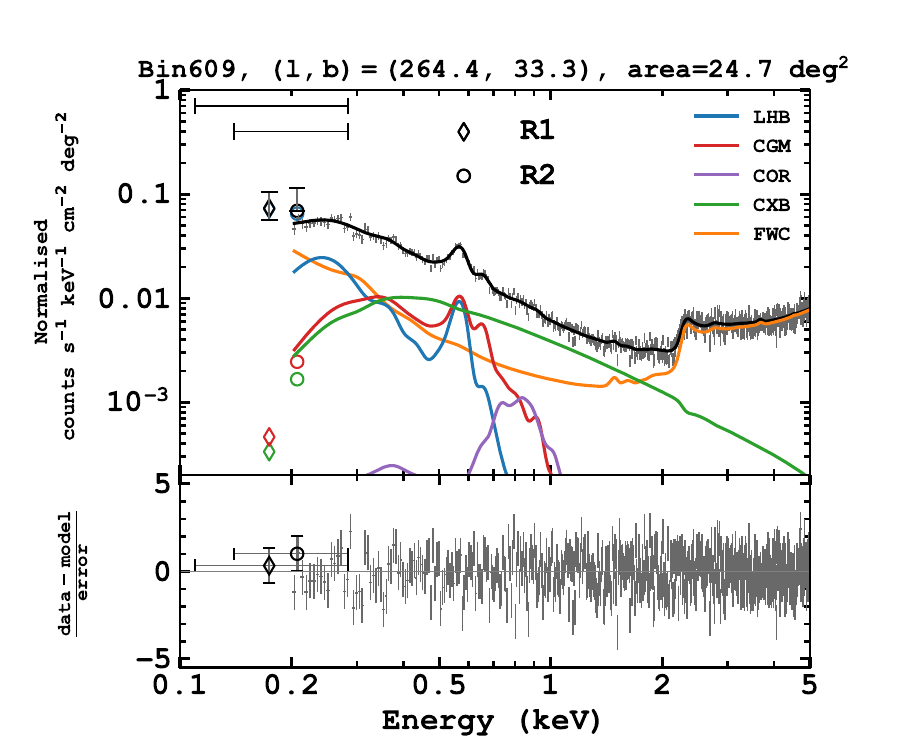}
    \includegraphics[width=0.45\textwidth]{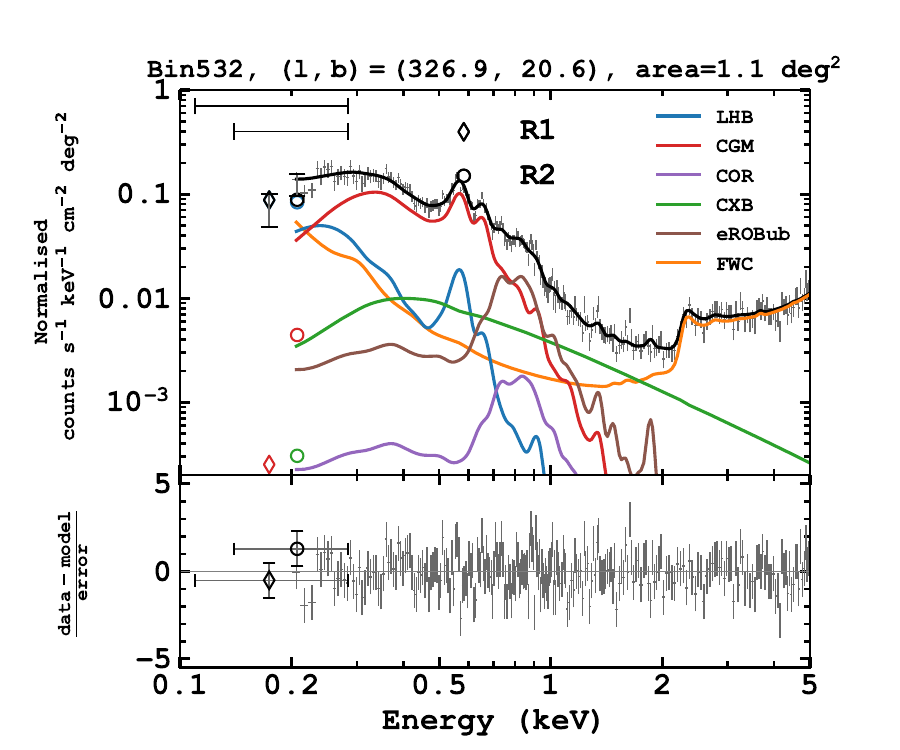}

    \caption{Example spectra outside (\textit{left}) and inside (\textit{right}) of the eROSITA bubbles, overlaid with the best-fit spectral models labelled in the legends. We modelled the eROSITA bubbles with an additional thermal component in brown. Both spectra have been divided by the effective area, aiming to bring the ROSAT R1 and R2 count rates into a reasonable range. This has the disadvantage of creating the fictitious jump of the instrumental background component above the gold absorption edge at $\sim 2\,$keV. The horizontal error bars at the top left corners of the figures reflect the width of the ROSAT R1 and R2 bands.}
    \label{fig:spec_eg}
\end{figure*}

Fig.~\ref{fig:spec_eg} shows two example spectra to illustrate our spectral templates, one outside the eROSITA bubbles and one inside. The discussion in the remainder of this Section deals with the detailed description of these model components.

\subsubsection{Local hot bubble} \label{subsubsec:LHB}
The LHB is a foreground component and was thus modelled as an unabsorbed optically thin plasma in collisional ionisation equilibrium (CIE) using the \texttt{apec} model \citep{apec} using \texttt{AtomDB} version 3.0.9 \citep{AtomDB}. Its temperature is allowed to vary freely only below $0.15$\,keV, for contour bins that have $\log_{10}{(N_{\rm H}/{\rm cm}^{-2})} < 20.5$, estimated from a combination of HI4PI and \textit{Planck} dust radiance map (see Sect.~\ref{subsubsec:abs}). The introduction of this bound was to prevent the LHB component from `switching' with the CGM component in low column density ($N_\mathrm{H}$) regions. We refer the reader to Sect.~\ref{subsec:degen} for a detailed discussion. While this effect was mitigated by introducing the ROSAT R1 and R2 bands into the spectral fitting, this degeneracy remained in some low $N_{\rm H}$ regions and necessitated the use of a hardbound. Its emission measure (EM) was left to vary in all locations. The abundance of the LHB is assumed to be solar.

In Appendix~\ref{app:nei}, we loosen our CIE prescription of LHB to see if there is evidence of non-equilibrium ionisation (NEI) in our data. In short, we did not find a clear indication of NEI despite it being the general expectation from simulations of the ISM \citep[e.g.][]{Avillez_2012,Breitschwerdt_2021}. We attribute this to the insufficient spectral resolution to resolve the emission lines and their ratios, which are crucial diagnostics of NEI.

\subsubsection{Absorption of background components} \label{subsubsec:abs}
For the X-ray background components (CGM, COR, CXB and eRObub), we adopted the simplifying assumption that they are absorbed by the same $N_\mathrm{H}$ layer within each bin. In addition, their absorption was modelled by the \texttt{disnht} model \citep{disnht}, which behaves identically to \texttt{tbabs} \citep{tbabs} but with a lognormal distribution with mean $\log{N_\mathrm{H}}$ and width $\sigma_{\log{N_\mathrm{H}}}$. We believe this treatment is more realistic than a single $N_\mathrm{H}$ as some of our bins cover a large sky area and a range of $N_\mathrm{H}$ within the field. While $\log{N_{\rm H}}$ is left to vary during the spectral fits, we fixed the value of $\sigma_{\log{N_\mathrm{H}}}$ in each contour bin according to the following estimation.

We estimated $N_{\rm H}$ independently from our X-ray spectral measurement assuming $N_{\rm H}=N_{\rm HI}+2N_{\rm H_2}$. We adopted the $N_{\rm HI}$ information from 21\,cm line measurement by \citet{HI4PI}, while $N_{\rm H_2}$ was estimated using the conversion given in \citet{Willingale13}: 
\begin{eqnarray}
    N_{\rm H_2} = 7.2\times10^{20} \left[1-\exp{\left(-\frac{N_{\rm HI}E(B-V)}{3\times10^{20}\,{\rm cm}^{-2}}\right)}\right]^{1.1}\;,
    \label{eq:HI4PI}
\end{eqnarray}
where $E(B-V)$ was taken from the conversion from the \textit{Planck} dust radiance ($R$) map using $E(B-V)/R=5.4\times10^5$ given in \citet{PlanckXI}. We refer to this $N_{\rm H}$ estimation simply by HI4PI $N_{\rm H}$ hereafter for brevity. We then estimated $\sigma_{\log{N_\mathrm{H}}}$ as the standard deviation of HI4PI $N_{\rm H}$ within the area of each contour bin. In other words, our estimated value of $\sigma_{\log{N_\mathrm{H}}}$ accounts only for the spatial variation in the column density and can be treated as a lower limit to the `genuine' variation, which should include the line-of-sight component. 

We have indeed attempted to let $\sigma_{\log{N_\mathrm{H}}}$ freely vary in the fits. However, this choice brought about two unforeseen issues:\\
1) We found the fitted $\log{N_\mathrm{H}}$ to be higher than total HI4PI $N_{\mathrm{H}}$ by $\gtrsim0.3$\,dex, usually at high $\sigma_{\log{N_\mathrm{H}}}$ areas ($\gtrsim0.6\,$dex) near ($l$,$b$)$\sim$($300\degr$,$-20\degr$), and consequently, boosted the CXB above a level allowed by cosmic variance. \\
2) A significant number of contour bins resulted in a vanishing LHB component, which we considered unphysical, especially since their occurrences appear random.\\
On the other hand, these issues were not present when $\sigma_{\log{N_\mathrm{H}}}$ was fixed using the aforementioned method. Therefore, we kept $\sigma_{\log{N_\mathrm{H}}}$ fixed throughout our spectral analysis.

\begin{figure}
    \centering
    \includegraphics[width=0.49\textwidth]{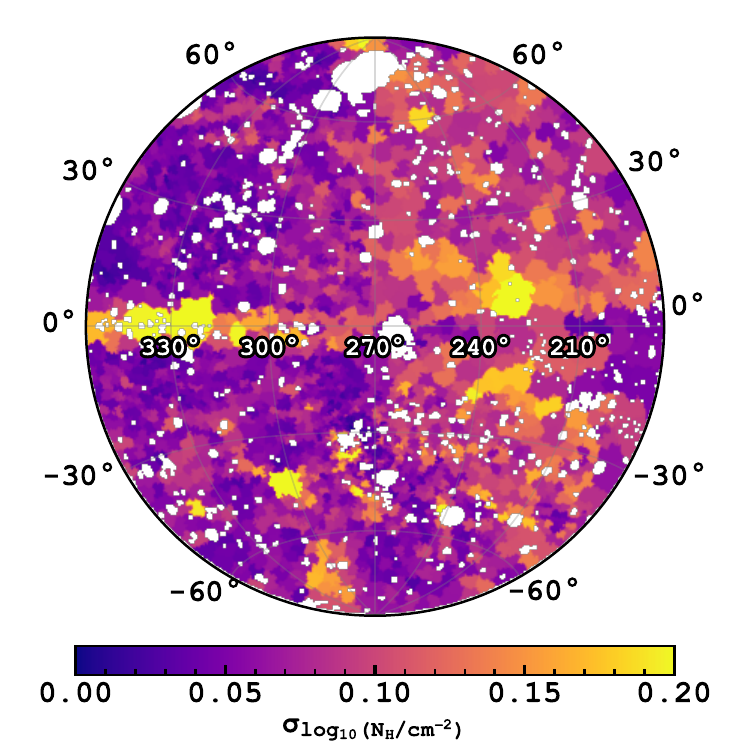}
    \caption{Values of $\sigma_{\log{N_{\rm H}}}$ used in our spectral fitting, considering only the spatial spread. They were computed using a combination of $N_{\rm HI}$ information from HI4PI and $N_{\rm H_2}$ information inferred from \textit{Planck}. (See Sect.~\ref{subsubsec:abs}.)}
    \label{fig:siglogNH}
\end{figure}

\begin{figure}
    \centering
    \includegraphics[width=0.49\textwidth]{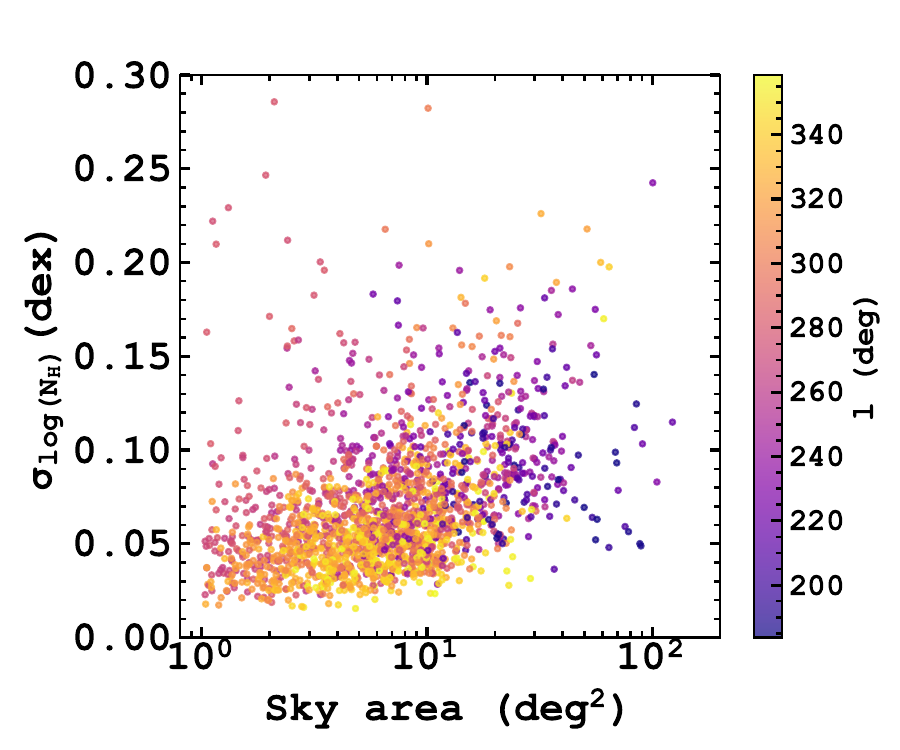}
    \caption{Scatter plot of $\sigma_{\log{N_{\rm H}}}$ against solid angle of contour bins. A positive correlation between $\sigma_{\log{N_{\rm H}}}$ and the contour bin area can be seen, as larger bins tend to capture a larger spread in $N_{\rm H}$. The points are coloured according to their Galactic longitudes.}
    \label{fig:siglogNH_area}
\end{figure}

Fig.~\ref{fig:siglogNH} shows the values of $\sigma_{\log{N_\mathrm{H}}}$ in all contour bins. A clear decreasing trend can be seen as a function of Galactic longitude. This is primarily caused by the larger contour bins away from the Galactic centre, capturing larger spatial spread in $N_{\rm H}$. The scatter plot in Fig.~\ref{fig:siglogNH_area} demonstrates this correlation. Of course, the areas of the contour bins are dictated by the S/N in the soft band, affected by both exposure time and soft band intensity in the sky, which happens to be lower between $250\degr \lesssim l \lesssim 180\degr$.

\subsubsection{Milky Way's circum-galactic medium}
The CGM component was modelled using an \texttt{apec} model with the abundance fixed at $0.1\,Z_\odot$. The low abundance was found to be required by the eFEDS data \citep{Ponti2023} and is shown to be stable against the choice of the emission models of optically thin plasma in CIE (for example, \texttt{Raymond-Smith, Mekal}) in Paper II (Ponti et al., in prep). The main impact of assuming a different abundance is reflected primarily in the emission measure. The behaviour is fairly linear; for instance, the conventional choice of $Z=0.3\,Z_\odot$ would lower the emission measure by about three times compared to $Z=0.1\,Z_\odot$.

Several recent publications have found a Galactic corona or a hotter CGM component necessary to reproduce the spectra of the SXRB \citep[e.g.][]{Gupta21,Bluem22,Ponti2023}. We refer the reader to Paper II for an in-depth discussion. In this paper, it suffices to note that this component was necessary for reproducing the spectra. We modelled the Galactic corona component using an \texttt{apec} model with solar abundance. Our choice of $\text{S/N} = 80$ at the $0.2$--$0.6$\,keV band did not usually provide sufficient constraints on the Galactic corona component in the spectral fitting. Hence, throughout this paper, we kept its temperature fixed at $kT_{\rm COR}=0.7\,$keV.

\subsubsection{Instrumental background} \label{subsubsec:FWC}
For the instrumental background, the spectrum for each TM was modelled by its own FWC model to account for any TM-specific features. The normalisation of the FWC model in the sky spectrum is related to that of the FWC spectrum by the ratio of their \texttt{BACKSCAL} header keywords and was fixed accordingly in the spectral fit. However, we noticed the \element{Al}-K$\alpha$ fluorescence line at $1.49$\,keV in the CXB regions (see Fig.~\ref{fig:CXB_reg} and Sect.~\ref{sec:CXB}) is always weaker than that in the FWC data for all TMs, by $\sim$20\%--60\% depending on the TM. The precise cause of this phenomenon is unknown. Still, it is suspected to be linked to the extra Aluminium put above the CCD when the filter wheel was rotated to the \texttt{CLOSED} position compared to the \texttt{FILTER} position. However, this explanation cannot explain the relatively large spread in the deficit between the TMs. To compensate for the Al-K$\alpha$ line deficit in the sky spectra, we singled out this line and refitted its normalisation of all TMs as the first step of all our spectral fits. It was then frozen in the subsequent optimisation of the rest of the free parameters.

\subsubsection{Summary of model parameters} \label{subsubsec:mod_param}
We elaborate on the description of the CXB in Sect.~\ref{sec:CXB}. In summary, two parametrisations (single ($\Gamma=1.7$) and broken power-law ($\Gamma_1=1.9$, $\Gamma_2=1.6$, $E_b=1.2\,$keV)) of the CXB component could reproduce the data equally well at high Galactic latitudes. Therefore, both were used in the spectral fits, and the differences between the two were considered to be systematic uncertainties.

% We have a special treatment for contour bins within the eROSITA bubbles, which will be described in Sect.~\ref{subsec:erobub_method}.

There are a total of 7 or 9 free parameters in the spectral fits: $kT_{\mathrm{LHB}}$, $\mathrm{EM}_{\mathrm{LHB}}$, $kT_{\mathrm{CGM}}$, $\mathrm{EM}_{\mathrm{CGM}}$, $\mathrm{EM}_{\mathrm{COR}}$, $\mathrm{norm}_{\mathrm{CXB}}$ and $\log{N_\mathrm{H}}$ for contour bins outside the eROSITA bubbles; and an addition of $kT_{\rm eRObub}$ and $\mathrm{EM}_{\rm eRObub}$ for bins inside the eROSITA bubbles.

%\textcolor{red}{Two things need to be added, on the topic of progressively more complicated models:
%1) We start with a fixed $kT_{\rm{LHB}}$ fit, 
%2) fixed template on half sky, leaving $kT_{\rm{LHB}}$ free.
%3) We need to add the description of the eROSITA bubble treatment here, either as a standalone subsection, or part of this subsection. We need to specify the priors on kT used and show the regions that we define as the eROSITA bubbles region.}

\subsection{Treatment of the cosmic X-ray background} \label{sec:CXB}
In our spectral analysis, detected sources with $F_{0.5\mathrm{-}2\,\mathrm{keV}} > 10^{-12}$\,erg\,s$^{-1}$\,cm$^{-2}$ as well as known clusters from X-ray cluster catalogues with $R_{500}\gtrsim 3\arcmin$ were masked \citep[][and references therein]{Liu_2022, DR1,Bulbul24}. The choice of the flux threshold is more than an order of magnitude higher than the eRASS1 flux limit. With this limit, we can assume eROSITA is complete in detecting sources above this threshold in all look directions, and the corresponding resolved fraction across the western Galactic hemisphere is uniform.
Masking of sources is expected to change the photon index of the cosmic X-ray background (CXB) from the canonical value of $\Gamma \sim 1.4$--$1.5$ \citep{Vecchi,Kushino,Hickox,Cappelluti17}; however, the high flux threshold guarantees this change is not spatially dependent.

\begin{figure}
    \centering
    \includegraphics[width=0.49\textwidth]{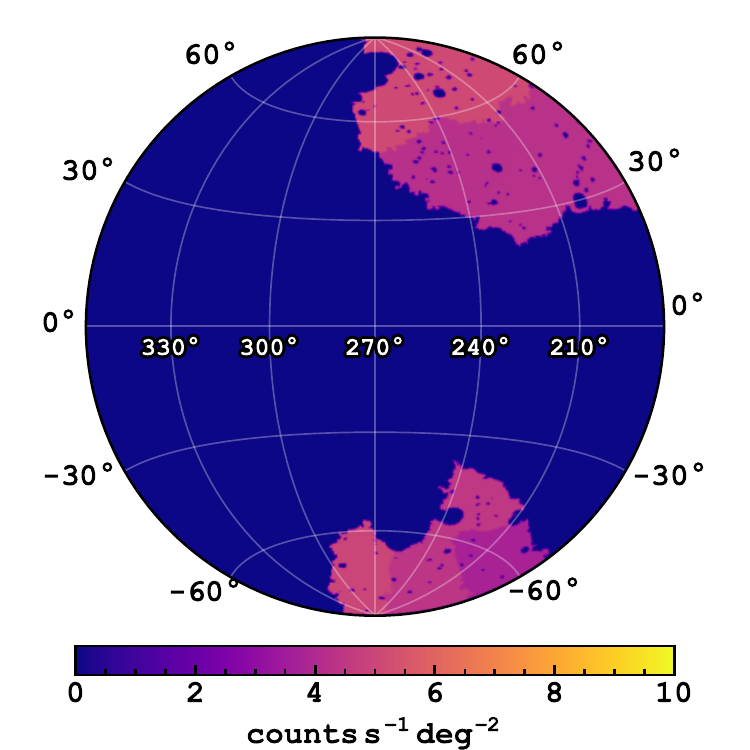}
    \caption{Similar to Fig.~\ref{fig:contbin} but showing only regions selected to determine the spectral shape of the CXB.}
    \label{fig:CXB_reg}
\end{figure}

Fig.~\ref{fig:CXB_reg} shows the regions we chose to determine the CXB model. These regions correspond to the spatial bins with centres located above $|b|>30\degr$, and are free of large-scale foreground structures upon binning the western Galactic hemisphere using \texttt{contbin} \citep{contbin} with a target S/N of 400.
In addition to the CXB model we focus on, we fitted the spectrum of each region independently with freely varying LHB, CGM, Galactic corona and absorption column densities similar to the description in Sect.~\ref{sec:comp}, with the omission of SWCX since the contribution of SWCX in eRASS1 is low (Dennerl et al., in prep).

From the spectral analysis of the regions shown in Fig.~\ref{fig:CXB_reg}, we found that simple power-law and broken power-law models perform equally well in reproducing the data, but a double broken power-law was unnecessary. Therefore, both simple and broken power-laws were adopted as our CXB models. We delay the details of this CXB analysis to Sect.~\ref{subsec:CXB_slope}.

\subsection{Fitting procedures} \label{sec:fit}
Spectral fitting was done in \texttt{PyXspec} version 2.1.0 \citep{xspec,pyxspec}. \citet{Lodders} was used as the reference for abundance, and absorption cross-sections from \citet{Verner} were assumed.

We repeated the following fitting procedure for all the 2010 contour bins. After determining the normalisation of the Al-K$\alpha$ line (Sect.~\ref{sec:comp}), we made a simultaneous fit of 5 eROSITA spectra (5 TMs) and 2 ROSAT band fluxes (R1 and R2) for each contour bin. The eROSITA spectra were fitted in the energy range $0.2$--$5$\,keV using Poissonian statistics (\texttt{cstat}) \citep{Cash1979}, and as the R1 and R2 data points were fluxes, $\chi^2$-statistics was used. We minimised the total statistics using the Levenberg-Marquardt algorithm. Then, we ran a MCMC with 64 walkers, each with 1.6\,ksteps, giving a total of 102.4\,ksteps, using the Goodman-Weare method \citep{GW}. The walkers were initially distributed in a Gaussian distribution around the best fit from the minimisation. We found discarding the initial 80\,ksteps, albeit aggressive, was a conservative and uniform way to ensure all parameters had moved past the burn-in phase and reached convergence in the vast majority of bins. In this paper, we report the median, 16$^\mathrm{th}$ and 84$^{\mathrm{th}}$ percentiles of the posterior distribution as the most probable value, the lower and the upper errors, respectively. If the posterior distribution is Gaussian, these reduce to the mean and $\pm1\,\sigma$ errors.

\subsection{Treatment of the eROSITA bubbles} \label{subsec:erobub_method}
%\textcolor{red}{Description of the treatment of the eROSITA bubble regions. We define the bins within the bubble by the shape increase in the surface brightness on the $0.6$--$1.0\,$keV band map. Refer the readers to the boundary defined at a later figure showing eROISTA bubbles parameters. A solar abundance apec model with temperature and emission measure free was used. Describe the latitude-dependent priors used for the other sky components except from ${\rm EM_{LHB}}$.}

The enigmatic eROSITA bubbles \citep{Predehl20} emit thermally in the soft X-rays at $\sim0.3\,$keV, and we modelled it using the \texttt{apec} model at solar abundance\footnote{We assumed solar abundance for simplicity, following \citet{Lallement_2016}. Indeed, there are reports of sub-solar abundances both from observations \citep[e.g.][]{Miller_2005, Kataoka_2013, Gu_2016} and as expectations from simulations \citep[e.g.][]{Mou_2023}. We defer the abundance measurement in eROSITA to an ongoing work (Yeung et al., in prep).}. Combining with the LHB and CGM, the multi-component spectral analysis in eRASS exposure depth and spectral resolution inescapably entails model degeneracies because of their similar spectral shapes (0.1, 0.2 and 0.3\,keV plasma in CIE). As such, we implemented a two-step approach in fitting spectra within the eROSITA bubbles region, where some spectral parameters outside the eROSITA bubbles were passed in the form of Gaussian priors into spectral fits within the eROSITA bubbles.

\begin{figure}
    \centering
    \includegraphics[width=0.49\textwidth]{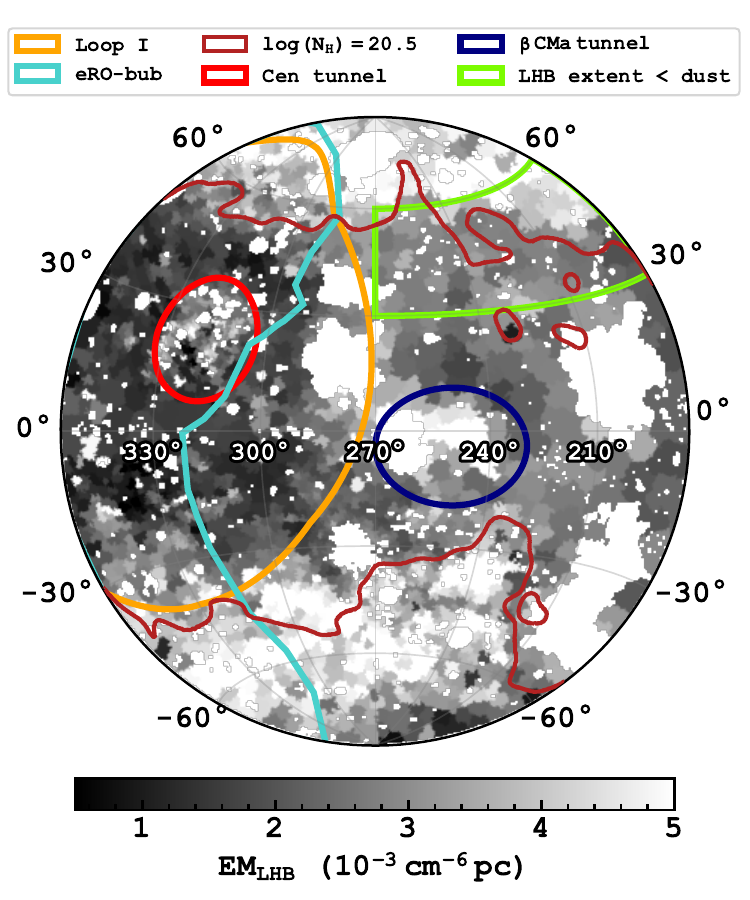}
    \caption{Finding chart for features discussed mainly in Sect.~\ref{subsec:erobub_method} and \ref{sec:EM_LHB}. The background image shows the emission measure of the LHB, which is a proxy of the extent of the LHB and is relevant for discussions (Sect.~\ref{sec:EM_LHB}) on interstellar tunnels and anti-correlation with dust. The demarcation of the eROSITA bubbles is shown by the turquoise line, based on the 0.6--1.0\,keV intensity map.}
    \label{fig:chart}
\end{figure}

We began by defining the demarcation of the eROSITA bubbles using the 0.6--1.0\,keV intensity map, where the eROSITA bubbles are the most prominent.
The turquoise line in Fig.~\ref{fig:chart} shows the demarcation.
% \footnote{The background image in Fig.~\ref{fig:chart} is not the 0.6--1.0\,keV intensity map.}.
% This demarcation matches well with the border of the eROSITA bubbles inferred from cross-matching the eROSITA and the SPASS polarised intensity maps \citep{Teng_2024}.
We fitted the spectra of the contour bins outside the eROSITA bubbles before those inside, following the procedure described in Sect.~\ref{sec:fit}. Subsequently, we fitted the contour bins within the eROSITA bubbles by passing Galactic latitude-dependent priors based on the fit results. More precisely, for a given contour bin within the eROSITA bubbles centred at ($l$,$b$), we created a Gaussian prior for each relevant parameter based on fitting results of all the bins outside the eROSITA bubbles centred within the range of $b\pm5\degr$. This implementation reflects our recurring observation with the eROSITA data that most spectral parameters are either constant (for example, the CXB) or exhibit primarily Galactic latitudinal dependence (for example, $kT_{\rm LHB}$, see Fig.~\ref{fig:LHB_kT_data}). These priors were only applied on 
$kT_{\rm LHB}$, $kT_{\rm CGM}$, $\mathrm{EM}_{\rm CGM}$, $\mathrm{EM}_{\rm COR}$, ${\rm norm_{CXB}}$.
We did not impose Galactic latitude-dependent priors on $\mathrm{EM}_{\rm LHB}$, $kT_{\rm eRObub}$, $\mathrm{EM}_{\rm eRObub}$ because in this paper, we primarily focus on inferring the 3-dimensional (3D) structure of the LHB. The eROSITA bubbles will be the focus of a forthcoming paper in the series (Yeung et al., in prep).

\subsection{Limitations of the current method}
At the end of this section, we would like to emphasise the data obtained by eROSITA is rich, especially for diffuse emission. A weakness of our method is that the inference of spectral parameters was independent of locations. However, some parameters, such as the LHB temperature, are likely correlated for bins in proximity based on the intuitive consideration that the local ISM would preferentially be in similar conditions the closer they are in space. There exist some potential methods that could take this into account. For instance, one idea is to use Bayesian hierarchical modelling and treat the proximity of bins as a hyperprior. Another potentially interesting method is to utilise an algorithm called the generalised morphological components analysis (GMCA) or its variants (sGMCA, pGMCA) to decompose the spectral components with the help of each component's intrinsically unique spatial distribution, which has recently been demonstrated to disentangle spectral components remarkably well in extended X-ray sources \citep{GMCA, Picquenot19, Picquenot21, Picquenot23, sGMCA}. 

\section{Results and discussion} \label{sec:result}
The sheer number of spectra we analysed prevents us from discussing them individually. Therefore, we provide a webpage hosted on the eROSITA Data Release 1 server that provides all the relevant information, including visualisations and the fitting results organised in downloadable tables, for readers who are interested in the spectra, their associated model parameters, uncertainties or fit quality of any our contour bins (Sect.~\ref{sec:website}).

\subsection{Evidence of a variable LHB temperature} \label{sec:LHB_kT}
%\textcolor{red}{Presentation of the fixed $kT_{\rm LHB}$ model. The content of this part remains uncertain because while the 'normal' version prefers a north-south gradient on $kT_{\rm LHB}$, what statistical tool (fit-stat/dof, or AIC, BIC) should we use to say a variable temperature model is better?
%State that we avoid the eROSITA bubbles region because we lack constraints there with a variable $kT_{\rm LHB}$ model. Present a map of the fitted temperatures without the eROSITA bubble region. Could consider plotting a $kT_{\rm LHB}$ vs $\sin{b}$ plot to quantify the gradient while taking into account the error.
%Contrast with the \citet{Snowden1990} finding using Wisconsin B, C band. Discuss possible reasons for the apparent discrepancy. For instance, B, C bands are low-energy filters, not sensitive above the carbon edge. B/C does not take into account of the eROSITA and Loop~I bubbles; as rightly put by the authors in 1990, B/C ratio cannot be directly translated into a $kT_{\rm LHB}$.}

\begin{figure}
    \centering
    \includegraphics[width=0.5\textwidth]{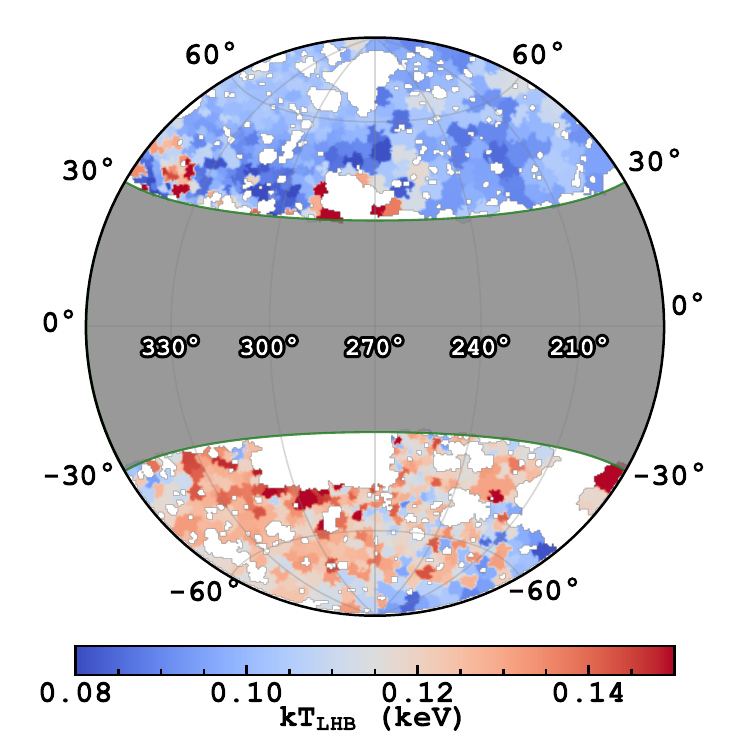}
    \caption{Spatial distribution of $kT_{\rm LHB}$ in the high latitude regions ($|b|>30\degr$).}
    \label{fig:LHB_kT_data}
\end{figure}

Fig.~\ref{fig:LHB_kT_data} shows the spatial distribution of $kT_{\rm LHB}$ in the high latitude regions, where the Galactic plane ($|b|<30\degr$) and regions overlapping with the Large Magellanic Cloud, known supernova remnants or superbubbles (Antlia, Orion-Eridanus, Monogem Ring, Vela; masked also in other parameter maps) and the recently discovered structure surrounding the LMC, dubbed the `Goat Horn Complex' \citep{Goat_horn}, were ignored. In addition, regions with $1\,\sigma$ LHB temperature or emission measure fitting uncertainty $\sigma_{kT_{\rm LHB}} < 5\times10^{-3}\,$keV or ${\sigma_{\rm EM_{LHB}} < 5\times10^{-5}\,{\rm cm^{-6}\,pc}}$ were removed. These criteria are possible indications of sub-optimal spectral fits; the former is to avoid regions with $kT_{\rm LHB}$ pegged at edges of the uniform prior, and the latter is useful for removing regions with ${\rm EM_{LHB}}$ pegged close to zero. A total of 788 valid bins remained following the screening (36 bins were screened out).

\begin{figure}
    \centering
    \includegraphics[width=0.5\textwidth]{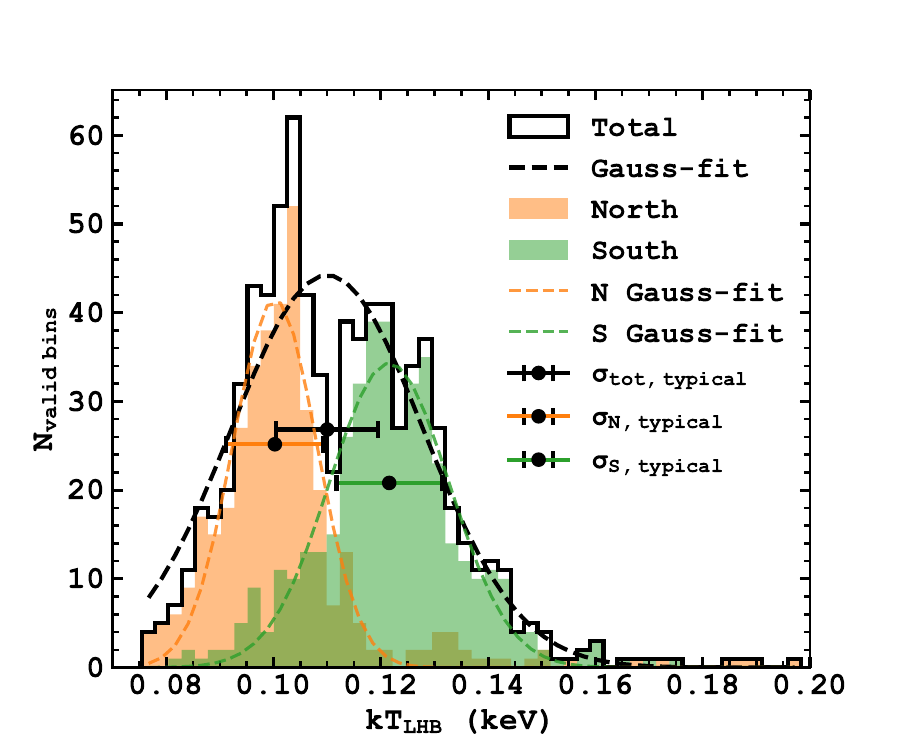}
    \caption{Distribution of $kT_{\rm LHB}$ in the high latitude regions ($|b|  > 30\degr$). The unfilled histogram outlined in black displays the distribution of all the high latitude bins displayed in Fig.~\ref{fig:LHB_kT_data}. The black dashed line shows the Gaussian best fit for the distribution. The black data point with error bars shows the typical (median) $1\,\sigma$ spectral fitting uncertainty, centred at the mean of the fitted Gaussian. It is plotted at a height of $\exp{(-1/2)}N_{\rm gauss, peak}=0.606\,N_{\rm gauss, peak}$ for proper comparison between the spectral fitting uncertainty and the width of the distribution. The former ($0.010$\,keV) is approximately half of the latter ($0.018$\,keV), demonstrating a genuine variation in the LHB temperature. The sample is divided into the northern (orange) and the southern (green)  Galactic hemispheres to demonstrate the primary source of variation is a large-scale gradient instead of small-scale fluctuations. The northern hemisphere is cooler ($0.100\,$keV) than the south ($0.122\,$keV), and comparing their distributions with the respective typical spectral fitting uncertainties shows that each hemisphere exhibits approximately constant LHB temperature.}
    \label{fig:LHB_kT_hist}
\end{figure}

The histogram of $kT_{\rm LHB}$ of all the valid bins is shown in the black line in Fig.\,\ref{fig:LHB_kT_hist}. We report a median (and the 0.16 and 0.84 quantiles) $kT_{\rm LHB}$ of $0.111_{-0.015}^{+0.018}$\,keV. The observation of the mean $1\,\sigma$ fitting uncertainty ($0.010\,$keV), as shown in the black error bar, being significantly smaller than the $\sigma$ of the distribution ($0.018\,$keV) indicates a variable $kT_{\rm LHB}$. An inspection of Fig.~\ref{fig:LHB_kT_data} shows the primary origin of this variation is a large-scale temperature gradient, with the northern Galactic hemisphere being cooler than the south by $\sim 0.02\,$keV. This is further demonstrated by dividing the histogram in Fig.~\ref{fig:LHB_kT_hist} into the north (orange) and south (green). The north-south temperature dichotomy is evident. By overlaying the typical spectral fitting uncertainties within each hemisphere on the figure, one can notice the width of the temperature distributions within both hemispheres could be attributed mainly to their respective spectral fitting uncertainties, especially in the north. Hence, the dominating factor of the spread in the LHB temperature is a large-scale temperature gradient, but not bin-to-bin fluctuations. To complement the systematic change in $kT_{\rm LHB}$ observed from the projected map and histogram, Appendix~\ref{app:highSN_spec} highlights the spectral signature that determines the LHB temperature using high-S/N spectra from each hemisphere.

We tested if the northern and southern temperature distributions could be drawn from the same underlying distribution using the two-sample Kolmogorov-Smirnov test. The test returned a statistic of 0.693 and a p-value in the order of $10^{-102}\ll 0.01$. Thus, we can safely reject the null hypothesis that the temperature dichotomy occurs by chance.

% We did a more detailed analysis using the maximum likelihood estimation method, with the main goal of quantifying our ability to constrain the mean temperatures and widths of the intrinsic temperature distributions while considering each region's fitting uncertainties.
To quantify the intrinsic temperature distributions in each hemisphere, we used a maximum likelihood approach to identify the mean and width of the Gaussian distributions, which best reproduce the observed temperature measurements, including their statistical errors.
% An assumption that went into this analysis is that the intrinsic temperature in each hemisphere follows a Gaussian distribution, and so is the fitting uncertainty of each region.
We constrained the mean temperature in the northern and southern Galactic hemispheres to be $kT_{\rm N}=100.8\pm0.5$\,eV and $kT_{\rm S}=121.8\pm0.6$\,eV, respectively. We also extracted the widths ($1\,\sigma$) of the intrinsic temperature distributions, which are $\sigma_{\rm N}=2.9^{+1.0}_{-1.3}\,$eV in the north and $\sigma_{\rm S}=8.5^{+0.7}_{-0.6}$\,eV in the south. This result reiterates the temperature dichotomy is highly significant, as is evident from the precision with which we can determine the mean temperature of each hemisphere. It is also clear that the southern hemisphere exhibits a larger intrinsic temperature scatter than the northern counterpart. We would like to emphasise that the statistical uncertainty of the means and the widths of the intrinsic temperature distributions should not be confused with the fitting uncertainty of each bin. The temperature dichotomy is highly significant because of the large number of bins that sampled each hemispheric distribution well despite the individual bin having a median fitting uncertainty ($\simeq 0.01\,$keV; see Fig.~\ref{fig:LHB_kT_hist}) larger than the intrinsic widths of the distributions.

\begin{figure}
    \centering
    \includegraphics[width=.5\textwidth]{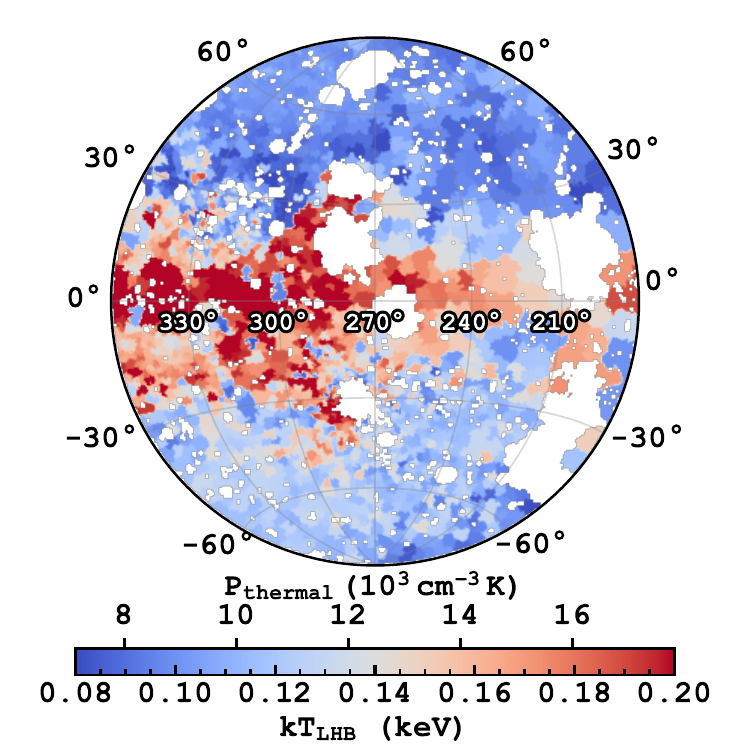}
    \caption{Map of $kT_{\rm LHB}$ including low latitude regions. We note that the colour bar is scaled differently from Fig.~\ref{fig:LHB_kT_data}. The thermal pressure is also shown under the assumption of constant $n_e=4\times10^{-3}\,{\rm cm^{-3}}$.}
    \label{fig:LHB_kT_all}
\end{figure}

\begin{figure*}
    \centering
    \includegraphics[width=0.49\textwidth]{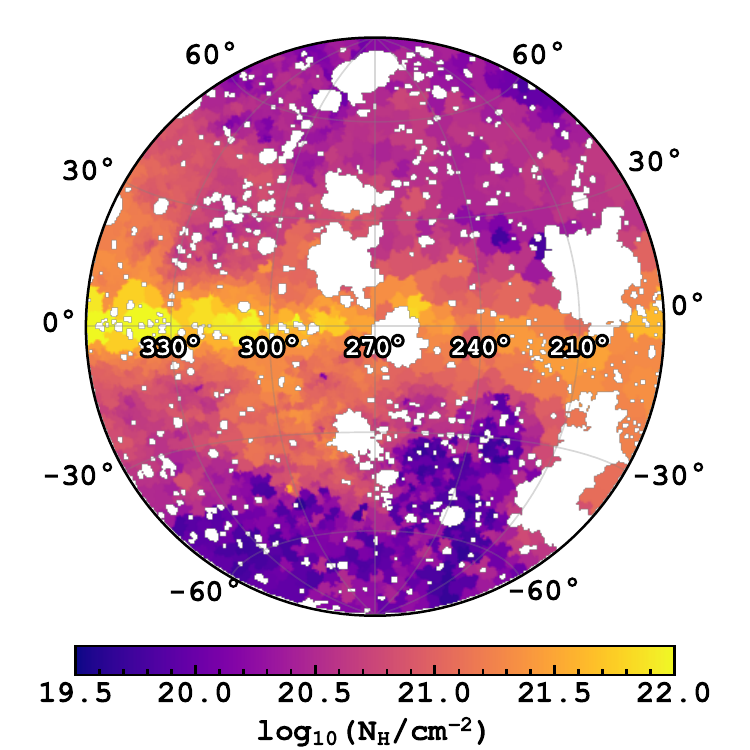}
    \includegraphics[width=0.49\textwidth]{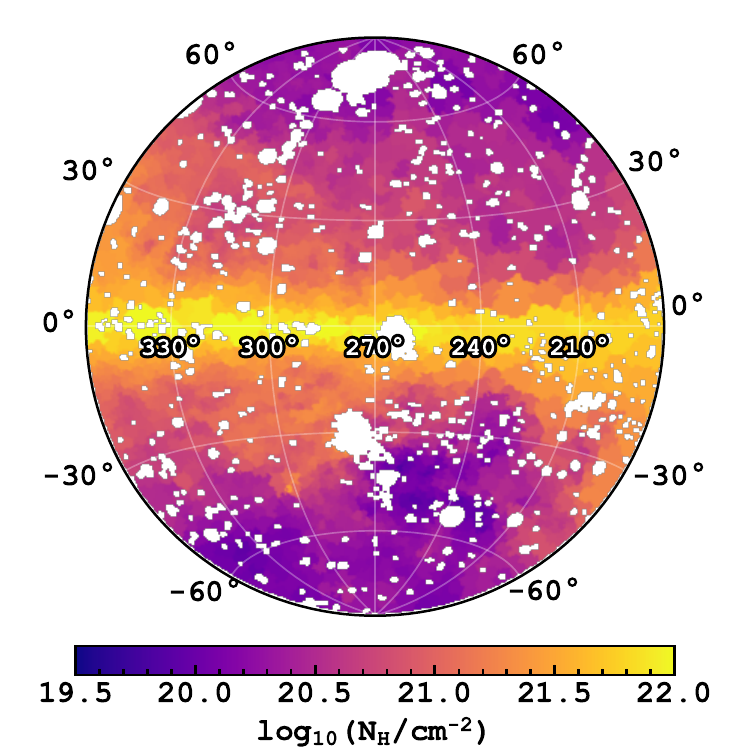}
    \caption{Comparison of fitted $N_{\rm H}$ and independent estimate of $N_{\rm H}$ from \ion{H}{I} and dust measurements. \textit{Left}: Fitted $N_{\rm H}$. \textit{Right}: Estimate of $N_{\rm H}$ combining neutral hydrogen information from HI4PI \citep{HI4PI} and $E(B-V)$ information derived from Planck radiance map \citep{PlanckXI}.}
    \label{fig:logNH}
\end{figure*}

Having established a temperature dichotomy at high Galactic latitudes, it is natural to ask if a smooth transition across the Galactic plane connects them. We separated the discussion of high latitude regions since we believe our measurement there is relatively secure. Still, it is unclear how close to the Galactic plane one can venture before one is heavily biased by complexities such as multiple line-of-sight emitting and absorbing components. We relax the $|b|>30\degr$ limit and plot all the valid bins, including those on the Galactic plane in Fig.~\ref{fig:LHB_kT_all}, using the same screening applied to Fig.~\ref{fig:LHB_kT_data}. Fig.~\ref{fig:LHB_kT_all} shows a remarkable temperature enhancement towards the Galactic plane, especially at $l\gtrsim270\degr$. Inspection of the spectra and their posterior distributions suggests the spectra have significant constraining power on $kT_{\rm LHB}$ down to at least $10\degr$ of the Galactic plane, albeit a subjective choice. The column densities in these regions ($\gtrsim10^{21}\,{\rm cm^{-2}}$; see Fig.~\ref{fig:logNH} for $N_{\rm H}$ information) are adequate and peak strongly at the first wall of LB absorption ($\sim\!100-200$\,pc) \citep{Lallement22,Edenhofer23}, enabling us to differentiate between the unabsorbed LHB and the absorbed components, primarily the CGM. Therefore, we argue the temperature enhancement towards the inner Galaxy is a real feature of the LHB.

\begin{figure}
    \centering
    \includegraphics[width=0.5\textwidth]{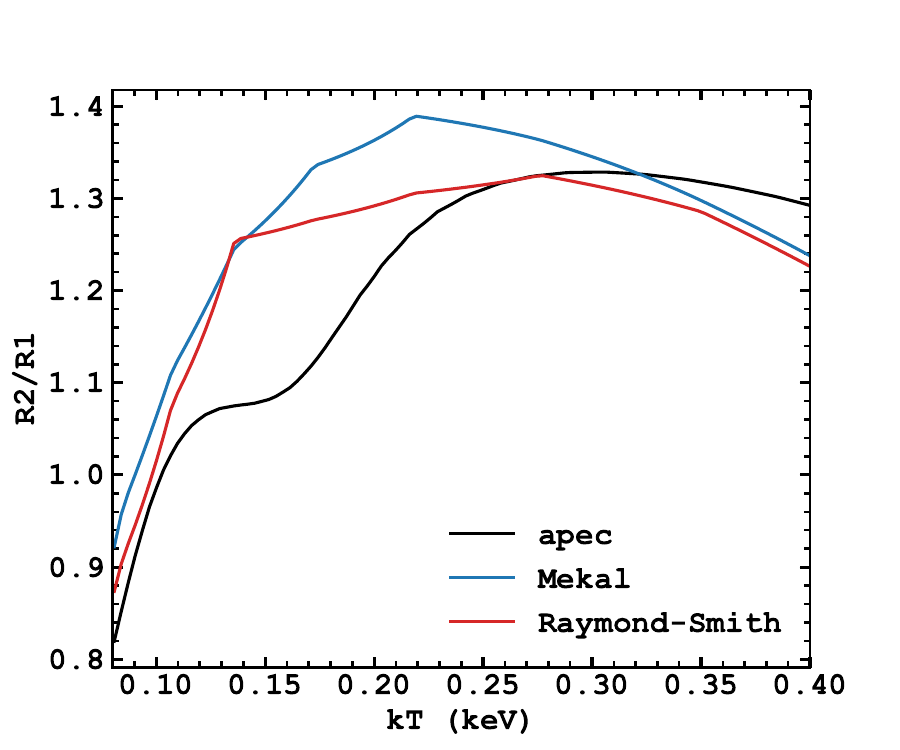}
    \caption{Calibration curves of R2/R1 band ratio of the \texttt{Raymond-Smith} \citep{Raymond_1977}, \texttt{Mekal} \citep{Mewe_1985,Mewe_1986,Liedahl_1995}, and \texttt{apec} \citep{apec} models as a function of the temperature of an unabsorbed plasma at solar abundance.}
    \label{fig:CIE_R2R1}
\end{figure}

First, we must address why earlier studies did not observe the north-south LHB temperature gradient using ROSAT data (see Sect.~\ref{subsec:past_obs} for a deeper literature discussion). The R2/R1 band ratio is the main tracer of LHB temperature using ROSAT data, commonly assumed to be only contributed by the LHB emission. Fig.~\ref{fig:CIE_R2R1} shows the R2/R1 band ratio as a function of plasma temperature at solar abundance, using three common plasma models, \texttt{Raymond-Smith} \citep{Raymond_1977}, \texttt{Mekal} \citep{Mewe_1985,Mewe_1986,Liedahl_1995} and \texttt{apec} \citep{apec}. The \texttt{Raymond-Smith} model was commonly used by studies before and around the millennium \citep[e.g.][]{Snowden1990,Snowden97,Snowden98} to infer the temperature of the LHB. The calibration curve of the \texttt{Raymond-Smith} model in Fig.~\ref{fig:CIE_R2R1} shows the R2/R1 ratio is insensitive to temperatures above $0.13\,$keV. This could be one of the reasons why the LHB temperature was not found to be $\sim 0.2\,$keV in earlier LHB temperature measurements using ROSAT.

\begin{figure*}
    \centering
    \includegraphics[width=.49\textwidth]{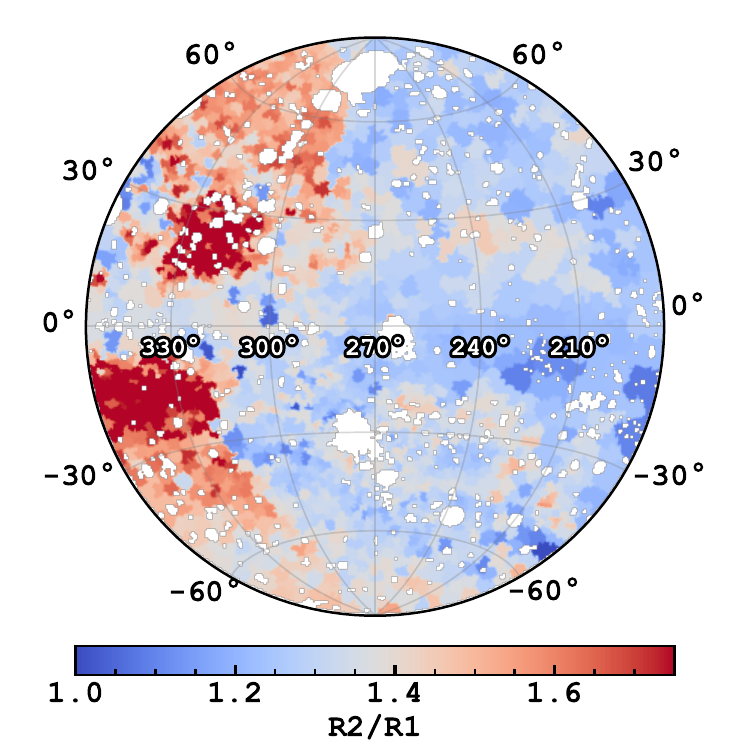}
    \includegraphics[width=.49\textwidth]{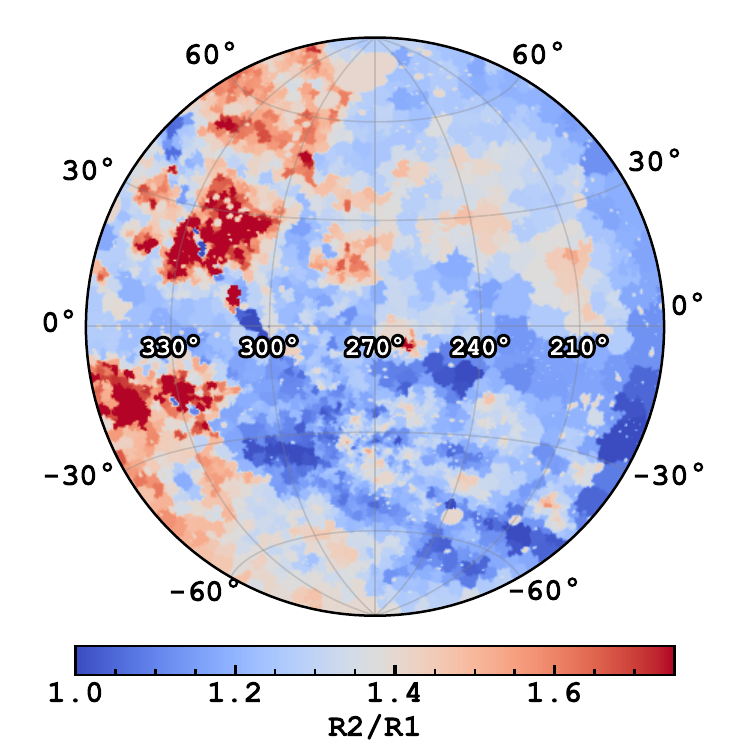}
    \caption{Comparison of our forward-modelled R2/R1 band ratio map with the observation. \textit{Left}: R2/R1 band ratio map calculated by folding our best-fit (median) spectral models with the ROSAT R1 and R2 band responses. \textit{Right}: R2/R1 data binned to the same contour-binning scheme \citep{Snowden97}.}
    \label{fig:R2/R1}
\end{figure*}

Another possibly more important reason is the flawed assumption that only the LHB emission contributes to the R1 and R2 bands. To demonstrate this, we forward-modelled the same models used to extract Fig.~\ref{fig:LHB_kT_all}, using the R1 and R2 band responses. The resulting R2/R1 band ratio map is shown in the left panel of Fig.~\ref{fig:R2/R1}. The eROSITA bubbles appear distinctly on the map, indicating that absorbed background structures can shine through and alter the R2/R1 ratio. The north-south LHB temperature dichotomy also disappears. It is fair to suggest the forward-modelled R2/R1 map does not capture the primary morphological features of the $kT_{\rm LHB}$ map, and conversely, $kT_{\rm LHB}$ directly inferred from R2/R1 without contributions from the background components could be biased. Furthermore, regions within the eROSITA bubbles in the forward-modelled map commonly show ${\rm R2/R1} > 1.4$, a limit according to the calibration curves in Fig.~\ref{fig:CIE_R2R1} that should not be crossed for an unabsorbed plasma. This is another evidence of the absorbed components contributing to the R1 and R2 bands, boosting the R2/R1 ratio. The right panel of Fig.~\ref{fig:R2/R1} displays the binned R2/R1 ratio of the original ROSAT maps presented in \citet{Snowden97}, which can be directly compared to our forward-modelled map. Their resemblance reiterates that our $kT_{\rm LHB}$ measurement is not at odds with the ROSAT data, and including background components in the modelling is essential to extract information on the LHB.

\begin{figure}
    \centering
    \includegraphics[width=0.5\textwidth]{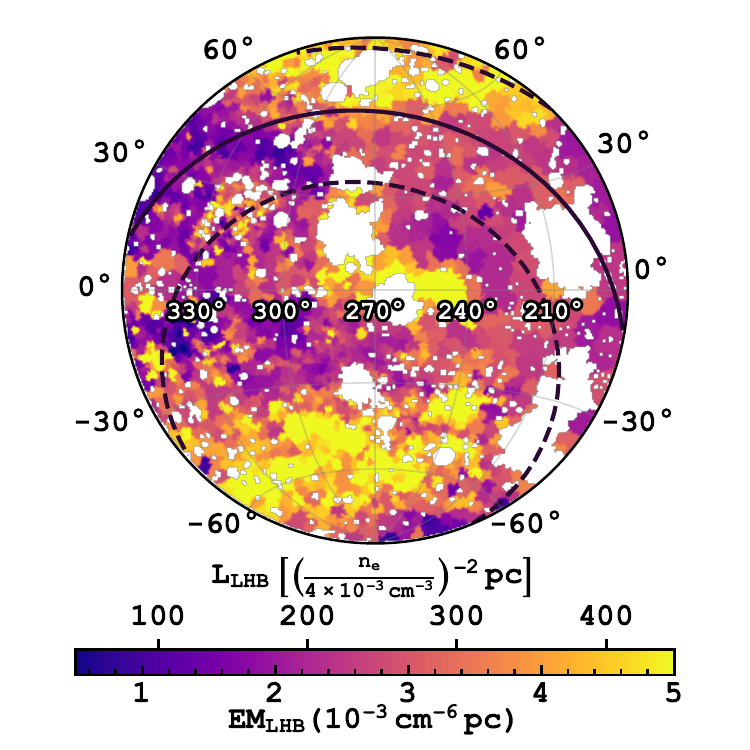}
    \caption{Spatial distribution of $\mathrm{EM}_{\rm LHB}$. Regions with $\mathrm{EM}_{\rm LHB}$ uncertainty $<5\times10^{-5}\,$cm$^{-6}$\,pc were also masked. The solid black line indicates the position of the ecliptic, and the two dashed lines represent a range of $\pm25\degr$ around it in ecliptic latitude where the solar wind density is expected to be high. The extent of the LHB under the assumption of $n_e=4\times10^{-3}\,{\rm cm^{-3}}$ is also shown.}
    \label{fig:LHB_EM}
\end{figure}

\begin{figure*}
    \centering
    \includegraphics[width=0.49\textwidth]{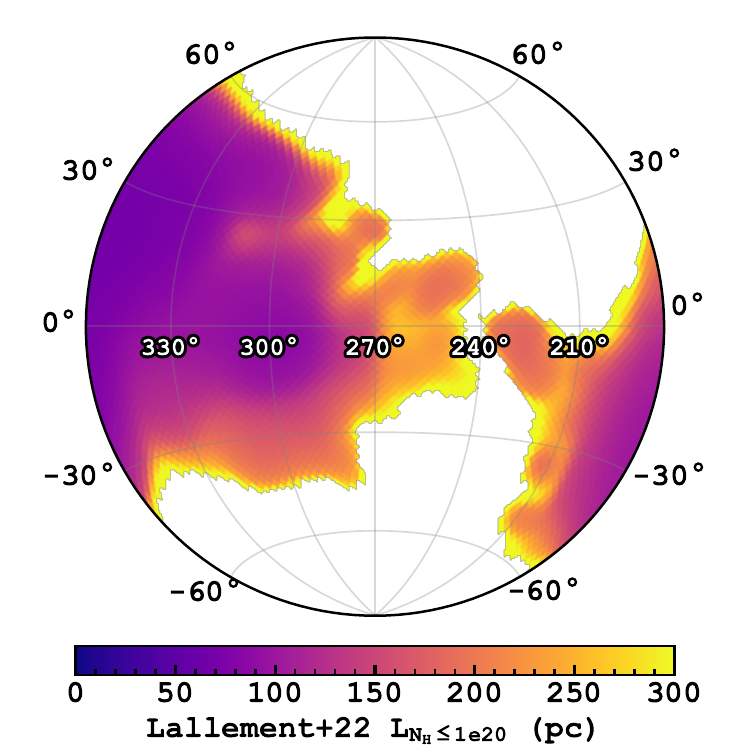}
    \includegraphics[width=0.49\textwidth]{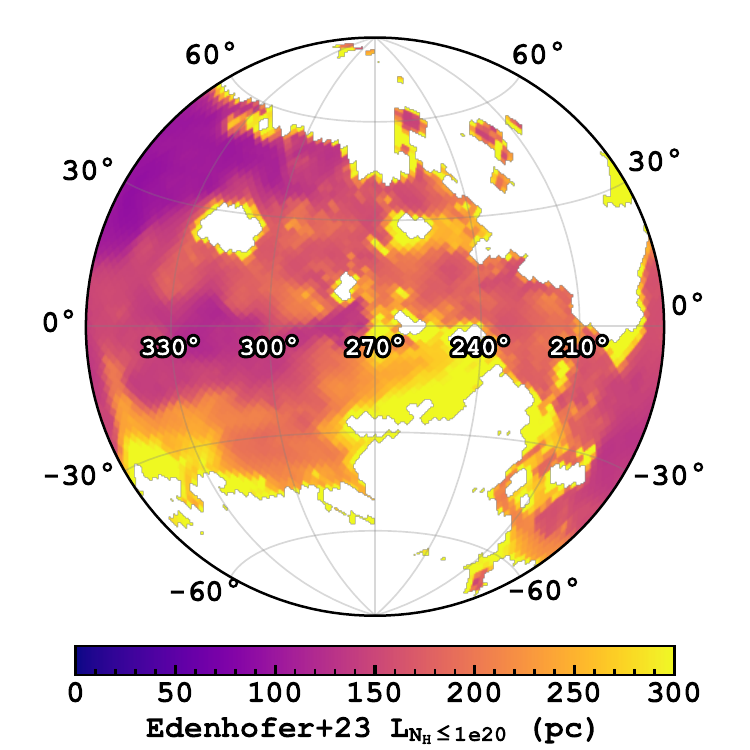}
    \caption{Distance at which the integration of $N_{\rm H}$ reaches $10^{20}\,{\rm cm}^{-2}$ in the \citet{Lallement22} (\textit{left}) and \citet{Edenhofer23} (\textit{right}) dust cubes, as proxies of the extent of the local bubble. The empty regions in the map indicate that the integration does not reach $N_{\rm H} \geq 10^{20}\,{\rm cm}^{-2}$ before $400\,$pc.
    }
    \label{fig:dustmap}
\end{figure*}

\begin{figure}
    \centering
    \includegraphics[width=0.49\textwidth]{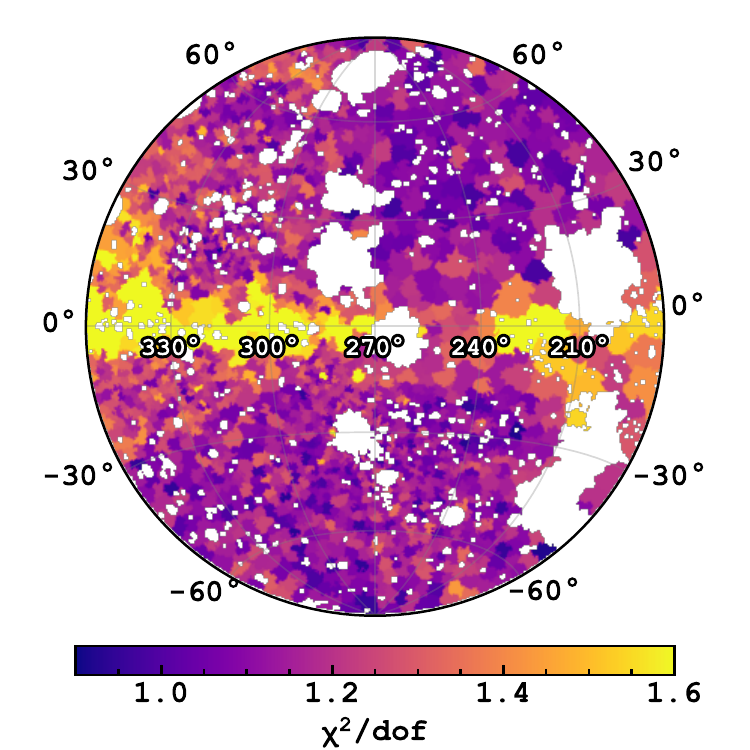}
    \caption{Map of reduced-$\chi^2$ ($\chi^2/{\rm dof}$) after rebinning each spectrum, imposing that each spectral bin receives at least ten counts.}
    \label{fig:fitstat}
\end{figure}

Merely showing our measurement is consistent with ROSAT does not necessarily mean it is robust, especially towards the inner Galaxy at low latitudes. We want to show that the LHB emission we measured at low latitudes must be local and not heavily contaminated by background ISM. Fig.~\ref{fig:LHB_EM} shows the corresponding ${\rm EM_{LHB}}$ map. The details of it are discussed in Sect.~\ref{sec:EM_LHB}, but for the current purpose, it suffices to note the ${\rm EM}$ at low latitudes within the Loop~I contour in Fig.~\ref{fig:chart} is one of the lowest in this hemisphere. Our LHB component is unlikely to capture extra emission from the background ISM and returns a lower ${\rm EM}$. Fig.~\ref{fig:dustmap} demonstrates the column density reaches $10^{20}\,{\rm cm^{-2}}$ at a very close distance ($\sim\!100\,$pc) and inspection of the local dust radial profiles suggests most of the total $N_{\rm H}$ is located within the first $250\,$pc, corroborating that very few background ISM photons can contaminate our LHB measurement. Indeed, because of the lower ${\rm EM_{\rm LHB}}$, the uncertainties of the $kT_{\rm LHB}$ is larger ($\sigma_{kT_{\rm LHB}} \sim 0.03\,$keV), which is a well-known anti-correlation that can also be readily observed from their posterior distributions.
These regions, excluding the ones that are exactly lying on the Galactic plane, do not show poorer fit statistics than the rest of the sky (see Sect.~\ref{sec:robust} and Fig.~\ref{fig:fitstat}). Even if they do, they are mostly due to larger residuals above $\gtrsim0.6$\,keV, and not at energies where the LHB dominates. Therefore, we conservatively consider our inferred LHB temperature credible at $|b|>10\degr$ and characterise it using spherical harmonics in the next Section.

\subsubsection{Spherical harmonic analysis} \label{subsec:sph_harm}
In Figs.~\ref{fig:LHB_kT_data} and \ref{fig:LHB_kT_all}, an apparent gradient or dichotomy in approximately the north-south direction and an enhancement towards the Galactic plane can be observed, respectively.
One way to quantify a gradient in the sky is by fitting a combination of spherical harmonics to the data given the uncertainties, and the simplest case of a gradient is a dipole.
We decided to separate the analysis of the high latitude regions ($|b|>30\degr$) from the `full sample' down to $|b|>10\degr$, because we want to tackle two different issues: 1) quantify the significance of the (high latitude) north-south gradient as modelled by a dipole, and 2) produce an empirical multipole model that captures the main $kT_{\rm LHB}$ profile.

The significance of the high latitude north-south gradient can be evaluated by independently fitting a monopole and dipole model to the data. Then, the F-test could determine if the improvement of using a dipole is statistically significant over the null hypothesis of a uniform $kT_{\rm LHB}$ model (monopole).

The expansion of any well-behaved functions (in our case, the LHB temperature)  on a sphere can be expanded into spherical harmonics $Y_{lm}$ up to degree $l_{\rm max}$ by
\begin{eqnarray}
    kT_{\rm LHB}(\theta,\phi) = \sum_{l=0}^{l_{\rm max}}\sum_{m=-l}^{m=+l}a_{lm}Y_{lm},
\end{eqnarray}
where $a_{lm}$ is the complex coefficient associated with each $Y_{lm}$.
The question is finding the set of $a_{lm}$ that minimises the $\chi^2$ between the observed data and spherical harmonic model $kT_{\rm LHB}(\theta,\phi)$, which can be written explicitly as
\begin{eqnarray}
    \chi^2 = \sum_{i}\left[\frac{d(\theta_i, \phi_i)-\sum_{l=0}^{l_{\rm max}}\sum_{m=-l}^{m=+l}a_{lm}Y_{lm}(\theta_i,\phi_i)}{\sigma(\theta_i,\phi_i)}\right]^2,
\end{eqnarray}
where $d(\theta_i,\phi_i)$ and $\sigma(\theta_i, \phi_i)$ represent the $i^{\rm th}$ data point and uncertainty associated with it. In our convention, $\phi_i=l_i$ and $\theta_i = 90-b_i$, where $l_i$ and $b_i$ are the Galactic longitude and latitude. For real-valued functions, the conjugate property of spherical harmonics means that $a_{l(-m)}=(-1)^{m}a_{lm}^*$, where the $*$ denotes the complex conjugate. It follows that $a_{l0}$ has no imaginary part, and the number of independent parameters required to describe each spherical harmonics of degree $l$ is $2l+1$ (each complex $a_{lm}$ requires two).

The most probable $a_{lm}$ coefficients and their associated uncertainties are found by running MCMC using the \texttt{emcee} package \citep{emcee}, with the walkers initialised (with a small spread) at the minimum $\chi^2$ position found by the Levenberg-Marquardt algorithm implemented in \texttt{lmfit} \citep{lmfit}.

%For the spherical harmonic analysis, regions within $10\degr$ of the Galactic plane were masked.
We begin with evaluating the significance of the north-south gradient, using the same regions ($|b|>30\degr$) shown in Fig.~\ref{fig:LHB_kT_data}.

\begin{figure*}
    \centering
    \includegraphics[width=0.49\textwidth]{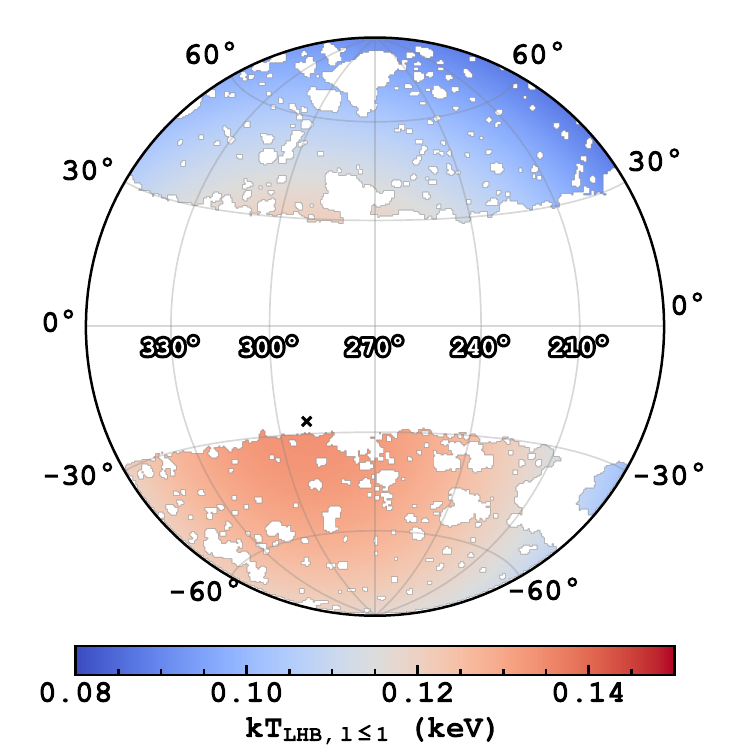}
    \includegraphics[width=0.49\textwidth]{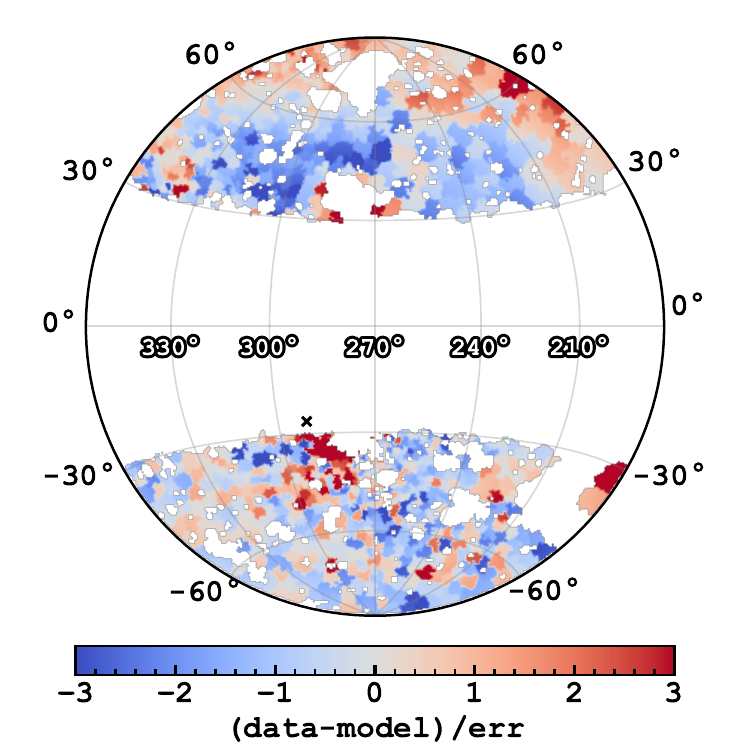}
    \caption{Dipole model and residual. \textit{Left}: Dipole model of ${kT_{\rm LHB}}$. \textit{Right}: Residual normalised by the $1\,\sigma$ fitting uncertainty of each valid bin. The black cross in both panels shows the direction of the dipole towards $(l,b)=(291\degr,-26\degr)$.}
    \label{fig:dipole}
\end{figure*}

We began by modelling $kT_{\rm LHB}$ as a constant (monopole) in the unmasked regions. We found a median $kT_{\rm LHB}$ of 
$0.1154\pm{0.0003}$\,keV with $\chi^2/{\rm dof} = 6.36$ (915\,dof), a unacceptable fit. Subsequently, we fitted the data with a dipole ($l_{\rm max}\leq 1$). The most probable model and the corresponding residual normalised by the bins' fitting uncertainty are shown in Fig.\,\ref{fig:dipole}.
The $\chi^2/{\rm dof}$ decreased to $4.01$ (912\,dof) compared to the monopole. Using the F-test and a significance level of 0.001, we deduced an F-statistic of $179.7 > {\rm F_{crit}} = 5.5$ with a $p$-value in the order of $10^{-16}$. This suggests the dipole model is strongly preferred over the constant model, and the presence of a north-south gradient is statistically significant.
However, one could still recognise systematic residuals in the northern hemisphere by inspecting the residual image, reflecting the north-south gradient is not simply a dipole. 

%1609 bins remained in the analysis.
% We increased the model complexity further to a quadrupole ($l\leq2$), aiming to reproduce the large-scale variation in $kT_{\rm LHB}$ even better to resolve the systematic residual in the southern hemisphere. The resulting model and residual are shown in Fig.\,\ref{fig:quadrupole}. The under-subtracted band in the dipole model in the southern hemisphere was largely resolved. This improved the $\chi^2/{\rm dof}$ to $2.49$ (771\,dof). Using the same significance level of 0.001, the F-test suggests the quadrupole model is preferred over the dipole model: F-statistic $=27.46>{\rm F_{crit}}=4.15$, $p$-value $\simeq{10^{-16}}$. Despite the relatively large $\chi^2/{\rm dof}$ of $2.49$, we decided not to extend to even higher multipoles because no more obvious large-scale structures can be seen in the residual map. We also show the distribution of the residual of the quadrupole model in Fig.\,\ref{fig:resid_hist}, in which the width of the distribution ($0.014\,$keV) is similar to that of the mean spectral fitting uncertainty ($0.012\,$keV). This means much of the temperature spread we see in Fig.\,\ref{fig:LHB_kT_data} can be attributed to the large-scale gradient in $kT_{\rm LHB}$ instead of small-scale fluctuations.

\begin{figure*}
    \centering
    \includegraphics[width=0.49\textwidth]{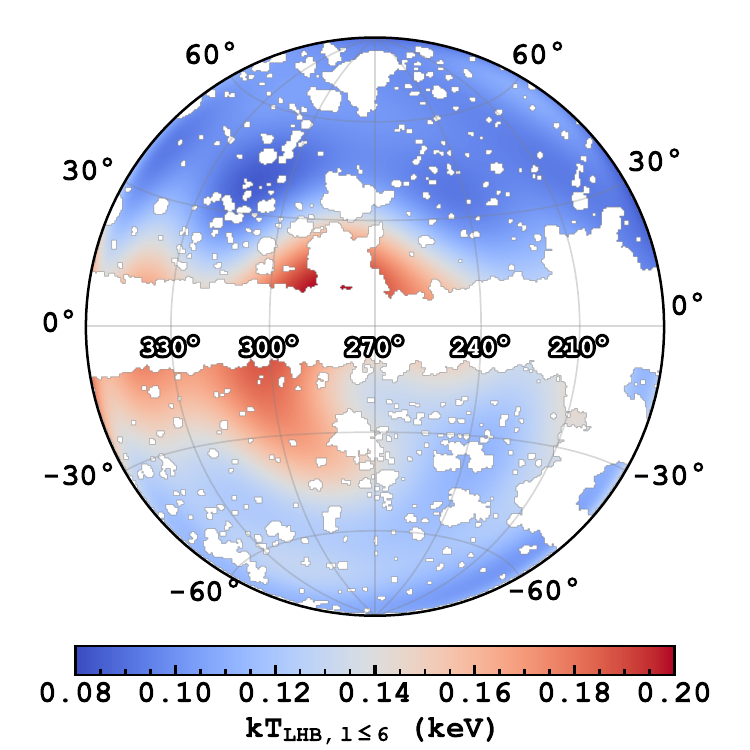}
    \includegraphics[width=0.49\textwidth]{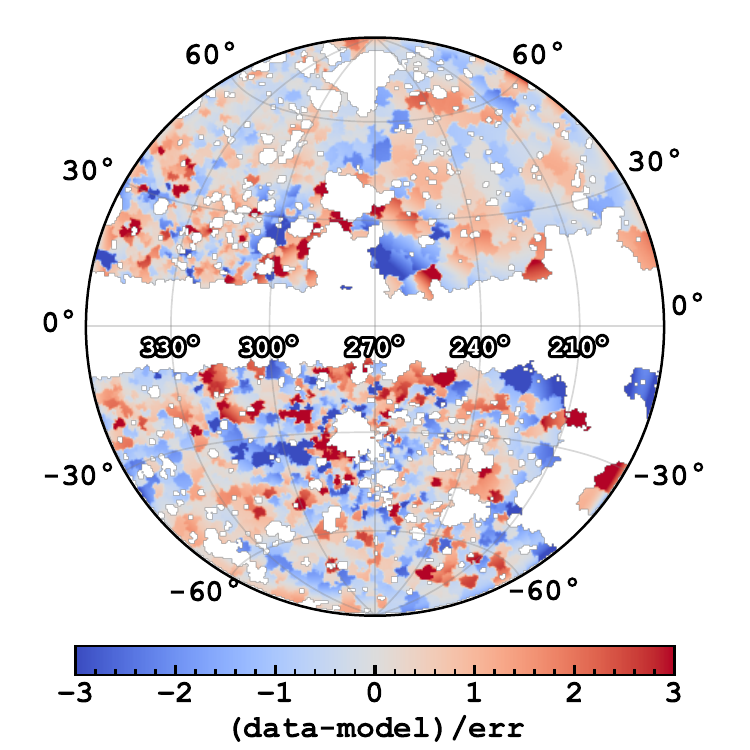}
    \caption{Similar to Fig.\,\ref{fig:dipole} but for the spherical harmonics model of $l_{\rm max}=6$. The data were fitted down to $10\degr$ from the Galactic plane. We note that the colour bar of the left panel was scaled up to $0.2$\,keV, identical to Fig.~\ref{fig:LHB_kT_all} for comparison.}
    \label{fig:lmax6}
\end{figure*}

Naturally, raising $l_{\rm max}$ can improve the fidelity of our spherical harmonics model to reproduce the data more closely. We included data as close as $10\degr$ from the Galactic plane for this empirical model. With multiple trials of different $l_{\rm max}$ values, we arrived at a $l_{\rm max}=6$ model that captures the main large-scale features in the data reasonably well, presented in Fig.~\ref{fig:lmax6}. It has a $\chi^2/{\rm dof}$ of 3.11 (1560 dof). We emphasise that we do not associate any physical interpretations with the multipoles. It merely serves as an empirical model for the LHB temperature profile. However, we speculate  on the origin of the gradient in Sect.~\ref{subsubsec:speculation}. The model parameters can be found in Appendix~\ref{app:sph_harm}. Appendix~\ref{app:lat_profile} presents the latitudinal profiles of $kT_{\rm LHB}$, where both the uncertainties and scatters of $kT_{\rm LHB}$ are shown to compare with the spherical harmonics models.

% \begin{figure}
%     \centering
%     \includegraphics[width=0.45\textwidth]{figures/LHB_kT_quad_resid_CORkT700eV_hist.pdf}
%     \caption{Similar to Fig.\,\ref{fig:LHB_kT_hist}, but showing the distribution of the $kT_{\rm LHB}$ residual of the quadrupole ($l\leq2$) model.}
%     \label{fig:resid_hist}
% \end{figure}

% We refer the readers to Appendix\,\ref{app:sph_harm} for the detailed fitting results of the spherical harmonics. In Appendix\,\ref{app:sph_harm}, we also fitted the dipole and quadrupole models to regions including the eROSITA bubbles. These are not presented here because Gaussian priors on $kT_{\rm LHB}$ were introduced to regions within the eROSITA bubbles, using fitting results of the clean areas outside the bubbles, making the regions inside the bubbles not truly independent. The treatment of the eROSITA bubbles was presented in Sect.~\ref{subsec:erobub_method}.

\subsubsection{Comparison with past observations} \label{subsec:past_obs}
The most relevant references on the large-scale temperature variations of the LHB are \citet{Snowden1990}, \citet{Snowden2000} and \citet{Liu2017}. \citet{Snowden1990} inferred a temperature gradient of the LHB for the first time using the Wisconsin B/C band intensity ratio. They reported a mean temperature of $0.097\,$keV ($10^{6.05}$\,K), and a dipole gradient pointing towards ($l$,$b$) = ($348\fdg7$,$-11\fdg2$) going from $0.064\,$keV ($10^{5.87}$\,K; near Galactic anti-centre) to $0.127\,$keV ($10^{6.17}$\,K; near Galactic centre). With the advent of ROSAT All-Sky Survey data (RASS), \citet{Snowden2000} compiled a catalogue of X-ray shadows at high Galactic latitudes ($|b|>20\degr$). With these X-ray shadows and taking the simplifying assumption similar to us, that all components except the LHB are absorbed by the total Galactic $N_{\rm H}$, they arrived at an LHB temperature dipole in a similar direction, but with the dipole spanning the range of $0.094$($10^{6.04}\,$K)--$0.116$ ($10^{6.13}$\,K), only $\sim1/3$ of that of \citet{Snowden1990}. We note that our dipole model has the dipole amplitude in between the two studies ($A_{\rm di}=0.0133\pm0.0004$\,keV\footnote{The full temperature range spans by the dipole model is $2A_{\rm di}$.}; Table\,\ref{tab:dipole}). \citet{Liu2017} make use of both the RASS R2/R1 band ratio as well as the estimation of the SWCX contribution from the DXL sounding rocket \citep{DXL} mission to conclude that $kT_{\rm LHB}$ is fairly uniform over the sky at $0.097\pm0.019\,$keV (\citet{Bluem22} recently lower this estimate to $0.084\pm0.019$\,keV using \texttt{AtomDB} version 3.0.9). Inspection of the $kT_{\rm LHB}$ map (left panel of their Fig.\,6) from \citet{Liu2017} shows enhanced temperature in the Galactic south pole, but the enhancement appeared more localised than we observed using eROSITA, possibly caused by the unsubtracted eROSITA bubbles component.

\citet{Snowden1990, Snowden2000} and our work show markedly different dipole direction of the LHB temperature. While a quantitative comparison of  \citet{Snowden1990, Snowden2000}'s dipole models with our dipole model is problematic because we found the LHB temperature profile not fully following a dipole, a qualitative comparison suggests our observed dipole axis is almost $\sim 60\degr$ away from  \citet{Snowden1990, Snowden2000}'s. In contrast, despite being similar to \citet{Snowden1990, Snowden2000} in using band ratio as a proxy of LHB temperature, \citet{Liu2017} reports a temperature map that resembles ours more closely. For \citet{Snowden1990}, Wisconsin B (0.13--0.188\,keV) and C (0.16--0.284\,keV) bands match the peak emitting energies of a $\sim0.1\,$keV plasma and are largely (but not completely) unaffected by the hotter emission from the CGM when taking the band ratio as a proxy of LHB temperature. However, it was completed by 10 sounding rocket flights in $\sim 8$ years, sampling various parts of a solar cycle. We note that the analysis was done before the realisation that the SWCX process could contaminate the SXRB. Thus, how much of the B/C band ratio genuinely traced the LHB emission is unclear. ROSAT R1 ($0.11$--0.284\,keV) and R2 (0.14--0.284\,keV) bands have similar energy coverage to the B and C bands and are similarly sensitive to the LHB temperature. Subtraction of the SWCX (comparing Fig.~11a of \citet{Snowden97} and Fig. 3 of \citet{Liu2017}) and background absorbed emission (see Sect.~\ref{sec:LHB_kT}) can significantly change the profile of the R2/R1 band ratio. The dipole gradient \citet{Snowden2000} found could be plagued by SWCX as RASS was conducted near solar maximum.

Despite only covering the western Galactic hemisphere, it is a positive sign that our LHB temperature profile shows compatible morphology to the \citet{Liu2017}'s map, but with adequate significance to suggest a gradient. However, the spherical harmonic models we presented are a partial view without information from the eastern Galactic hemisphere. The low-order spherical harmonics would almost certainly change significantly should this information become available, as they are most sensitive to features spanning large angular scales by definition. Nevertheless, we suspect the change is unlikely to reconcile our model with \citet{Snowden1990, Snowden2000}'s dipole as there is only a very weak sign of longitudinal temperature dependence from the eROSITA data.

\subsubsection{Temperature anti-correlation with absorption column density}
The left panel of Fig.~\ref{fig:logNH} shows the distribution of $\log_{10}(N_{\rm H}/{\rm cm}^{-2})$ inferred from our spectral fitting.

\begin{figure}
    \centering
    \includegraphics[width=0.49\textwidth]{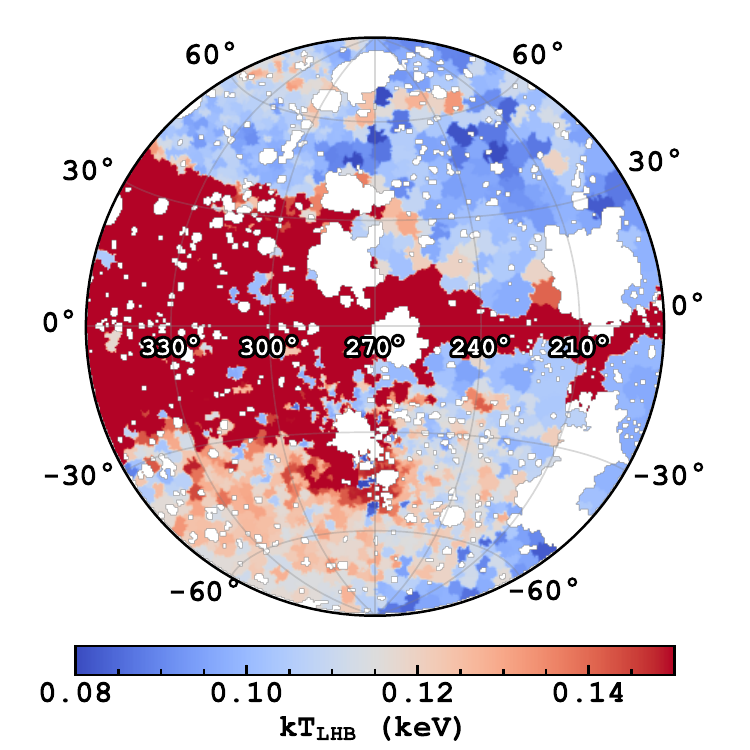}
    \caption{Map of $kT_{\rm LHB}$ upon fixing the $N_{\rm H}$ to the HI4PI $N_{\rm H}$. The colour scaling conforms to Fig.~\ref{fig:LHB_kT_data} to highlight the high latitude temperature dichotomy.}
    \label{fig:LHB_kT_fixNH}
\end{figure}

Neglecting the Galactic plane where both our and HI4PI $N_{\rm H}$ measurements are not particularly accurate due to modelling simplification and self-absorption, the two $N_{\rm H}$ maps possess almost identical morphology, with the X-ray absorption inferring a lower $N_{\rm H}$ at high Galactic latitudes. By comparing the $kT_{\rm LHB}$ maps in Fig.~\ref{fig:LHB_kT_data} and Fig.~\ref{fig:LHB_kT_all} with the $N_{\rm H}$ maps, one can immediately notice the anti-correlation between the $kT_{\rm LHB}$ and $N_{\rm H}$, especially in the southern Galactic hemisphere. The hottest regions of the LHB above $b<-30\degr$ are where the $N_{\rm H}$ are the lowest. Interestingly, as is described in more detail in Sect.~\ref{sec:EM_LHB}, the hottest regions correspond to the largest $\mathrm{EM}_{\rm LHB}$.

Given the morphological similarity of the southern hot patch and the low $N_{\rm H}$ region and the observation that the X-ray absorption column density there is consistently lower than the HI4PI measurement, we suspected the cause of the hot patch was the result of our inaccuracies in fitting the $N_{\rm H}$, biasing $kT_{\rm LHB}$ in the process. Therefore, we reran the spectral modelling of all contour bins and imposed the condition that the $N_{\rm H}$ must be fixed at the HI4PI values.

We show the resulting map on $kT_{\rm LHB}$ in Fig.~\ref{fig:LHB_kT_fixNH}. The $kT_{\rm LHB}$ gradient remains, with similar amplitude and direction. The anti-correlation with $N_{\rm H}$ remains clear. Not surprisingly, the quality of the spectral fits became increasingly worse towards the Galactic plane, as the assumption of using total $N_{\rm H}$ alone the line-of-sight becomes questionable. The result of this test indicates the LHB temperature gradient is unlikely to be caused by our $N_{\rm H}$ determination, and in addition, leaving $N_{\rm H}$ free during spectral fitting was appropriate and necessary.

\subsubsection{Speculations on the source of temperature gradient}  \label{subsubsec:speculation}
The mechanism that set up this temperature gradient is unclear. Still, it is not completely unexpected. \citet{Schulreich23} demonstrate in a sophisticated numerical study that sequential supernova explosions could have created the LHB, with some possibly exploding in the last $\sim 1$--$2$\,Myr. In their simulation, the LHB temperature is not uniform in the present day; instead, it shows a large-scale gradient that could span approximately an order of magnitude. Therefore, the scenario of recent off-centre supernova explosions can, in principle, explain our measured temperature contrast. However, their simulated temperature gradient direction differs from our measurement. They back-traced the trajectories of the stellar populations, which most likely hosted massive stars that expanded the LB using \textit{Gaia} EDR3 data. Further, they predicted the massive stars' explosion times and positions by considering the initial mass function and stellar isochrones. They used these informed explosion sites and times as input parameters of their simulation; thus, the simulation's gradient direction is not arbitrary. Judging from their Fig.~4, the simulated temperature gradient is primarily along the Galactic centre-anticentre line, in contrast to our measurement in the north-south direction.
Whilst the uncertainty of the stellar traceback does not allow for the explosion sites in the south, shock reflections from the thick LB shell and gas sloshing following any explosions could easily change the direction of the temperature gradient (Pacicco, M. \& Schulreich, M., priv. comm.). Pressure gradients in the LHB are washed out following the sound crossing timescale (a few $0.1$--$1$\,Myr). Transient shocks within the LHB exist in the simulation of \citet{Schulreich23}. These shocks create hotter and denser plasma layers and can be preferentially picked up in their X-ray emissions as emission measure scales as $n^2$. The temperature dichotomy we observe could be caused by these shocks in the south. It is likely a matter of fine-tuning the simulations and choosing a correct time stamp to reproduce the current LHB observables so that these shocks appear in the right place and time.
%We note that the stellar trajectories shown in \citet{Schulreich23} are just the most probable solution. It might be well within the uncertainty of the stellar orbit calculation to recreate the temperature gradient we observe. 
Last but not least, the density of the plasma also displays a similar gradient as the temperature in their simulation, which we do not see or have not seen with the current instrumental sensitivity in shadowing experiments \citep[e.g.][]{GMC_shadow} that probed only very few sight lines. Future observations in this direction will be essential to confirm or reject this scenario.

The second possible scenario hinges on more assumptions, including the density of the LHB and magnetic pressure within it do not vary significantly in different directions. Consider a simple static scenario where the LHB is not expanding and is in pressure equilibrium with the surrounding ISM. The temperature gradient, in this case, reflects a pressure gradient set up by the surrounding ISM, where its profile can be traced by the thermal pressure of the LHB, given by
\begin{eqnarray}
 P_{\rm thermal} &=&  nkT_{\rm LHB}\\
     &=& \left(n_e + \sum_A{n_A}\right)kT_{\rm LHB}\\
    &\simeq& 1.92n_e kT_{\rm LHB}, \label{eq:pressure}
\end{eqnarray}
where $n_A$ is number density of the $A^{\rm th}$ element and $n=1.92n_e$ is a common estimation of the total particle density\footnote{The factor of $1.92$ assumes the abundance in \citet{Anders_1989}. Opting for the abundance of \citet{Lodders} (the abundance reference we use for spectral fitting) results in a factor closer to $1.9$. The difference is dominated by the difference in Helium abundance: $\left(n_{\rm He}/n_{\rm H}\right)_{\rm AnGr} = 0.0977$ versus $\left(n_{\rm He}/n_{\rm H}\right)_{\rm Lodd} = 0.0792$.} \citep[e.g.][]{Galeazzi_2007,Snowden14}. The colour bar in the $kT_{\rm LHB}$ map in Fig.~\ref{fig:LHB_kT_all} shows also the resulting thermal pressure, assuming $n_e = 4\times10^{-3}\,{\rm cm}^{-3}$ \citep{GMC_shadow}. An enhancement of pressure near the Galactic disc can be observed, and the pressure decreases rapidly away from it. It makes intuitive sense that the Galactic disc exerts a larger pressure on the LHB. Indeed, the vertical pressure profile in the Solar neighbourhood is approximately an exponentially decaying function from the midplane, with a scale height of $\sim 500\,$pc \citep{Cox_2005}.
The smaller high latitude temperature or pressure gradient could also be explained if the initial surrounding medium was not uniform before the formation of the LHB. This scenario does not necessarily conflict with the off-centre supernova explosions scenario in the simulation of \citet{Schulreich23} and could both be at work.

% \begin{figure}
%     \centering
%     \includegraphics[width=0.5\textwidth]{figures/Pressure_pow_fixsiglogNH_varLHBkT_CORkT700eV_freeLHBEM_ZEA.pdf}
%     \caption{Estimated thermal pressure map of the LHB, assuming a constant electron density of $n_e=4\times10^{-3}$. A clear enhancement near the Galactic plane can be seen.}
%     \label{fig:Pressure}
% \end{figure}

\subsection{Emission measure and extent of the local hot bubble} \label{sec:EM_LHB}

Fig.~\ref{fig:LHB_EM} shows the spatial distribution of the emission measure of the LHB, $\mathrm{EM}_{\rm LHB}$. In addition to masking known large superbubbles and supernova remnants, we also excluded regions with $\sigma_{\mathrm{EM}_{\rm LHB}}<5\times10^{-5}\,{\rm cm^{-6}\,pc}$. Regions that make up the latter are usually biased, possibly spectral fits with vanishing LHB components or MCMC chains that did not converge well. Before delving into the correlation or anti-correlation  of $\mathrm{EM}_{\rm LHB}$ with other parameters, it is best to discuss $\mathrm{EM}_{\rm LHB}$ in the context of the extent of the LHB.

$\mathrm{EM}_{\rm LHB}$ is directly related to the extent of the LHB in a given look direction if one knows the line-of-sight density profile, which is given by the following equation:
\begin{eqnarray}
    \mathrm{EM}_{\rm LHB} = \int{n_e(l) n_{\rm H}(l)dl}.
\end{eqnarray}
By adopting the assumption of a fully ionised solar abundance plasma \citep[see][for caveats of this assumption]{Leahy23}, $n_e/n_{\rm H} \simeq 1.2$ and further assuming a constant density profile, we obtained a simple relation between $\mathrm{EM}_{\rm LHB}$ and the extent of the LHB $L$:
\begin{eqnarray}
    \mathrm{EM}_{\rm LHB} = \frac{n_e^2L}{1.2}. \label{eq:dist}
\end{eqnarray}
One could calibrate $n_e$ under the constant density assumption using sight lines through various molecular clouds on the surface of the LHB. Recent work suggests $n_e\simeq(4\pm0.5)\times10^{-3}\,{\rm cm}^{-3}$, moderately independent of the look direction \citep{GMC_shadow}, which is the number we adopted in converting $\mathrm{EM}_{\rm LHB}$ to the extent of the LHB. This is smaller than $n_e = (4.68\pm 0.47)\times 10^{-3}\,{\rm cm}^{-3}$ inferred by \citet{Snowden14}, but within the uncertainties. The assumed electron density gives rise to the largest systematic uncertainty in our 3D LHB model. The difference between the two numbers alone entails a $37\%$ difference in estimating the LHB distance.

\begin{figure*}
    \centering
     \includegraphics[width=\textwidth]{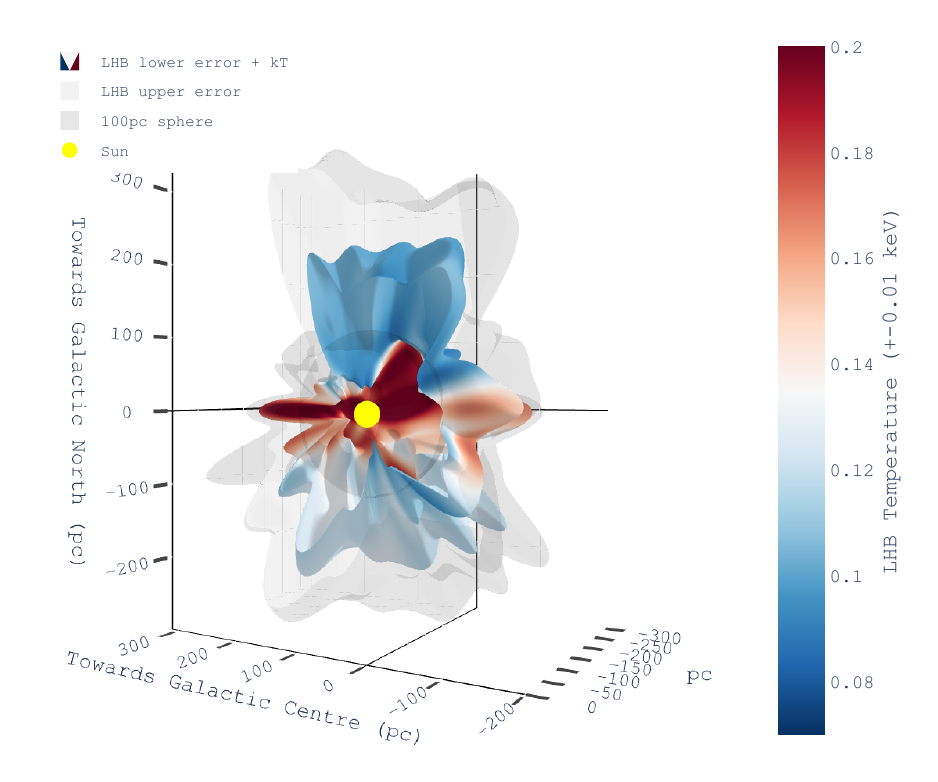}
    \caption{Three-dimensional structure of the LHB in the western Galactic hemisphere assuming a constant density of $4\times10^{-3}\,{\rm cm^{-3}}$. The inner (opaque, coloured) and the outer (grey, translucent) surfaces show the $\pm1\,\sigma$ uncertainty bounds of the distance under the constant $n_e$ assumption. We note that the two surfaces only reflect the uncertainty in the spectral fitting but not in $n_e$. We note that $kT_{\rm LHB}$ is also colour-coded on the inner surface. A sphere of 100\,pc radius is shown around the Sun (yellow) as a ruler. An interactive version of this figure, including the comparison with dust maps (not shown), can be accessed from \href{https://www.aanda.org/articles/aa/olm/2024/10/aa51045-24/aa51045-24.html}{online} or the accompanying website (Sect.~\ref{sec:website}).}
    \label{fig:3D}
\end{figure*}

Figs.~\ref{fig:LHB_EM} and \ref{fig:3D} show the structure of the LHB assuming $n_e=4\times10^{-3}\, {\rm cm}^{-3}$, in the ZEA projection and a 3D-rendered surface respectively. For the latter, the LHB surface is smoothed and interpolated into the masked regions using radial basis function interpolation (\texttt{scipy.interpolate.rbf(*, smooth=0.3)}) implemented in the \texttt{Python} package \texttt{scipy} \citep{scipy}. 

Despite its irregular shape, the LHB is systematically more extended away from the disk, presumably because the denser medium permeated along the Galactic plane prohibits its expansion. The conventional picture of $1\,$MK plasma displacing colder ISM phases in the solar neighbourhood first put forward by \citet{Sanders1977} appears to explain our result excellently, especially in the southern Galactic hemisphere. An almost perfect anti-correlation with Galactic $N_{\rm H}$ can be observed in the southern Galactic hemisphere, slightly less so in the northern hemisphere.

For a more comprehensive view of the multiphase ISM in the solar neighbourhood, we also compared our spatial model of the hot phase of the ISM to the local dust maps, inferred primarily from Gaia extinction data by \citet{Lallement22} and \citet{Edenhofer23} independently. A sophisticated method was used by \citet{Pelgrims20} to trace the inner surface of the local bubble (LB) from 3D dust maps, involving identifying inflexion points in the differential extinction radial profiles and an iterative refining process. In our analysis, we took a simpler approach by identifying the LB extent at which the integration of dust maps reaches a column density of $N_{\rm H}=10^{20}\,{\rm cm}^{-2}$, where the optical depth is approximately unity at $0.2$\,keV. Both the extinction cubes of \citet{Lallement22} and \citet{Edenhofer23} were converted from $A_V$ to $N_{\rm H}$ cubes using the conversion $N_{\rm H}=2.21\times10^{21} A_V$ \citep{Guever2009}. The extent of the LB inferred from these two cubes is shown in Fig.~\ref{fig:dustmap}.

We could see excellent agreement in a few areas (see the finding chart in Fig.~\ref{fig:chart}):\\
1) in the general direction of the Loop~I superbubble or the eROSITA bubbles, both the LHB plasma and dust cubes show distances in the order of 100\,pc. The LHB emission appears to be absorption-bounded.\\
2) The LHB is much more extended in the low $N_{\rm H}$ regions in the southern Galactic hemisphere.\\
3) Around $240\degr\lesssim l \lesssim 270\degr$ on the Galactic plane ($\sim$ $\beta$ Canis Majoris interstellar tunnel \citep{Gry85,welsh91}), one could match the larger extent of the LHB well with the dust maps, both in position and morphology, especially when compared with \citet{Edenhofer23}.\\
4) In the northern Galactic polar cap ($l\gtrsim60$\degr), the LHB is more extended when there is low $N_{\rm H}$.\\
5) Towards the constellation Centaurus at ($l$, $b$) $\simeq$ ($315\degr$, $25\degr$), one can see a hint of an extended tunnel, possibly connecting to the Loop~I superbubble. This feature can also be seen when integrating the \citet{Edenhofer23}'s cube in the right panel of Fig.~\ref{fig:dustmap}.\\
6) On the other hand, we could see in the region of ($180\degr \lesssim l \lesssim 240\degr$, $15\degr \lesssim b \lesssim 45\degr$), little absorbing material is present within 400\,pc, yet the LHB does not extend as freely as in regions towards the LoopI/eROSITA bubbles and the Galactic poles. This could be a hint that the hot plasma is not completely volume-filling, or its density is lower in this general direction, or the dust wall of the LB in this direction is very low in density. Indeed, a peak of low-density dust at $\sim 150\,$pc could be present there as found by the recent work by \citet{Oneill_24}.
We elaborate on the interstellar tunnels in Sect.~\ref{subsec:tunnel}.

\begin{figure}
    \centering
    \includegraphics[width=0.49\textwidth]{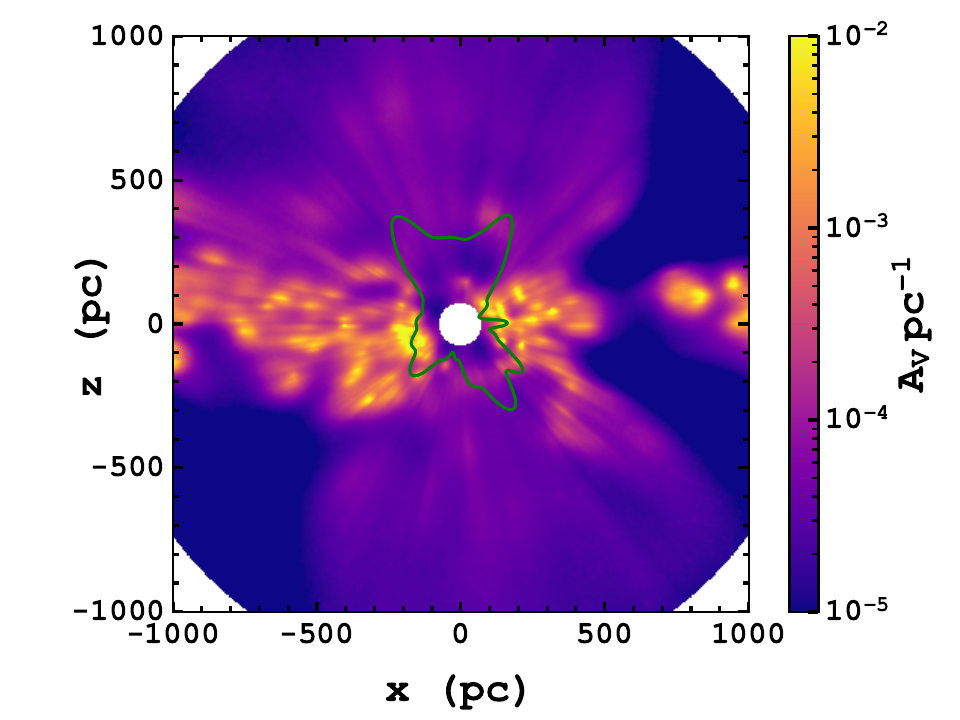}
    \caption{A slice of the \citet{Edenhofer23} cube at the $x$-$z$ plane overplotted with the silhouette of the LHB.}
    \label{fig:xz}
\end{figure}

The anti-correlation of ${\rm EM_{LHB}}$ and dust can also be appreciated by looking at slices of 3D dust cubes. Fig.~\ref{fig:xz} shows the shape of the LHB shell overlaid on the $y=0$ slice in the \citet{Edenhofer23} dust cube. The extent of the LHB matches the onset of extinction extremely well, filling gaps of low dust density.

\citet{Liu2017} produced the latest X-ray LHB model before our work. Consulting their $\mathrm{EM}$ map and great-circle cuts (their Figs.~6 and 7), one can see despite the difference in methodology (spectral-fitting versus band-ratio) and instrument, the inferred $\mathrm{EM}$ and shape of the LHB are remarkably similar. Nevertheless, we have identified some areas that are interestingly different:\\
i) at $l\sim240\degr$ we found a more distinct transition in LHB extent between the Galactic plane ($b\sim0$, region of the $\beta$ Canis-Majoris tunnel) and a large-scale protrusion in the southern Galactic hemisphere ($b\lesssim-30\degr$) than \citet{Liu2017} .\\
ii) The aforementioned channel around ($l$,$b$)=($315\degr$, $25\degr$) did not appear in the \citet{Liu2017}'s $\mathrm{EM}$ map. It was because that region was categorised as contaminated by the background Loop~I superbubble and was excluded from the analysis. In fact, the appearance of the tunnel was seen in the ROSAT R1+R2 band map even after the SWCX subtraction \citep[][their Fig.~2]{Liu2017}.

\subsection{Degeneracy between the local hot bubble and the Milky Way's circum-galactic medium components} \label{subsec:degen}
This Section discusses the degeneracy between the LHB and CGM components and our mitigation methods. We begin by laying out the observations that demonstrate this degeneracy. In summary, this led us to incorporate ROSAT R1 and R2 bands into our analysis, as well as imposing a uniform prior to limit $kT_{\mathrm{LHB}} <0.15$\,keV for regions of low $N_{\rm H}$.

\begin{figure}[ht]
    \centering
    \includegraphics[width=0.49\textwidth]{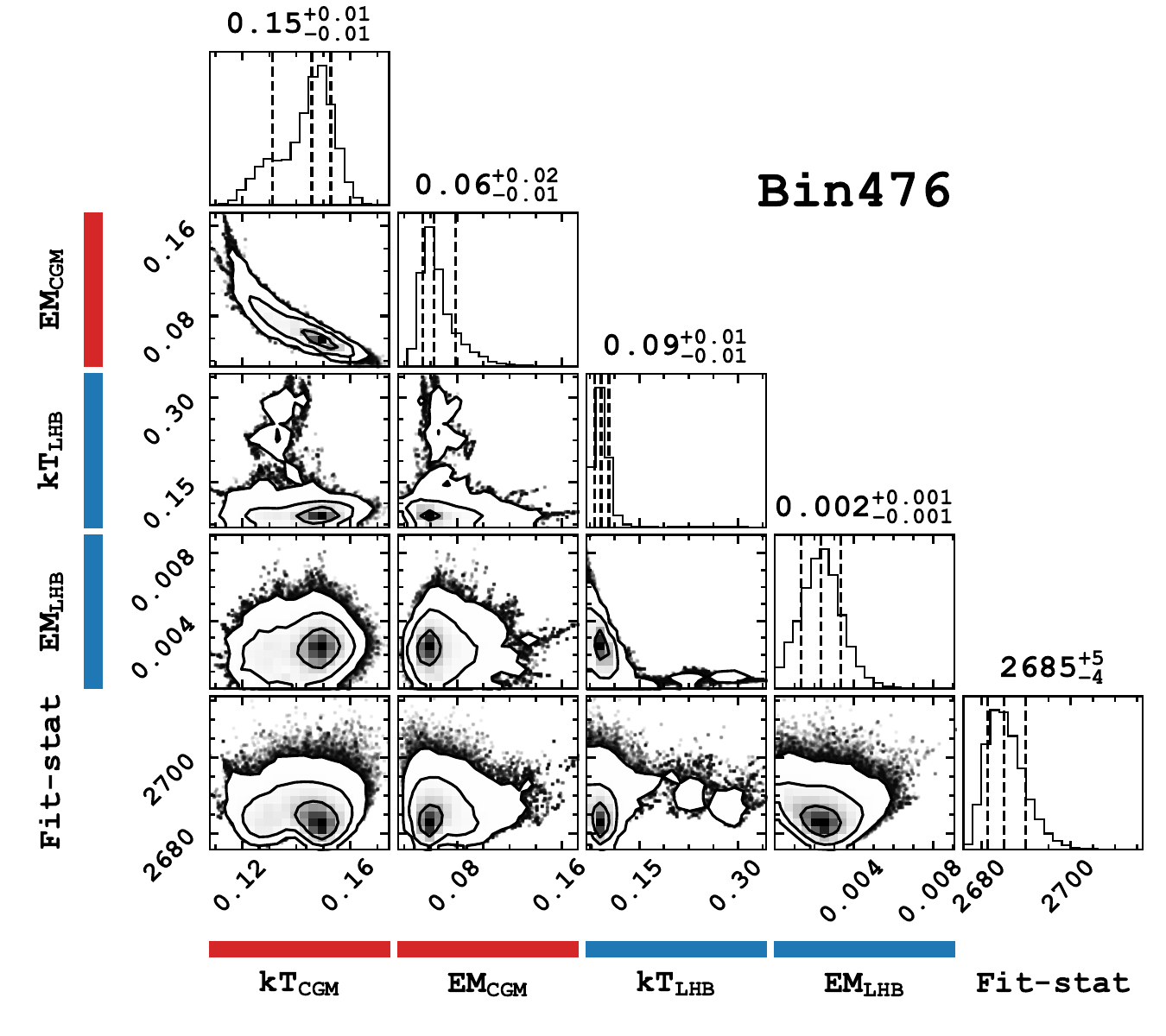}
    \includegraphics[width=0.49\textwidth]{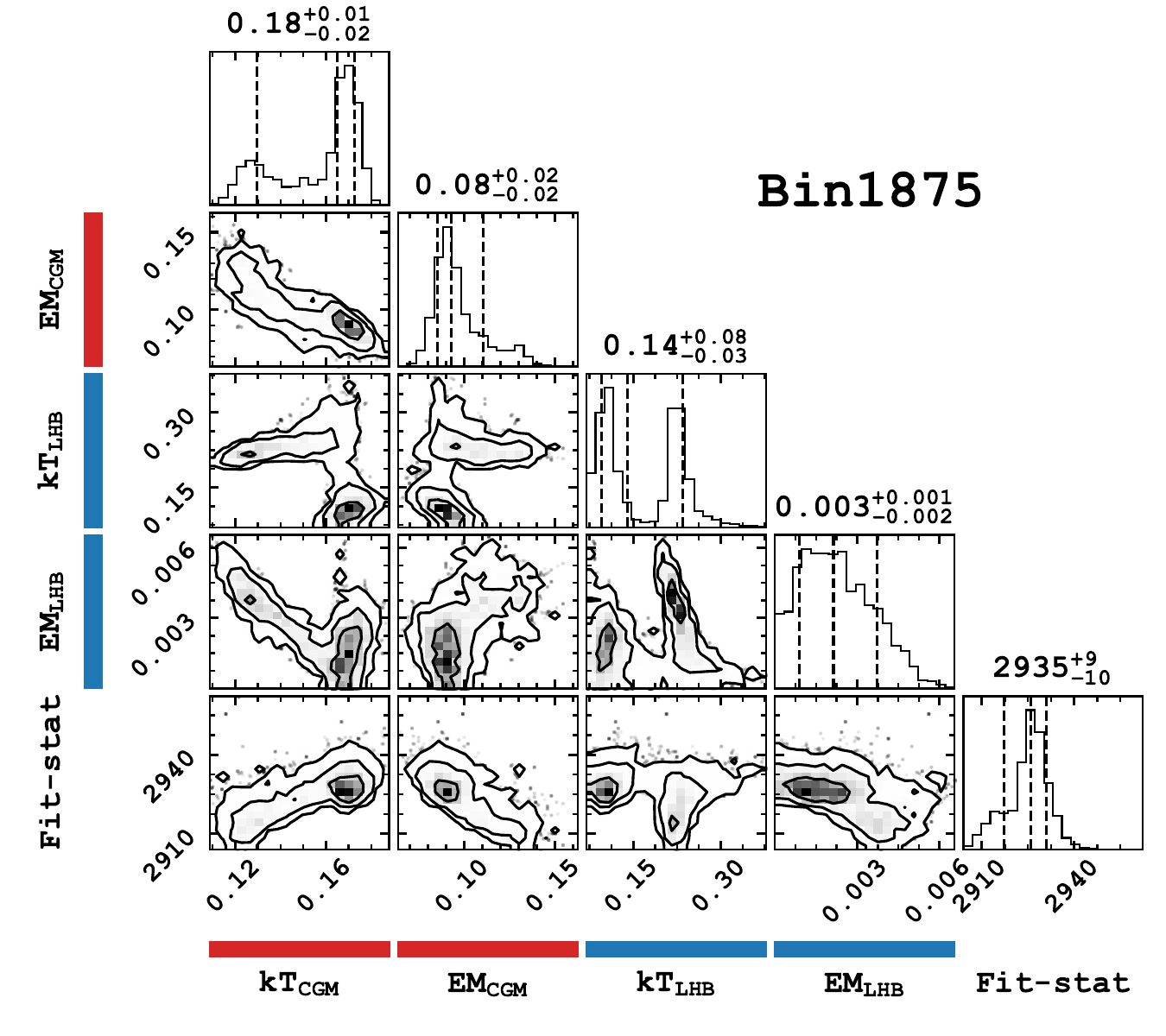}
    \caption{Corner plots showing LHB and CGM parameters are degenerate in regions of low $N_\mathrm{H}$. \textit{Top:} Bin476 at ($l$, $b$)=(257\fdg9, 59\fdg6). \textit{Bottom:} Bin1875 at ($l$, $b$)=(296\fdg7, 47\fdg6).}
    \label{fig:noR12}
\end{figure}

Inspecting the spectral fitting results with only eROSITA data revealed a degeneracy between the LHB and the CGM components in some contour bins. An example of this degeneracy can be seen in Fig.~\ref{fig:noR12}, where we show the parameter correlations in the spectral fits of Bin476 and Bin1875, located at ($l$, $b$)=(257\fdg9, 59\fdg6) and (296\fdg7, 47\fdg6) respectively. Both regions have $\log{(N_\mathrm{H}/\mathrm{cm^{-2}})}\simeq20.5$ as traced by the HI4PI survey \citep{HI4PI} using Eq.~(\ref{eq:HI4PI}). One can see the posterior distributions of $kT_{\mathrm{CGM}}$ are bimodal in both bins. The lower $kT_{\mathrm{CGM}}$ peak is contributed by either partially in Bin476, or completely in Bin1875, by samples that have $kT_{\mathrm{LHB}} > kT_{\mathrm{CGM}}$. Similarly, the posterior distributions of $kT_{\mathrm{LHB}}$ show either an extended tail in Bin476 or a second peak in Bin1875, at temperatures at least doubled of the main peak. We believe these solutions are unlikely to represent the real picture, but are caused by the similarity of the LHB's and CGM's \texttt{apec} models observed at eROSITA's energy resolution. The main differentiating factor of LHB and CGM is the absorption that causes CGM component to drop off at the low energy end. Therefore, one would expect the two to exhibit some levels of degeneracy at regions of low $N_{\mathrm{H}}$. To help break the degeneracy, one would ideally go to energies lower than 0.2\,keV, but eROSITA's effective area there is small. However, ROSAT maintains significant effective area until $\sim0.1$\,keV, and is therefore more sensitive to detect the drop off caused by absorption in the CGM component. The details of how the ROSAT data were used were described in Sect.~\ref{sec:ana}. We note that ideally, one should subtract the heliospheric SWCX contributions from the ROSAT R1 and R2 band maps as in \citet{Uprety}, who reported an all-sky average SWCX contribution of $30\pm8\%$ in R1 and $8\pm10\%$ in R2. However, given that the average relative uncertainties in R1 and R2 are $37\%$ and $31\%$ in our contour bins and that they only contributed to two spectral bins in the fitting, we believe the bias introduced by neglecting SWCX in ROSAT data is vastly subdominant to the resulting fit parameters of our full spectral analysis. Perhaps a more pertinent question would be whether a pair of data points from ROSAT possesses sufficient capability to resolve the degeneracy between the LHB and CGM components.

\begin{figure}
    \centering
    \includegraphics[width=0.49\textwidth]{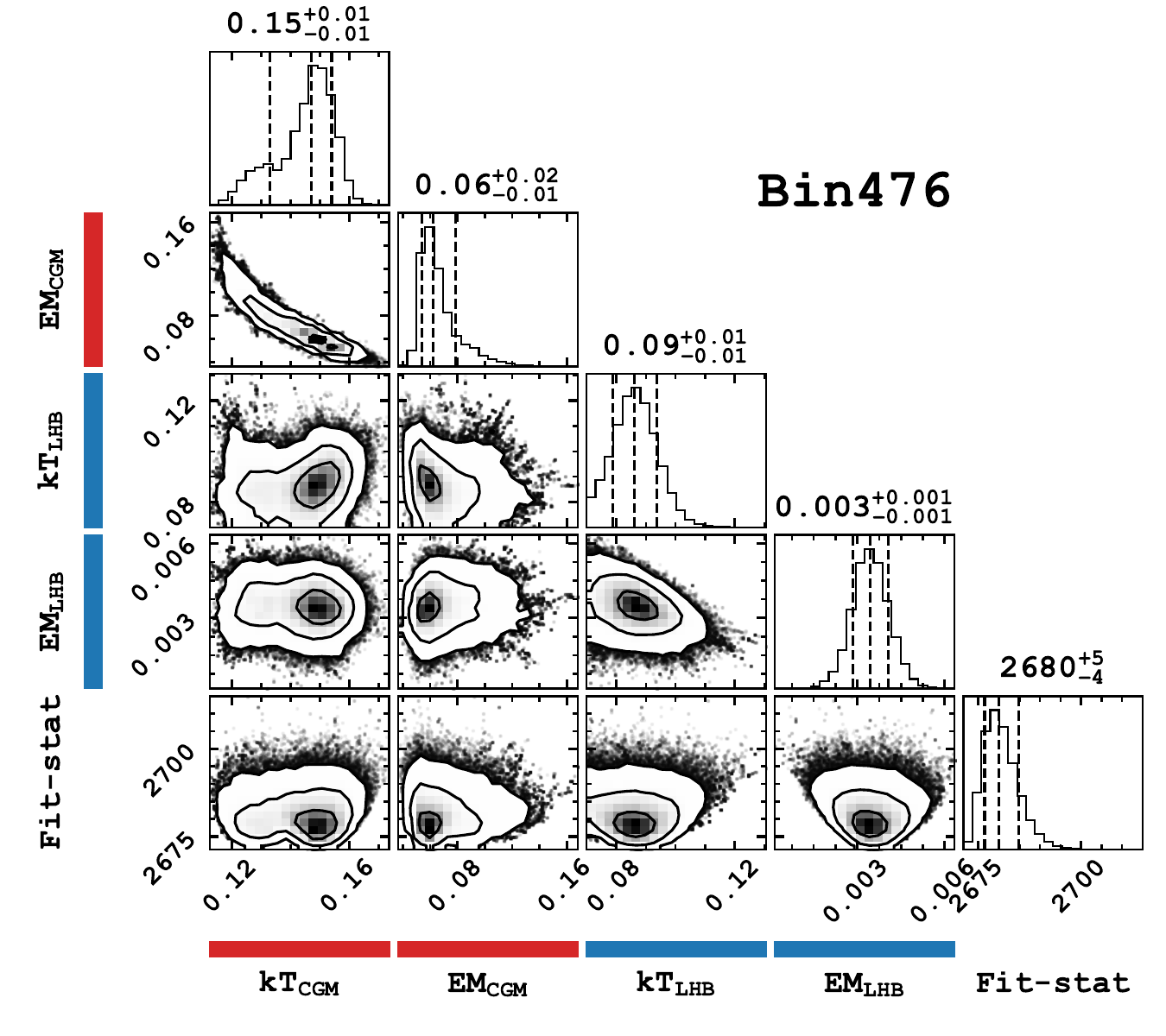}
    \includegraphics[width=0.49\textwidth]{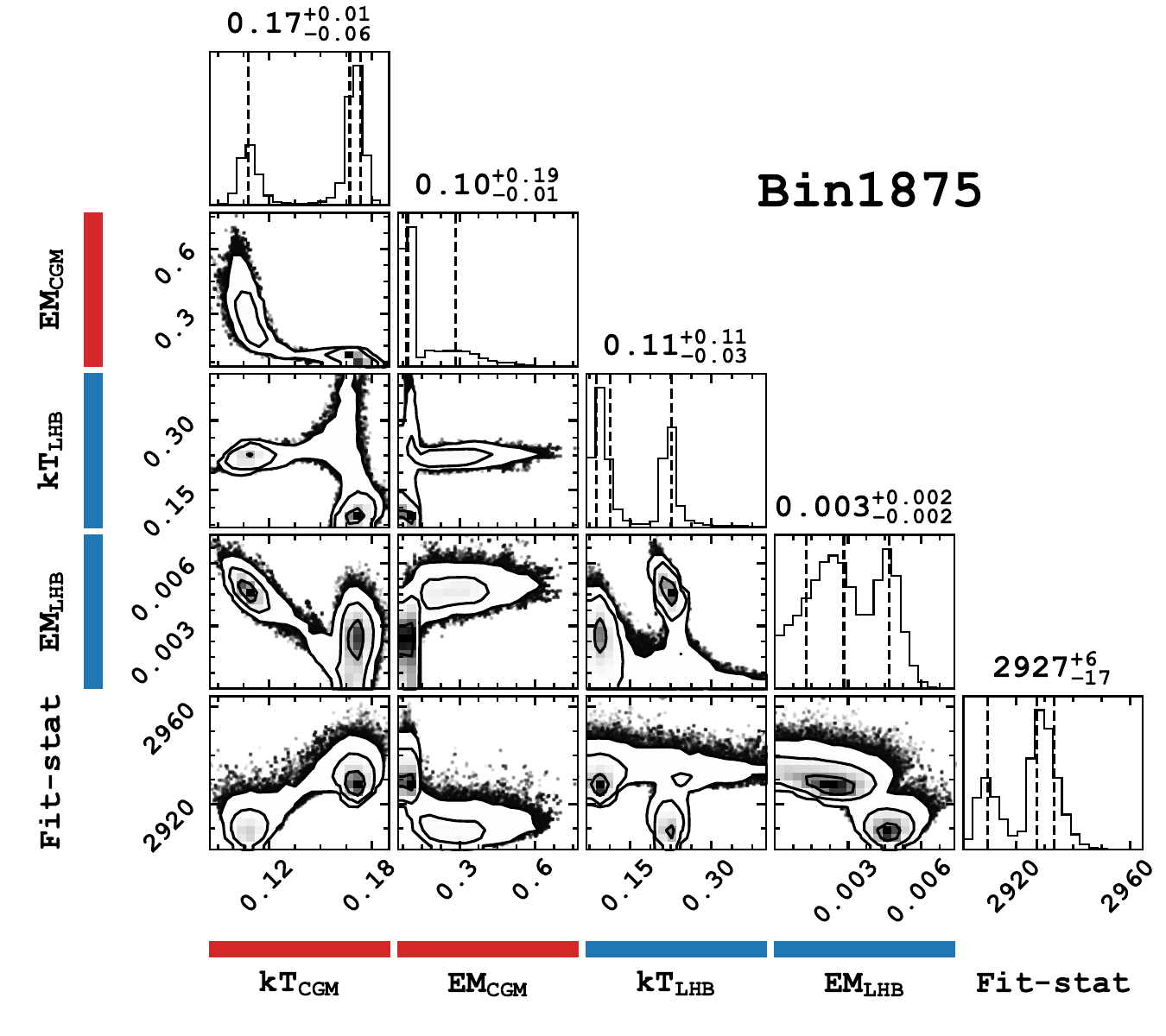}
    \caption{The same plots and panel configuration as Fig.~\ref{fig:noR12} but with ROSAT R1 and R2 data. }
    \label{fig:R12}
\end{figure}

The answer is largely affirmative in many contour bins plagued by the degeneracy. For a specific example and a direct comparison, the top panel of Fig.~\ref{fig:R12} shows the corner plot of Bin476 after including the ROSAT data. The high temperature tail in the posterior of $kT_{\mathrm{LHB}}$ vanished and samples with $kT_{\mathrm{LHB}} > kT_{\mathrm{CGM}}$ disappeared. Motivated by the positive results, we incorporated the ROSAT R1 and R2 bands into the spectral analysis.

\begin{figure}
    \centering
    \includegraphics[width=0.49\textwidth]{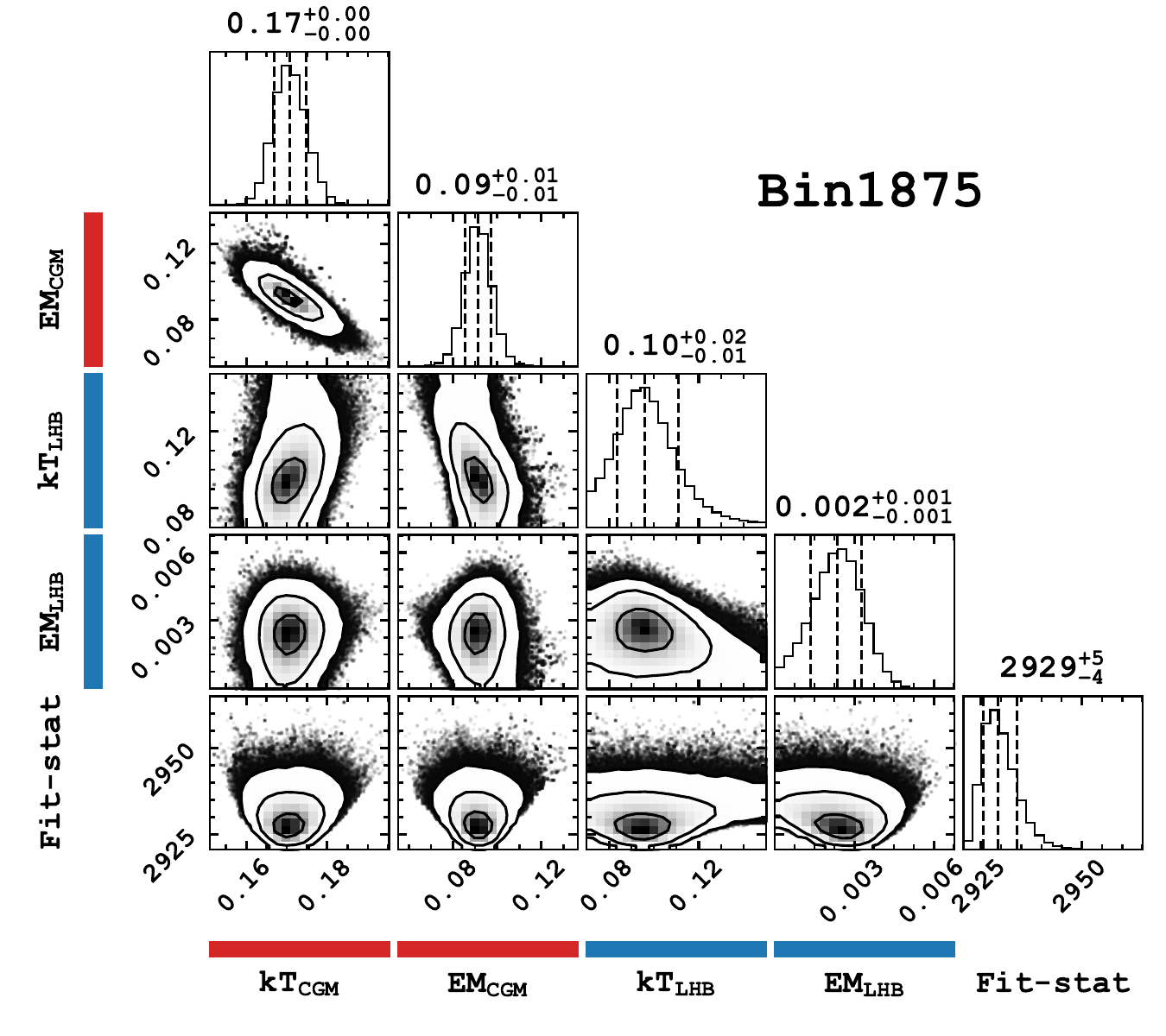}
    \caption{Corner plot of Bin1875 after using ROSAT data and imposing a uniform prior on $kT_{\rm{LHB}}$ below $0.15\,\rm keV$.}
    \label{fig:cons_prior}
\end{figure}

Nonetheless, cases remained where this was inadequate, especially at low $N_{\rm H}$ regions. For instance, in Bin1875 (lower panel of Fig.~\ref{fig:R12}), the degeneracy remained. The posterior distributions remained highly bimodal, albeit a longer MCMC chain was needed to sample the second peak ($kT_{\mathrm{LHB}} > kT_{\mathrm{CGM}}$). Unfortunately, we could not break this degeneracy using the currently available data. Therefore, we decided to forbid $kT_{\mathrm{LHB}}$ to go above $0.15$\,keV by imposing a uniform prior between $0.07$--$0.15$\,keV on $kT_{\mathrm{LHB}}$. Not surprisingly, the resulting posterior distributions no longer show a second peak, as shown in Fig.~\ref{fig:cons_prior}. This prior was used in the fitting of all the contour bins  where $\log{(N_{\rm H}/{\rm cm^{-2}})} < 20.5$. To our knowledge, the LHB temperature has seldomly been measured to be higher than $0.15\,\rm keV$ \citep[e.g.][]{Snowden1990,Snowden2000,McCammon02,Liu2017}, and conversely, the CGM seldomly below $0.15\,\rm keV$ \citep[e.g.][]{McCammon02,Yoshino09,Gupta21,Bluem22,Ponti2023}, justifying our choice of prior in the low $N_{\rm H}$ regions.

\subsection{Solar wind charge exchange} \label{subsec:SWCX}
The halo orbit of SRG/eROSITA around the Sun-Earth Lagrangian point L2 \citep{Freyberg20} avoids much of the magnetospheric SWCX that originates from the interaction of the solar wind ions with neutrals in the Earth's exosphere \citep{Kuntz19}. An image of the Earth's exosphere in Ly-$\alpha$ taken from 2348\,$R_{\Earth}$ away shows that it is contained within 94\,$R_\Earth$ \citep{Kameda}, while SRG never ventured within 175\,$R_\Earth$ from the Earth \citep{Freyberg20}, effectively immune to the magnetospheric SWCX emission since it is always pointing perpendicular to the Sun-Earth-L2 line. Even avoiding the bulk of the exosphere, in principle, SWCX can also occur on the surface of the magnetotail, but we expect its effect to be small since such a thin interaction and emission layer would incur brightness variation in short timescales (minutes to hours), and we do not observe obvious stripes along the ecliptic longitude lines in the soft band in eRASS1 (eROSITA completes scanning a great circle primarily along this direction every four hours).

We further ignored the effect of heliospheric SWCX in our spectral analysis, which can indeed contaminate eROSITA data in general. We verify that this is at least a satisfactory assumption by looking at the $\mathrm{EM}_{\rm LHB}$ map a posteriori. It is well-known that the solar wind density is higher at low heliographic latitudes than the poles, especially around solar minima \citep[e.g.][]{Porowski22}. Therefore, unlike pointed observations, eRASS provides a way of estimating the heliospheric SWCX contribution from detecting extra emission preferentially along the ecliptic plane. In Fig.~\ref{fig:LHB_EM}, we overlaid the ecliptic in black on the $\mathrm{EM}_{\rm LHB}$ map, and the dashed black lines mark the range of $\pm25\degr$ within the ecliptic, where the denser, slow solar wind dominates during solar minima \citep[e.g.][]{McComas98,McComas03}. The fact that there is no obvious enhancement within this ecliptic latitude range suggests the heliospheric SWCX contamination in our results is minor. Indeed, as we gradually approached solar maximum, eRASS3 and 4 data clearly show enhancement in this band, but the presentation and the detailed analysis of this effect will be presented in another work (Dennerl et al., in prep).

\subsection{Energetics and pressure balance within the local hot bubble}
%\textcolor{red}{An important topic in the local interstellar medium is the stable-coexisting of the molecular phase (local interstellar clouds; LIC) and the hot phase. \citep{Snowden14} showed from ROSAT 1/4\,keV band that the pressure within LHB is $10^{4}\,{\rm cm^{-3}\,K}$. What is our measurement in different positions? Could the measured magnetic field in the local ISM explain the discrepancy between LIC pressure and LHB pressure? We also need to reference Nicola's work on the CGM to discuss the pressure balance between the LHB and the CGM, in the context of whether LHB is open or closed in the caps.}

We estimated the thermal energy of the LHB as follows:
\begin{eqnarray}
    E_{\rm thermal} &=& \sum P_{\rm thermal}\Delta V \\
                    &=& 1.92n_e\sum kT_{\rm LHB}(l,b)\Delta V,
\end{eqnarray}
where $\Delta V$ is the volume occupied by the LHB plasma within a contour bin. The $\mathrm{EM}_{\rm LHB}$ information is implicitly passed into $\Delta V$ in the form of the extent of the LHB. We have incomplete coverage of the sky. Hence, we scaled up our estimation of $E_{\rm thermal}$ to a total of $4\pi$ solid angle. Finally, we estimated $E_{\rm thermal} = 1.3^{+0.9}_{-0.5}\times10^{51}\,$erg, where we have assumed $n_e=4\times 10^{-3}\,{\rm cm}^{-3}$. This is in the order of the energy released by a supernova explosion. However, one must consider that the LHB's size (radius) is $>100\,$pc, which is too large for a single explosion.  Therefore, the LHB is more likely to have been episodically reheated and expanded by successive supernova explosions and simultaneously radiated its energy away to produce the current energy content.

We presented the thermal pressure map in Fig.~\ref{fig:LHB_kT_all}, using $P_{\rm thermal}/k = nT$. We found the mean pressure of the LHB is $P_{\rm thermal}/k=10100^{+1200}_{-1500}\,{\rm cm^{-3}\,K}$. These numbers are consistent with pressure measured in the sight line of ($l$,$b$)=(144\degr,0\degr) by \citet{Snowden14}, hence supporting their finding of the LHB being in pressure equilibrium with the local interstellar clouds (LICs), after accounting for magnetic pressure as measured by \textit{Voyager~I} outside of the heliosphere \cite{Burlaga14}. But we also point out that \cite{Snowden14}'s sight line lies on the Galactic plane, which we found to have systematically higher thermal pressure.
This pressure is also fully consistent with measurements based on X-ray shadowing using eRASS:4 data \citep{GMC_shadow}.

Compared with the latest numerical simulation of the LHB by \citet{Schulreich23}, our measured pressure is consistent with their simulated present-day pressure of $10100\,{\rm cm^{-3}\,K}$. However, this results from combining their $n$ and $T$, which are lower and higher by roughly an order of magnitude compared to our results, respectively. Despite the apparent coincidence, their simulation shows how successive supernova explosions in the Sco-Cen complex can lead to the large-scale temperature gradient within the LHB, as we found in this work (Sect.~\ref{sec:LHB_kT}).

The estimated Galactic midplane total ISM pressure is $P/k \sim (2.8\,\pm0.7)\times10^{4}\,{\rm cm^{-3}\,K}$, about two to three times the thermal pressure we estimated from the LHB on the plane \citep{Boulares_1990,Cox_2005}. As already put forth at the time, one can approximate the cosmic rays, magnetic and dynamical (possibly thermal) pressure as equipartition. This yields a thermal pressure from the ISM about $P_{\rm thermal}/k\sim10^4\,{\rm cm^{-3}\,K}$, consistent with our measurement.
Finally, we would like to highlight the LHB pressure of $P_{\rm thermal}/k\sim10^4\,{\rm cm^{-3}\,K}$ is lower than the typical values seen in superbubbles, supernova remnants or wind-blown bubbles \citep[e.g.][]{Oey_2004,Sasaki_2011,Kavanagh_2012,Sasaki_2022}. This may indicate the LHB being open towards higher Galactic latitudes.
%\subsection{Interface with Loop~I}
%\textcolor{red}{Depending on the results from the spectral fits with priors, we look at the $\mathrm{EM}_{\rm LHB}$ again in these regions. Do we see extra path lengths associated with Loop~I, or do we see less because both bubbles are stopping the expansion of each other (we do not see through because there is compression of dust in the interface? The boring result of similar $\mathrm{EM}_{\rm LHB}$ in this region is also interesting because this would mean a strange scenario that LHB expands unobstructed in the assumed presence of Loop~I. It would mean all its emission is completely attributed to the CGM or the eROSITA bubble components. It raises the question of whether what we get for eROSITA bubbles is really just from the eROSITA bubbles. How do we tackle the necessary discussion on the eROSITA bubbles?}

\subsection{Interstellar tunnel network} \label{subsec:tunnel}
\citet{Cox74} famously postulated supernovae can generate and maintain an interstellar tunnel network in the Galaxy, filled with $\sim 0.1\,$keV gas.
Mapping the hot phase structure of the LB/LHB in emission has more limitations than mapping the cold phase of the local ISM with stellar extinction or line absorption because one has to assume a model of the emitting components and the absorption they are subject to. This makes inference of the LHB shape on the Galactic plane difficult as one expects many line-of-sight emitting and absorbing components. Also, extensions and possible connecting tunnels of the LHB that are not aligned with the radial direction are difficult to infer. Nevertheless, as mentioned in Sect.~\ref{sec:EM_LHB}, we found evidence of hot gas filling nearby channels that lack neutral material. 

The most convincing evidence of such a channel is the $\beta$\,CMa tunnel ($l\sim250\degr$,\,$b\sim0\degr$), where even the projected morphology of the tunnel in $\mathrm{EM}_{\rm LHB}$ (see Fig.~\ref{fig:LHB_EM}) is similar to the dust map (especially, the right panel of Fig.\,\ref{fig:dustmap}). Historically, ROSAT did not provide enough evidence of the tunnel being filled by hot gas \citep[][and references therein]{Welsh09}.\footnote{Closer inspection of the R12 band count rate map suggests there is indeed a weak enhancement, independent of the subtraction of SWCX \citep{Snowden97,Uprety}.} However, \citet{Dupin98} and \citet{Gry01} found the intervening absorbing clouds to the stars $\beta$\,CMa and $\epsilon$\,CMa to show absorptions from high ionisation species such as \ion{C}{IV} and \ion{Si}{III}, but comparison with the cloud turbulent temperature ($\sim 7000$\,K) demonstrated that they are likely partially ionised due to photoionisation of the hot stars and/or the interaction with the hot LHB gas in the surrounding.
Our observation confirms the presence of the surrounding hot gas in the tunnel. Indeed, both the UV and our soft X-ray observations corroborate the scenario where the warm gas in LICs forms a conductive layer between themselves and the surrounding hot gas \cite[see e.g.][]{Cowie79,Slavin89}. \citet{welsh91} suggested this tunnel is connected to the Gum nebula \citep{Gum_1952}, which lies in the same direction at $400\,$pc \citep{Brandt_1971}. The Gum nebula is likely part of a large superbubble GSH~238+00+09 \citep{Heiles_1998}, which the LHB could also connect to. 

\begin{figure*}
    \centering
    \includegraphics[width=0.49\textwidth]{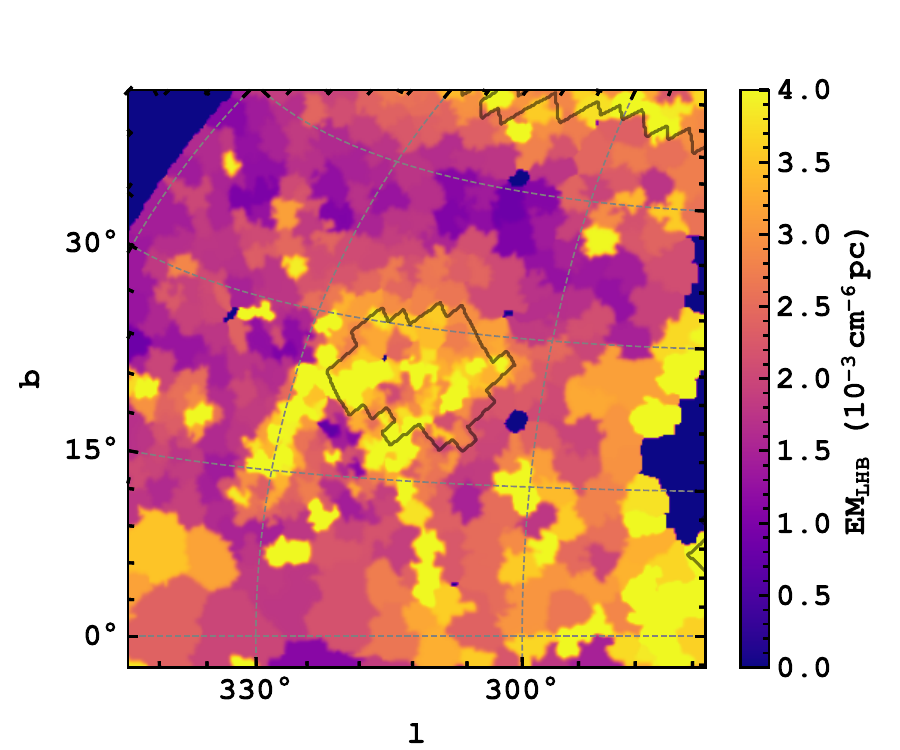}
    \includegraphics[width=0.49\textwidth]{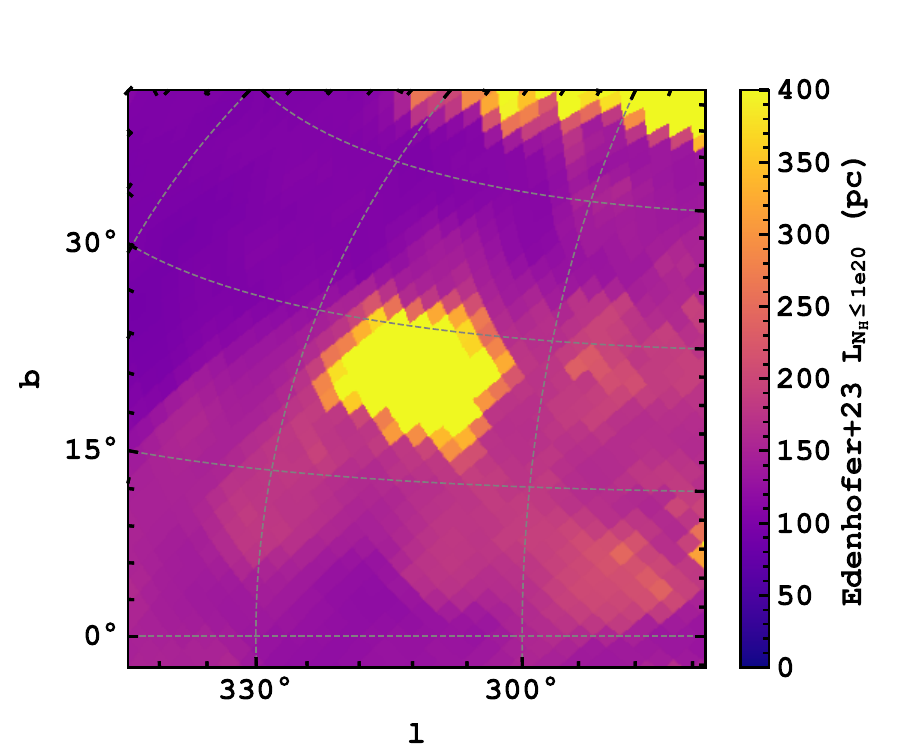}
    \caption{Zoom-in of the Centaurus tunnel in the smoothed $\mathrm{EM}_{\rm LHB}$ map (\textit{left}) and dust distance map by integrating the $dA_V$ map up to $10^{20}\,{\rm cm}^{-2}$ (\textit{right}) \citep{Edenhofer23}. The contour line outlining the region where the $N_{\rm H}$ does not reach $10^{20}\,\rm cm^{-2}$ in $400\,\rm pc$ is overlaid on the $\mathrm{EM}_{\rm LHB}$ map.}
    \label{fig:cen_tun}
\end{figure*}

In addition, we report on a possible channel filled by hot plasma towards the direction ($l\sim 315\degr$, $b\sim25\degr$), in the constellation Centaurus. A zoom-in view of the $\mathrm{EM}_{\rm LHB}$ and dust map tracing the neutral matter in the region is shown in Fig.~\ref{fig:cen_tun}. There is an enhancement in $\mathrm{EM}_{\rm LHB}$ in this direction, which anti-correlates with the amount of dust extinction. This implies the presence of extra path length of the LHB plasma, possibly filling the tunnel. It might be another channel connecting to the Loop~I superbubble \citep{Egger95}, in addition to the Lupus tunnels first discovered from \ion{Na}{I} absorption lines \citep{Welsh94,Lallement03}. The latter are, unfortunately, located on the Galactic plane and, thus, are difficult to identify in our spectral analysis. The Centaurus tunnel region is located on the edge of the eROSITA bubbles, further complicating the spectral fitting. A dedicated spectral analysis, with a tailored spectral extraction region, of this region will likely help disentangle the emission from the Loop~I superbubble (its nature or existence has become unclear after the discovery of the eROSITA bubbles) from the eROSITA bubbles, in terms of spectral properties and distance. This analysis is currently ongoing.

The Antlia supernova remnant ($l$, $b$)$\sim$($275\degr,15\degr$) \citep[e.g.][]{Fesen_21}, Monogem Ring ($l$, $b$)$\sim$($ 200\degr, 8\degr$) \citep[e.g.][]{Knies_24} and Orion-Eridanus superbubble ($l$, $b$)$\sim$($205\degr,-30\degr$) \citep[e.g.][]{Pon_16} are other nearby bubbles that our data suggest the LHB could be touching or connected to. Monogem Ring is arguably even more interesting because there is a lack of dust \citep{Lallement22, Edenhofer23} in the line-of-sight up to its distance of $\simeq300\,$pc \citep[][we refer to the closest components in the whole Gemini-Monoceros X-ray enhancement]{Knies_24}, consistent with the eROSITA spectral analysis of it \citep{Knies_24}. The LHB might be currently merging or about to interact with it.

On the other hand, there is the presence of channels of low $N_{\rm H}$ that do not seem to be filled with soft X-ray-emitting plasma. The more obvious one is located at ($l$, $b$)$\sim$($260\degr, 29\degr$), which can be seen in the right panel of Fig.~\ref{fig:dustmap}. The ${\rm EM_{\rm LHB}}$ map in Fig.~\ref{fig:LHB_EM} does not show an anti-correlated enhancement.

In summary, our data demonstrates that the displacement model in which the LHB plasma fills the LC works well overall. Only a few regions seemingly do not abide by the displacement model. 

\subsection{Beyond constant density of the local hot bubble} \label{subsec:LHB_vs_LB}
We have discussed the close anti-correlation between the local dust map and the emission measure of the LHB in Sect.~\ref{sec:EM_LHB}. The dust maps (Fig.~\ref{fig:dustmap}) we compared our LHB shape to were inferred from integrating the $N_{\rm H}$ up to $10^{20}\,{\rm cm^{-2}}$ (an optical depth for a $0.2\,$keV photon) from our position. 
An alternative way of locating the shell of the LB is by finding the first peak of the extinction as adopted by \citet{Pelgrims20} and \citet{Oneill_24}. This might be more physically motivated as the expansion on the LB is expected to create a shell of denser material at its boundary that is not necessarily dense enough to be opaque to soft X-ray.

\begin{figure*}
    \centering
    \includegraphics[width=0.49\textwidth]{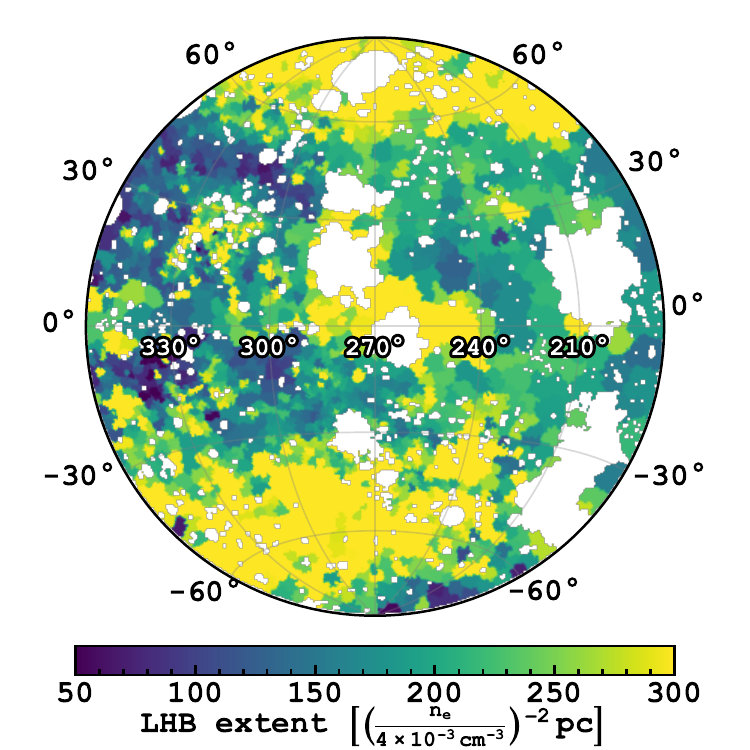}
    \includegraphics[width=0.49\textwidth]{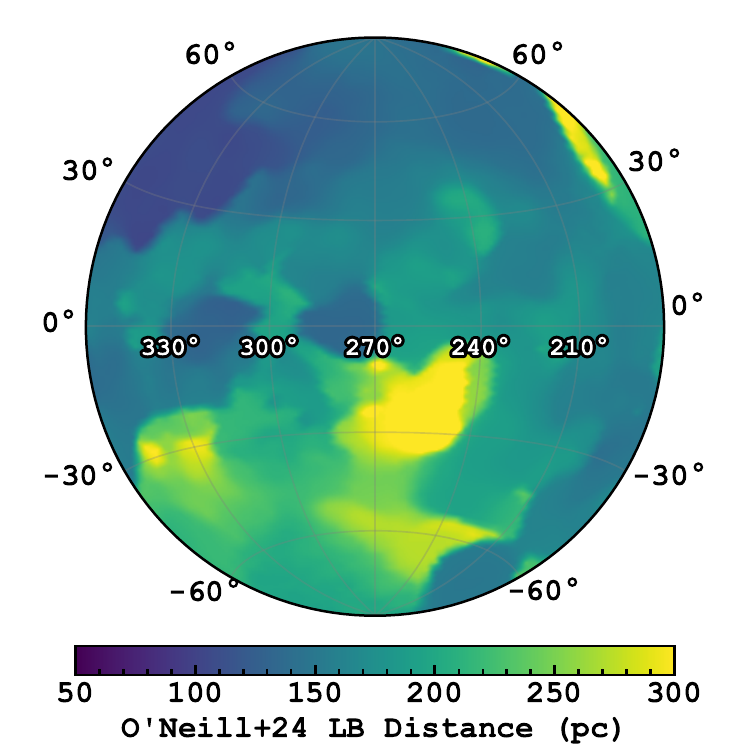}
    \caption{Comparison of the local (hot) bubble extent from X-ray and dust. \textit{Left:} Extent of the local hot bubble under the constant $n_e=4\times10^{-3}\,{\rm cm^{-3}}$ assumption. \textit{Right:} Local bubble shell tracing the closest extinction peak inferred by \citet{Oneill_24}. An interactive comparison in 3D can be accessed from the accompanying website (Sect.~\ref{sec:website}).}
    \label{fig:oneill}
\end{figure*}

A comparison of the inferred LB shell from \citet{Oneill_24} using \citet{Edenhofer23} data to our constant density LHB model is shown in Fig.~\ref{fig:oneill}. For this comparison, we adopted their $A'_{0.5}$ peak density distance as the extent of the LB. Their shapes are significantly different, most notably at $b>60\degr$, where the apparent enhancement in ${\rm EM_{LHB}}$ (hence extent) lacks a counterpart in the LB. Below $b\lesssim30\degr$, the LHB and the LB are both more extended but with a fairly different morphology. The clear tunnel in the LB at ($l$, $b$)\,$\simeq$\,($250\degr$, $-20\degr$) is not shared by the LHB. There is a reasonable agreement for regions between $-30\degr \lesssim b\lesssim 60\degr$, ignoring the various tunnels discussed in Sect.~\ref{subsec:tunnel}. The simplest reason for the discrepancy between the two maps is the constant density assumption is inaccurate, given $L_{\rm LHB}\,\propto\,{\rm EM_{LHB}}n_e^{-2}$, which means the extent of the LHB is more sensitive to $n_e$ than ${\rm EM_{LHB}}$. 

\begin{figure*}
    \centering
    \includegraphics[width=0.49\textwidth]{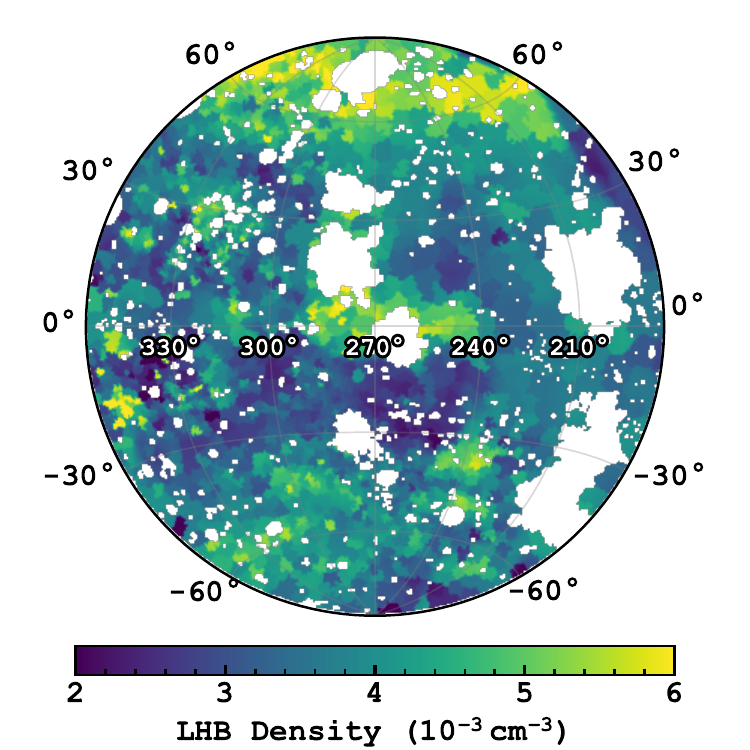}
    \includegraphics[width=0.49\textwidth]{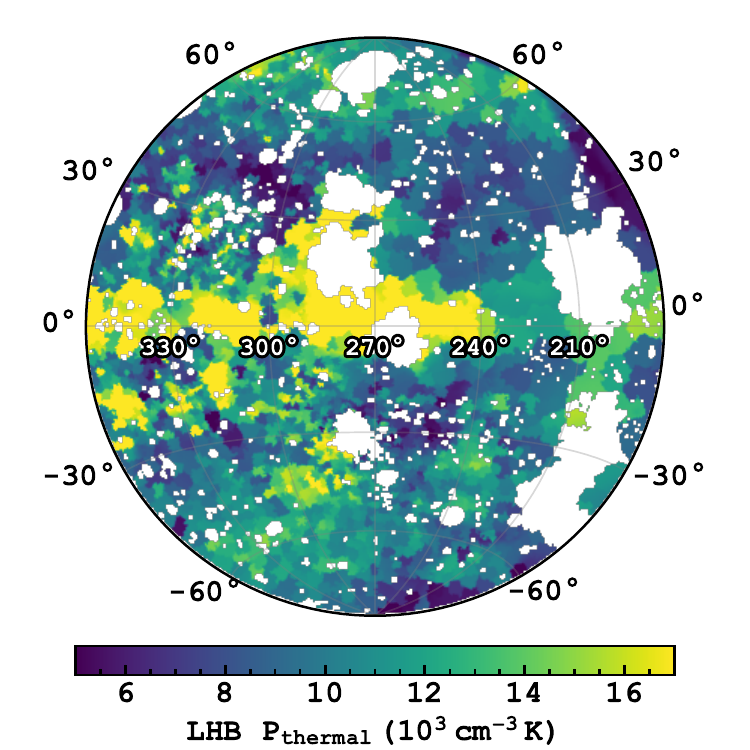}
    \caption{Electron density (\textit{left}) and thermal pressure (\textit{right}) of the local hot bubble assuming it extends up to the local bubble shell \citep{Oneill_24}.}
    \label{fig:ne}
\end{figure*}

One can work conversely, assuming the LB shell is the true extent of the LHB and estimate the electron density of the LHB plasma using Eq.~(\ref{eq:dist}), as well as its thermal pressure by Eq.~(\ref{eq:pressure}) subsequently. We show the resulting electron density and thermal pressure maps in Fig.~\ref{fig:ne}. Clearly, much of the modulation in ${\rm EM_{LHB}}$ is transferred to the variation in density, especially in regions where the LHB extent derived from the constant density assumption differs much from the extent of the LB. This causes the average LHB density at $b>60\degr$ to be higher than the rest of the areas by $\sim 40\%$. The extensions towards $\beta$\,CMa (at least the part on the Galactic plane) and Centaurus tunnels mentioned in Sect.~\ref{subsec:tunnel} do not have direct counterparts in the LB map, and thus they are taken as high-density regions of the LHB. The thermal pressure map slightly differs from the constant density case shown in Fig.~\ref{fig:LHB_kT_all}, where the former displays a weaker north-south gradient because of the higher electron density in the northern polar cap balancing the lower LHB temperature, and it produces a more uniform pressure profile on the largest scale. Indeed, being in pressure equilibrium with its surroundings would be the expectation for an old bubble such as the LHB. From the two maps, the half-sky median electron density and thermal pressure  are $n_e=3.75^{+0.87}_{-0.69}\times10^{-3}\,{\rm cm^{-3}}$ and $P_{\rm thermal}/k=10.5^{+3.4}_{-2.5}\times 10^{3}\,{\rm cm^{-3}\, K}$.

Assuming the LHB extends up to the first extinction peak is a sound assumption, but it is not without challenges. The hardest part to reconcile is the presence of tunnels that seem independent of the LB shell. We suggest it is unlikely a coincidence that ${\rm EM_{LHB}}$ shows a better anti-correlation to the local dust column density (Fig.~\ref{fig:dustmap}) than the extent of the LB shell, especially when the LB shell peak density is the lowest in these regions \citep[see Fig.~4 in][]{Oneill_24}. Not accepting it as a coincidence entails hot plasma located beyond the first wall of absorption. They could have the same origin as the LHB or had other heating mechanisms. One possibility is shown by the LB simulation by \citet{Schulreich23} (third and fourth columns of their Fig.~4), in which pockets of hot plasma could be formed beyond the main LB cavity because of the LHB anisotropic expansion into the inhomogeneous surrounding materials that were already stirred up during the wind-driven phase.

\subsection{Cosmic X-ray background} \label{subsec:CXB}

\subsubsection{Determination of the cosmic X-ray background photon indices} \label{subsec:CXB_slope}
\citet{Ponti2023} adopts a double-broken power-law model to describe the CXB in the eFEDS field, following the CXB synthesis model by \citet{Gilli}, which produces a steepening CXB slope $\lesssim1$\,keV from galaxy groups and clusters. The double-broken power-law\footnote{Three models of CXB were used in \citet{Ponti2023}. We refer to their `CXB' model here.} (\texttt{bkn2pow} in \texttt{Xspec/PyXspec}) can be written as
\begin{equation}
f(E)=
    \begin{cases}
    K(E/\mathrm{keV})^{-1.9} & \text{if $E \leq0.4$\,keV}\\
    K0.4^{-0.3}(E/\mathrm{keV})^{-1.6} & \text{if $0.4\,\mathrm{keV} \leq E \leq 1.2\,\mathrm{keV}$}\\      
    K\frac{0.4^{-0.3}}{1.2^{0.15}}(E/\mathrm{keV})^{-1.45} & \text{if $E \geq 1.2\,\mathrm{keV}$},
    \end{cases}
    \label{eq:bkn2pow}
\end{equation}
where $K$ is the normalisation of the power-law at 1\,keV in the unit of photons\,s$^{-1}$\,cm$^{-2}$\,keV$^{-1}$ and the first and second break energies are located at $0.4$ and $1.2$\,keV.

From the spectral fitting of the regions shown in Fig.\,\ref{fig:CXB_reg}, we found that the first break energy at $0.4$\,keV is redundant as the photon indexes below ($\Gamma_1=1.93^{+0.22}_{-0.21}$) and above ($\Gamma_2=1.91^{+0.09}_{-0.16}$) this energy are consistent within 1\,$\sigma$ when we left all three photon indexes free to vary in the spectral fitting. Therefore, we subsequently replace the double-broken power-law model with a single-broken power-law model, which can be treated as a special case of Eq.~(\ref{eq:bkn2pow}). It has the form
\begin{equation}
f(E)=
    \begin{cases}
     K(E/\mathrm{keV})^{\Gamma_1} & \text{if $E\leq E_b$}\\
     K(E_b/\mathrm{keV})^{\Gamma_2-\Gamma_1}(E/\mathrm{keV})^{-\Gamma_2} & \text{if $E \geq E_b$},
    \end{cases}
\end{equation}
where $E_b$ is the break energy and $\Gamma_1$ and $\Gamma_2$ are the photon indexes below and above $E_b$.
Another iteration of the spectral fitting with the single-broken power-law model indicates $E_b=1.19^{-0.26}_{-0.17}$, $\Gamma_1=1.81^{+0.24}_{-0.21}$ and $\Gamma_2=1.61^{+0.13}_{-0.07}$. This result suggests a change of slope in the CXB at $1.2$\,keV, below which the slope is possibly steeper from the contributions from active galactic nuclei, galaxy groups and galaxy clusters \citep{Hasinger93,Smith07,Yoshino09}. Nonetheless, the transition of the CXB slope is only significant on $\sim 1\,\sigma$ level\footnote{Fixing $E_b$ at $1.2$\,keV would decrease the statistical uncertainties of $\Gamma_1$ and $\Gamma_2$ to $1.87^{+0.16}_{-0.11}$ and $1.63^{+0.08}_{-0.06}$, respectively.} and is not strictly required by the data. Additionally, the uncertainties of the parameters are dominated by the spread from multiple regions instead of the statistical uncertainties within each region. A simple power-law fit yields $\Gamma=1.68^{+0.08}_{-0.10}$, in good agreement with $\Gamma_2$ in the single-broken power-law model. This is steeper than the canonical photon index of $1.4$--$1.5$. Calibration issues could partly cause this, as eROSITA indeed tends to measure cooler cluster temperature than \textit{Chandra} and \textit{XMM-Newton} for massive clusters \citep{Liu_2023,Migkas24}\footnote{The currently developing eSASS pipeline processing version \texttt{c030} will improve the calibration of the effective area and redistribution matrix above $1\,\rm keV$.}. Both treatments of the CXB reproduce the data almost equivalently, with the simple power-law and single-broken power-law resulting in an average C-statistic/dof in the Markov chain Monte Carlo (MCMC) runs of all regions of $3123^{+249}_{-99}/2595$ and $3092^{+228}_{-99}/2588$, respectively.
Therefore, we adopted both in our full western Galactic hemisphere analysis and discuss the systematic uncertainties associated with either choice in Sect.~\ref{subsec:sys_err}. Based on this analysis of the CXB, we froze $\Gamma_1$, $\Gamma_2$ and $E_b$ to be $1.9$, $1.6$ and $1.2$\,keV of the single-broken power-law model, $\Gamma$ to  $1.7$ for the simple power-law model as the CXB model in the full spectral analysis.
\subsubsection{Normalisation}
\begin{figure}
    \centering
    \includegraphics[width=0.49\textwidth]{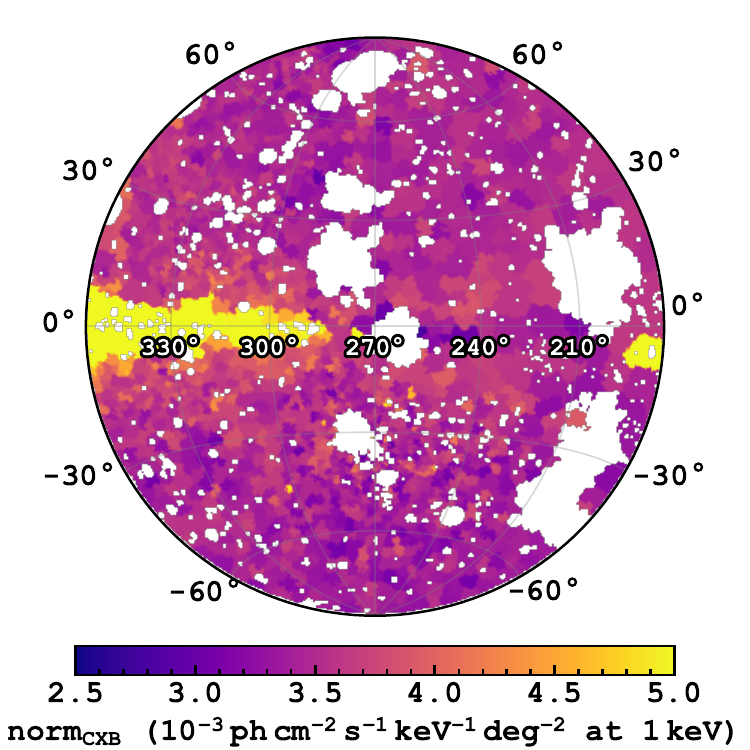}
    \caption{Normalisation of the cosmic X-ray background assuming a simple power-law of $\Gamma=1.7$.}
    \label{fig:CXB}
\end{figure}
Fig.~\ref{fig:CXB} shows the normalisation of the CXB, taken from the simple power-law model of $\Gamma=1.7$. Ignoring the Galactic plane, the normalisation of the CXB is extremely uniform. At $|b|>30\degr$, it has a median of ${\rm norm_{CXB}}=3.54^{+0.24}_{-0.17}\times10^{-3}\,{\rm photons\,cm^{-2}\,s^{-1}\,keV^{-1}\,deg^{-2}}$ at 1\,keV. This means the fluctuation of the CXB is $\lesssim10\%$, on the angular scales of $\sim1\fdg5$--$4\degr$ that we probe set by the contour binning scheme. The enhancement on the Galactic plane is likely due to the Galactic ridge emission.

Our main finding on the CXB is that eROSITA observes a steeper photon index than the conventional value of 1.45 \citep{Cappelluti17} (Sect.~\ref{subsec:CXB_slope}). It might be due to calibration issues \citep{Migkas24} or caused by the steepening of the CXB slope at soft energies. A newer processing version of eROSITA data will potentially mitigate or resolve the calibration issue. Given the steeper slope of $\Gamma\sim$1.6--1.7, the normalisation we found is also slightly higher compared to \textit{Chandra} ($(3.32\pm0.05)\times10^{-3}\,{\rm photons\,cm^{-2}\,s^{-1}\,keV^{-1}\,deg^{-2}}$, \citet{Cappelluti17}). In addition, the uniformity of the CXB demonstrates the robustness of our spectral fitting to differentiate multiple sky components.

\subsubsection{Systematic uncertainties associated with the choice of cosmic X-ray background model} \label{subsec:sys_err}
As described in Sect.~\ref{subsec:CXB_slope}, the power-law and the broken power-law model could reproduce the diffuse X-ray spectra equally well. So far, we have been discussing the results using a CXB described by a simple power-law model. To test the systematic effect caused by the CXB model, we fitted all spectra again in the same way but using a broken-power-law CXB component.
We then plotted the 2D maps and inspected the histograms of the other components. We observed that the distributions of most of the components were unaffected except for the LHB.

\begin{figure*}
    \centering
    \includegraphics[width=0.49\textwidth]{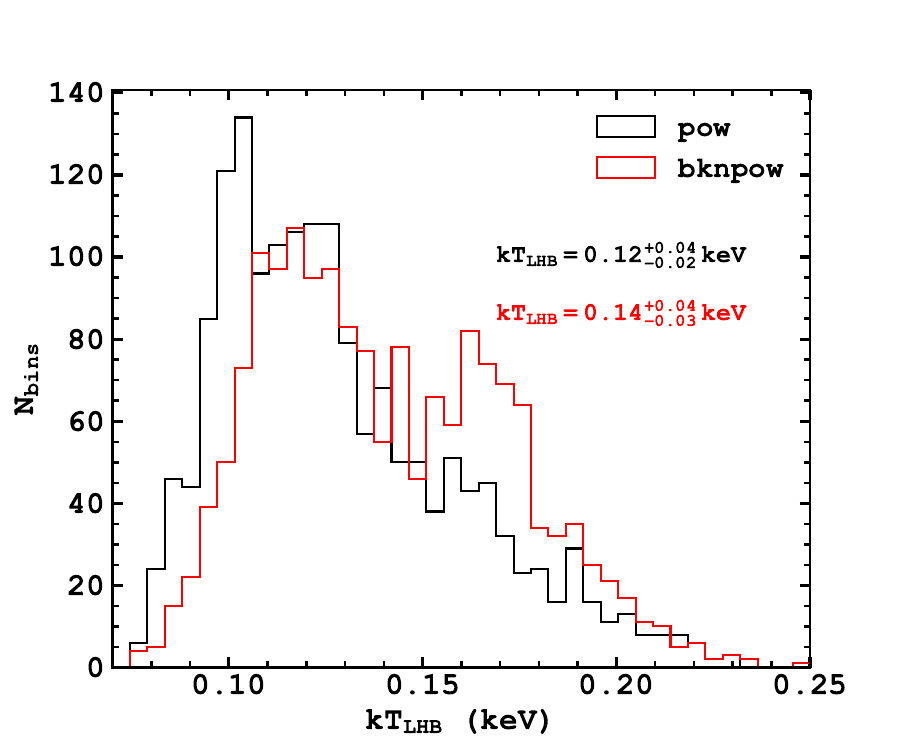}
    \includegraphics[width=0.49\textwidth]{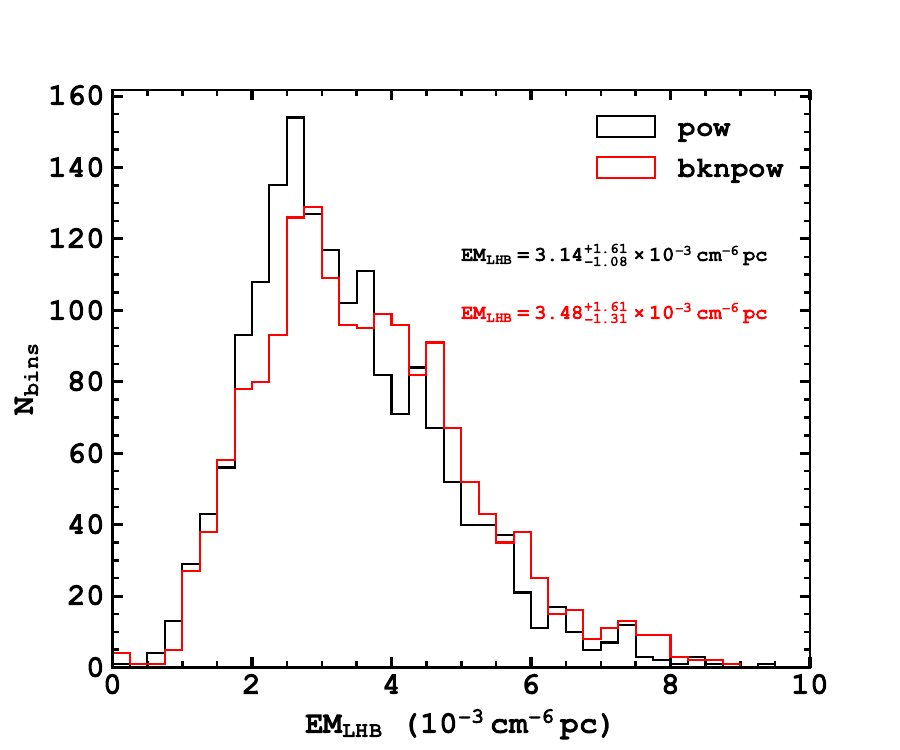}
    \caption{Distributions of $kT_{\rm LHB}$ (\textit{left}) and $\mathrm{EM}_{\rm LHB}$(\textit{right}) using either a simple power-law (pow) or a broken power-law (bknpow) parametrisation of the CXB.}
    \label{fig:LHB_pow_bknpow}
\end{figure*}

Fig.~\ref{fig:LHB_pow_bknpow} shows the difference in distributions of $kT_{\rm LHB}$ and $\mathrm{EM}_{\rm LHB}$ by using the two CXB models for regions above $|b|>10\degr$. The broken power-law model appears to raise the temperature of the LHB systematically by $0.02\,$keV in terms of the median. On the other hand, the impact of such temperature change on the EM is small ($\sim0.3\times10^{-3}{\rm \,cm^{-6}\,pc}$) compared to the width of the $\mathrm{EM}_{\rm LHB}$ distribution ($\sim1.5\times10^{-3}{\rm \,cm^{-6}\,pc}$). Inspection of the spectra suggests this is the result of the intricate balancing at the softest end of the spectra ($\sim 0.2\,\rm keV$). By increasing the temperature slightly, the LHB could lower its flux density at this energy to accommodate the steepening slope of the CXB.
% This effect is subtle. 
% and if one is only focussing on one spectrum, the statistical uncertainty is often adequately large ($\sim 0.01\,\rm keV$) for this effect to be overlooked.

\begin{figure}
    \centering
    \includegraphics[width=0.49\textwidth]{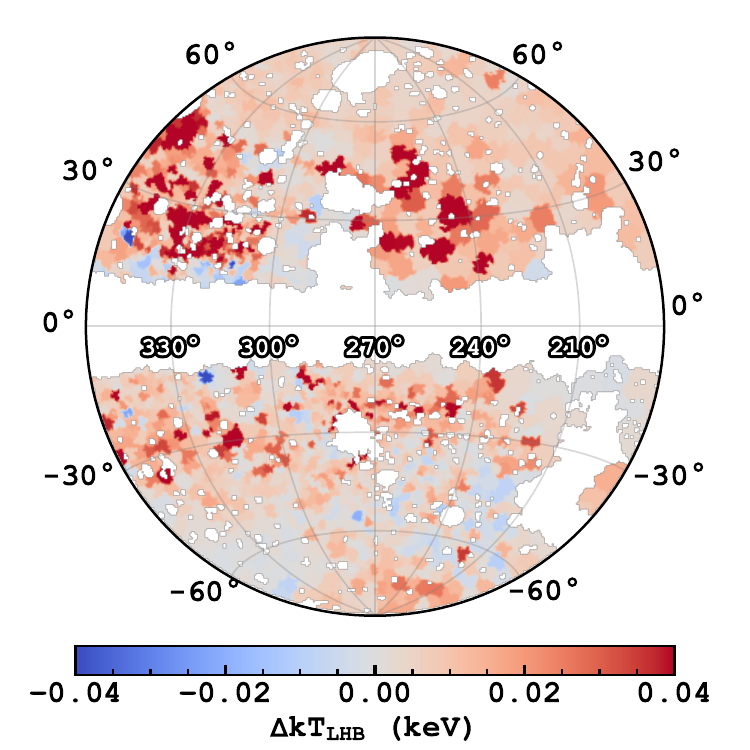}
    \caption{Difference in the LHB temperature in assuming a simple power-law CXB model and a broken power-law model, namely, $kT_{\rm LHB, bknpow}-kT_{\rm LHB, pow}$.}
    \label{fig:pow_bknpow_resid}
\end{figure}

Fig.~\ref{fig:pow_bknpow_resid} demonstrates this is indeed a systematic effect by showing the ubiquitous under-subtraction in the residual map made from subtracting the LHB temperature map determined by the simple power-law model from that of the broken power-law model. The broken power-law model consistently predicts a higher LHB temperature than the simple power-law model, and the difference is more enhanced towards the Galactic plane. We computed the median differences below and above $|b|=30\degr$. For $|b|>30\degr$, the median difference between the two CXB models is $0.009$\,keV, about the same as the $1\,\sigma$ statistical fitting uncertainty in this region. Between $10\degr < |b| < 30\degr$, the median difference rises to $0.029\,$keV, about 1.5 times the statistical uncertainty in this region. We consider these to be the systematic uncertainties of the absolute values of the temperature of the LHB. However, we emphasise that the temperature gradient reported in Sect.~\ref{sec:LHB_kT} is still present, as this systematic effect only steers the absolute value of $kT_{\rm LHB}$ in a single direction and is only very weakly dependent on location.
% From the uniformity of the systematic increase, we assign a $0.007\,$keV systematic uncertainty in all our measurements related to LHB temperature.
% In addition, given the uniformity of this systematics, we do not revise the spherical harmonic models again for the broken power-law model since they would largely be different in the monopole, by the aforementioned amount.

% $0.009$\,keV above 30\degr ($\sim1\sigma$ fitting uncertainty). Rises to 0.029\,keV just between 10-30\degr ($\sim1.5$ times the $1\sigma$ fitting uncertainty)

\subsection{Robustness of the spectral fits} \label{sec:robust}
% The robustness of the spectral fits at different regions can be estimated from the fit statistics divided by the degrees of freedom (stat/dof). We present this map in Fig.~\ref{fig:fitstat}. There are 2695 and 2693 dof outside and inside the eROSITA bubbles, respectively.

The use of \texttt{cstat} in evaluating the goodness of fit is possible but not straightforward, as the expected value and the variance of \texttt{cstat} depend on the number of counts in each spectral bin \citep{Kaastra_2017}. Instead of the rule-of-thumb $\chi^2/{\rm dof}\sim1$ indicating a good fit, one needs to compare the fitted \texttt{cstat} with the expected value and variance of \texttt{cstat} in each spectrum, which is a variable number dependent on counts. Therefore, while the spectral fitting was done using \texttt{cstat} (the eROSITA part), we report the $\chi^2/{\rm dof}$ after rebinning each spectrum to have at least 10\,counts in each spectral bin using the same model. The rebinning was done following the scheme presented by \citet{Kaastra_2016}. We present the map of the resulting $\chi^2/{\rm dof}$ values in Fig.~\ref{fig:fitstat}.

We would like to remind the reader that the relatively small and narrow range of $\chi^2/{\rm dof}$ results partly from fitting the spectra up to $5\,$keV, where the instrumental background dominates with a small CXB contribution. The wide energy range between $\sim 2$--$5$\,keV can almost always be well-modelled by the dominating, but fixed, FWC model, thus inclined to result in low $\chi^2/{\rm dof}$. However, this energy range has strong constraining power on the subdominant CXB component, which is the only contributing component and was thus included in our spectral fitting.

\begin{figure*}
    \centering
    \includegraphics[width=0.49\textwidth]{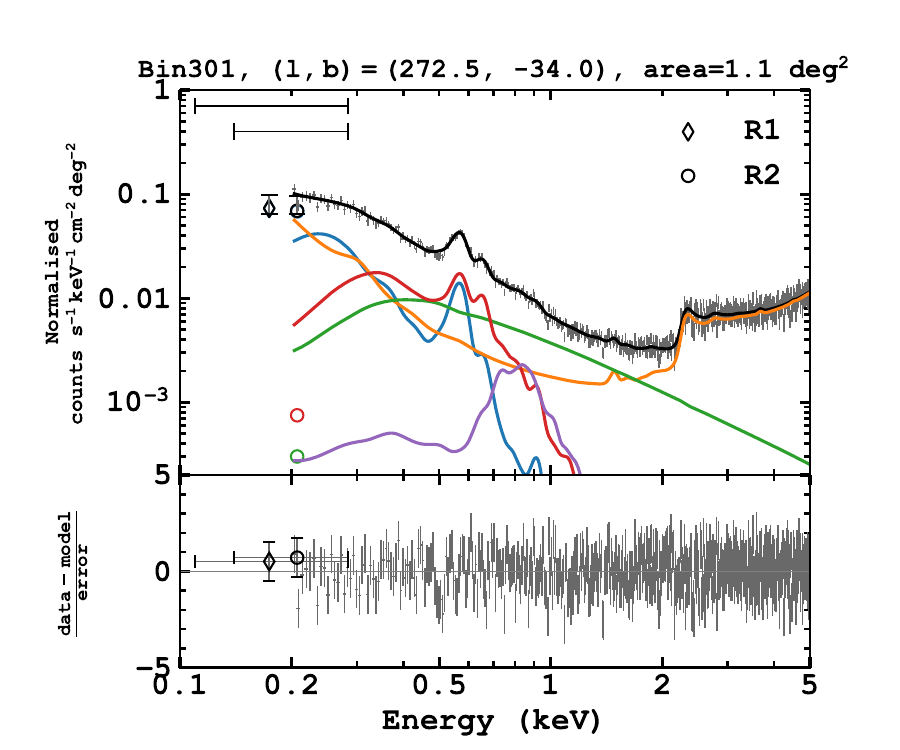}
    \includegraphics[width=0.49\textwidth]{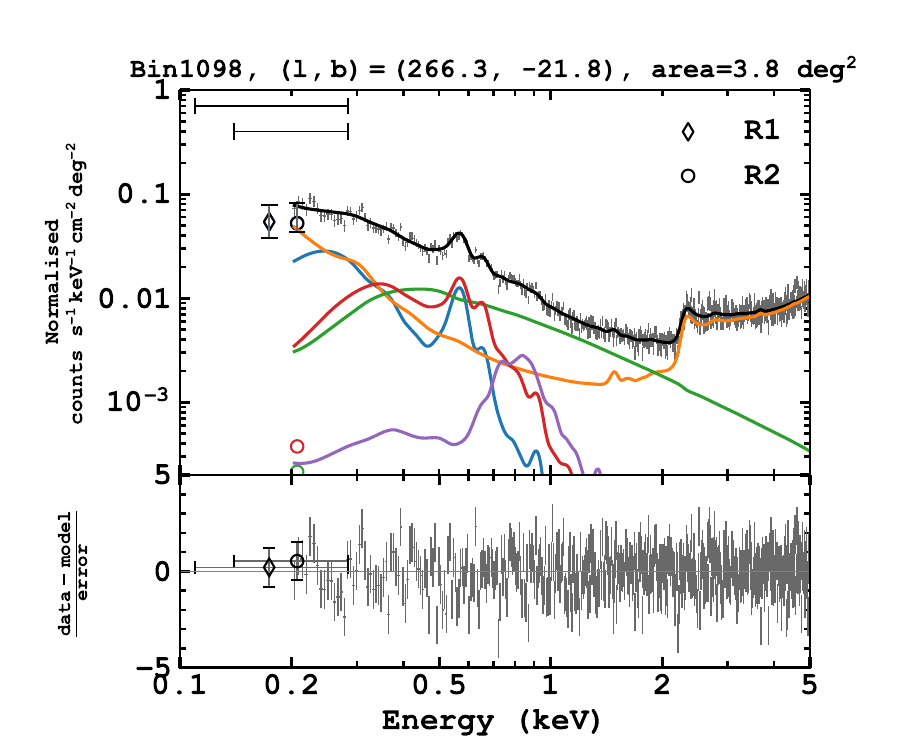}
    \caption{Example spectra having $\chi^2/{\rm dof}$ = 0.99 and 1.10.}
    \label{fig:chisq1}
\end{figure*}

\begin{figure*}
    \centering
    \includegraphics[width=0.49\textwidth]{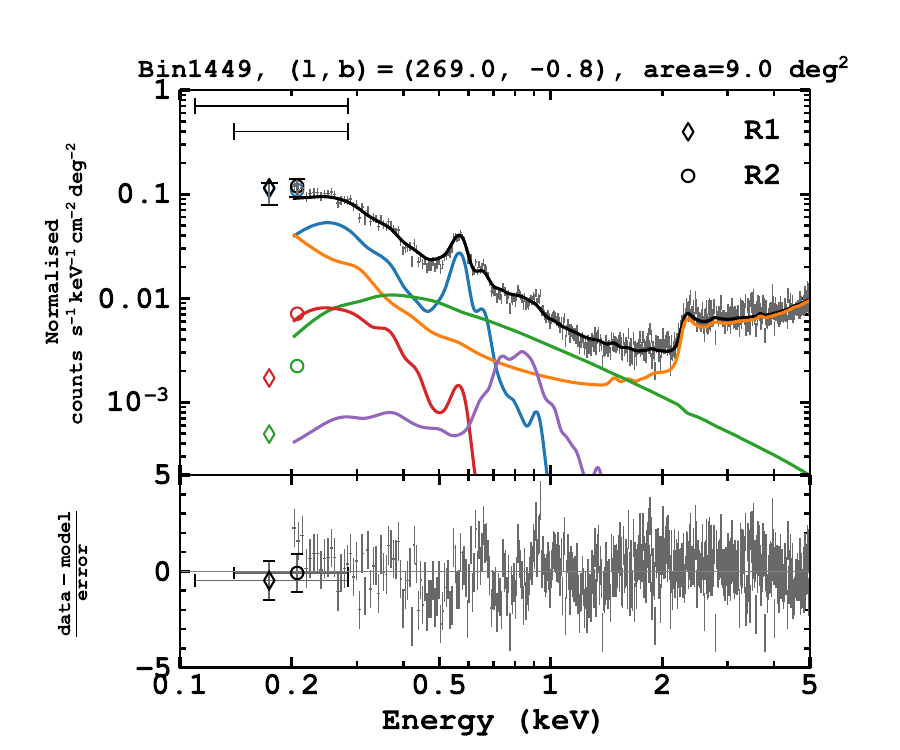}
    \includegraphics[width=0.49\textwidth]{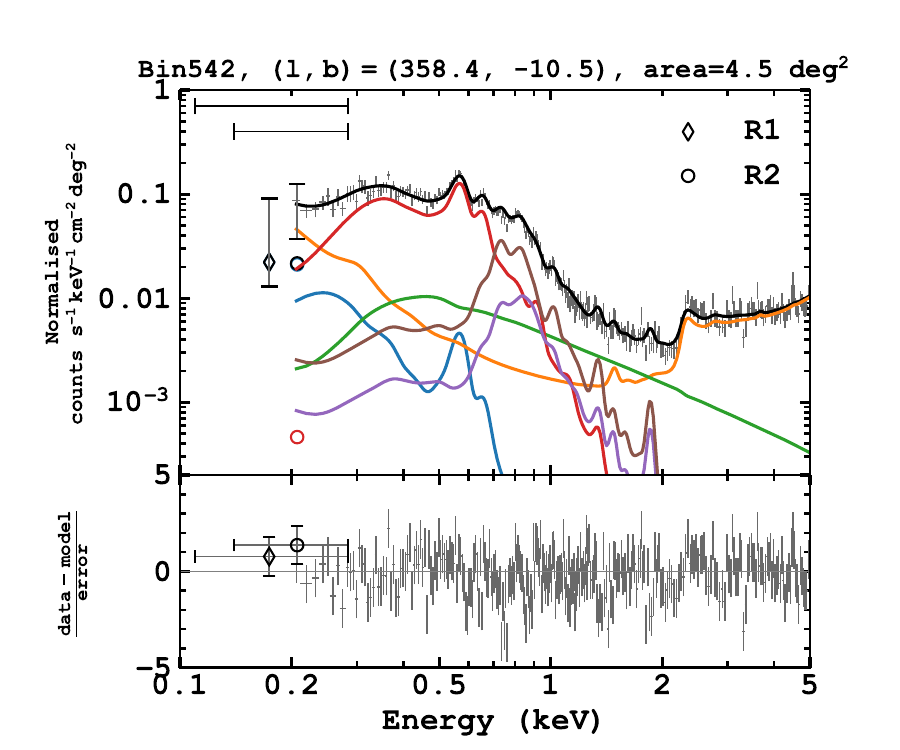}
    \caption{Example spectra having $\chi^2/{\rm dof}$ of 1.50 and 1.70, respectively. They are commonly found near the Galactic plane, towards the inner part of the Galaxy and the base of the eROSITA bubbles. The models usually cannot predict the spectrum between $\sim 0.7$--$1$\,keV well.}
    \label{fig:chisq2}
\end{figure*}

To better understand the robustness of the spectral fits, we show four spectra with increasing values of $\chi^2$/dof in Fig.~\ref{fig:chisq1} and \ref{fig:chisq2}, from $\sim1.0$--$1.7$. Fig.~\ref{fig:chisq1} shows spectra in the range of $1.0 \lesssim \chi^2\text{/dof} \lesssim1.1$, where no major discrepancies between data and model could be seen. Fig.~\ref{fig:fitstat} demonstrates these values are typical of spectra away from the Galactic plane and the central part of the Galaxy. On the Galactic plane, poorer spectral fits are expected because of the definite presence of multiple line-of-sight structures, both in emission and absorption, which are not part of our model template. Towards the inner part of the Galaxy ($310\degr\lesssim l \lesssim 360\degr$), this scenario exacerbates, and poorer fits begin to be prevalent to even high Galactic latitudes around $-20\degr\lesssim b \lesssim 30\degr$. There are many ways to obtain poor fits, but one of the most common ways is the poor reproduction between $\sim0.7$--$1$\,keV, where the \ion{Fe}{XVII}, \ion{Ne}{IX} and \ion{Ne}{X} lines dominate. The reproduction is usually imperfect when these lines are bright, especially at the base of the eROSITA bubbles. However, it is currently difficult to distinguish between calibration imperfections and real variations in the astrophysical source. The upcoming pipeline processing version \texttt{c030} will address the former slightly in terms of including a more accurate energy resolution in the redistribution matrix.

%This is a disadvantage of not imposing priors  on $\mathrm{EM}_{\rm LHB}$ within the eROSITA bubbles region, as it is degenerate with the other two thermal components. However, this common pitfall can be easily screened out by masking regions that have vanishing $\mathrm{EM}_{\rm LHB}$, or equivalently, vanishing uncertainty in $\mathrm{EM}_{\rm LHB}$ obtained from the MCMC chain as we did in Sect.~\ref{sec:LHB_kT}. 

% The left spectrum has a vanishing contribution from the LHB. Despite describing the eROSITA spectrum nicely, this model severely under-predicts the R1 and R2 band count rates. The right spectrum indicates a possible tension between the ROSAT and eROSITA data, as the model under-estimates the eROSITA count rates below $\sim 0.3\,$keV, but over-estimates the ROSAT count rates.

% Our spectral fitting results are visualised and can be consulted for each contour bin individually through the accompanying website. We also provide a table recording each contour bin's coordinates and fit results.
The robustness of each region's fit and MCMC chain could be inspected by its associated corner plot and steps plot in the choropleth maps on our website (Sect.~\ref{sec:website}).

\section{Conclusion} \label{sec:conclusion}
In this work, we binned the western Galactic hemisphere into $\sim 2000$ regions. We extracted their spectra at solar minimum (eRASS1), where the heliospheric SWCX emission was negligible (Dennerl et al. in prep). These spectra were fitted with a model template consisting of the LHB, the Milky Way's CGM, the Galactic corona, the CXB, and, depending on the location, the eROSITA bubbles. This resulted in maps of the parameters, primarily the temperature and emission measure maps for the thermal plasma. We focused on the results regarding the LHB. We summarise our main findings below:

\begin{itemize}
    \item We found the median temperature of the LHB to be $0.111^{+0.018}_{-0.015}\,$keV. Much of the spread comes from the approximately north-south gradient above Galactic latitudes of $30\degr$, with the south being hotter ($121.8\pm0.6\,$eV) than the north ($100.8\pm0.5\,$eV). Venturing closer to the Galactic plane, there seems to be an increase in LHB temperature, especially at $l\gtrsim270\degr$. The origin of the temperature gradient is unclear, but it could be set up by more recent supernova explosions within the LHB. The enhancement in temperature towards the Galactic plane could be due to the enhanced thermal pressure needed in the Galactic plane for the LHB to maintain pressure equilibrium with the surroundings.

    \item The emission measure of the LHB is higher towards high latitudes in both hemispheres. It entails that the LHB is more extended towards high latitudes, assuming a constant electron density. The emission measure is also spatially anti-correlated with the local dust column density, consistent with the displacement model put forth by \citet{Sanders1977}. We produced a 3D model of the LHB in the western Galactic hemisphere.

    \item We found two tunnels with a low local column density that appear to be filled by hot plasma. One is the well-known $\beta$~CMa tunnel, and the other is towards ($l$,$b$)$\sim$($315\degr$, $25\degr$), in the constellation Centaurus. This hints at the possibility of a widespread tunnel network connecting regions filled by the hot phase of the ISM.

    % \item Apart from the LHB, we found the eROSITA bubbles exhibiting a cooler shell at $0.2\,$keV around $5\degr$ in width, enveloping hotter ($\sim 0.3\,$keV) plasma in the interior. Interestingly, one cannot identify the same shell structure from the emission measure map. It might represent a shell of under-ionised plasma, yet to reach equilibrium.

    \item Both a simple and broken power law could fit the cosmic X-ray background well, but with a steeper slope ($\Gamma\simeq1.6$--$1.7$ above $1.2\,$keV) than the conventional value ($\Gamma=1.45$). Whether this represents a genuine steepening of the CXB slope below $\sim 1\,$keV or is caused by calibration inaccuracies is uncertain. It will likely become clear with a forthcoming new calibration version.

    From this study, it is clear that eROSITA and its all-sky surveys provide a valuable dataset for studying the SXRB. Future papers will follow to discuss other aspects of the SXRB. 

\end{itemize}
\section{Data availability} \label{sec:website}
The results of the spectral fits can be accessed and visualised as choropleth maps via the website \url{https://erosita.mpe.mpg.de/dr1/AllSkySurveyData_dr1/DiffuseBkg/} hosted on the eROSITA DR1 server. An interactive version of the 3D model of the LHB (Fig.~\ref{fig:3D}) and structures in the solar neighbourhood can also be accessed there, as well as the spherical harmonics models of LHB temperature presented in Sect.~\ref{subsec:sph_harm} and Appendix~\ref{app:sph_harm}. \texttt{Xspec} model files are also available through the website for readers interested in particular regions in the western Galactic hemisphere. The 3D interactive map in Fig.~\ref{fig:3D} is also available at \href{https://www.aanda.org/articles/aa/olm/2024/10/aa51045-24/aa51045-24.html}{https://www.aanda.org}.

\begin{acknowledgements}
MY thanks Mattia Pacicco and Michael Schulreich for their in-depth explanations of their LB simulations during the EAS meeting 2024 in Padova, which has greatly improved the interpretation of the results. MY extends his gratitude to Susanne Friedrich and Harald Baumgartner for their guidance and help in setting up the accompanying website on the eROSITA DR1 server. MY appreciates the helpful discussion on cluster masks and measurement of cluster temperature provided by Ang Liu. 
MY and MF acknowledge support from the Deutsche Forschungsgemeinschaft through the grant FR 1691/2-1.
GP acknowledges funding from the European Research Council (ERC) under the European Union’s Horizon 2020 research and innovation programme (grant agreement No 865637). MS acknowledges support from the Deutsche Forschungsgemeinschaft through the grants SA 2131/13-1, SA 2131/14-1, and SA 2131/15-1.
This work is based on data from eROSITA, the soft X-ray instrument aboard SRG, a joint Russian-German science mission supported by the Russian Space Agency (Roskosmos), in the interests of the Russian Academy of Sciences represented by its Space Research Institute (IKI), and the Deutsches Zentrum für Luft- und Raumfahrt (DLR). The SRG spacecraft was built by Lavochkin Association (NPOL) and its subcontractors, and is operated by NPOL with support from the Max Planck Institute for Extraterrestrial Physics (MPE).

The development and construction of the eROSITA X-ray instrument was led by MPE, with contributions from the Dr. Karl Remeis Observatory Bamberg \& ECAP (FAU Erlangen-Nuernberg), the University of Hamburg Observatory, the Leibniz Institute for Astrophysics Potsdam (AIP), and the Institute for Astronomy and Astrophysics of the University of Tübingen, with the support of DLR and the Max Planck Society. The Argelander Institute for Astronomy of the University of Bonn and the Ludwig Maximilians Universität Munich also participated in the science preparation for eROSITA.

The eROSITA data shown here were processed using the eSASS/NRTA software system developed by the German eROSITA consortium. 

MY thanks the developers of the following software and packages that made this work possible:  \texttt{numpy} \citep{numpy}, \texttt{matplotlib} \citep{matplotlib}, \texttt{astropy} \citep{astropy:2013, astropy:2018, astropy:2022}, \texttt{PyXspec} \citep{xspec}, \texttt{HEAsoft} \citep{ftools2}, \texttt{FTOOLS} \citep{ftools1}, \texttt{contbin} \citep{contbin},  \texttt{lmfit} \citep{lmfit}, \texttt{emcee} \citep{emcee}, \texttt{corner.py} \citep{corner}, \texttt{plotly} \citep{plotly}, \texttt{folium} \citep{folium}, \texttt{K3D-jupyter} \citep{k3d}, \texttt{OpenCV} \citep{opencv} and \texttt{scipy} \citep{scipy}.
\end{acknowledgements}

\bibliographystyle{aau}
\bibliography{main}

\appendix

\section{Test for non-equilibrium ionisation} \label{app:nei}
Non-equilibrium ionisation (NEI) can happen in the LHB if there were recent rapid heating or cooling events. Electrons respond quickly to temperature change via collisions, but ionisation and recombination of the ions lag behind \citep[e.g. see a review by][]{Breitschwerdt_2021}. As a result, the line intensities under the assumption of collisional ionisation equilibrium (CIE) do not reflect the real temperature of the gas. 
If a recent supernova exploded within the LHB, parts or all of LHB could be under-ionised due to rapid shock-heating. On the other hand, if the LHB spawned from a dense cloud, it could undergo rapid adiabatic cooling when it bursts out of the cloud, resulting in an over-ionised plasma \citep{Breitschwerdt_1994}. \citet{Henley_2007} investigated both scenarios using a combination of \textit{XMM-Newton}/EPIC-MOS and ROSAT/PSPC data and concluded the data could not distinguish between CIE and under-ionised plasma but disfavoured the over-ionised scenario. 

We used a simple test to evaluate whether our data (eRASS1+ROSAT R12) are sensitive to the signature of NEI. We chose the same regions that were used to determine the spectral shape of the CXB (Fig.~\ref{fig:CXB_reg}) for this analysis. They have high S/N (400) in the soft band. This minimises the statistical uncertainty and boosts any potential signatures of NEI. We refitted the spectra by replacing the CIE LHB model (\texttt{apec}) with a NEI LHB model (\texttt{nei}).

\begin{figure*}
    \centering
    \includegraphics[width=0.49\textwidth]{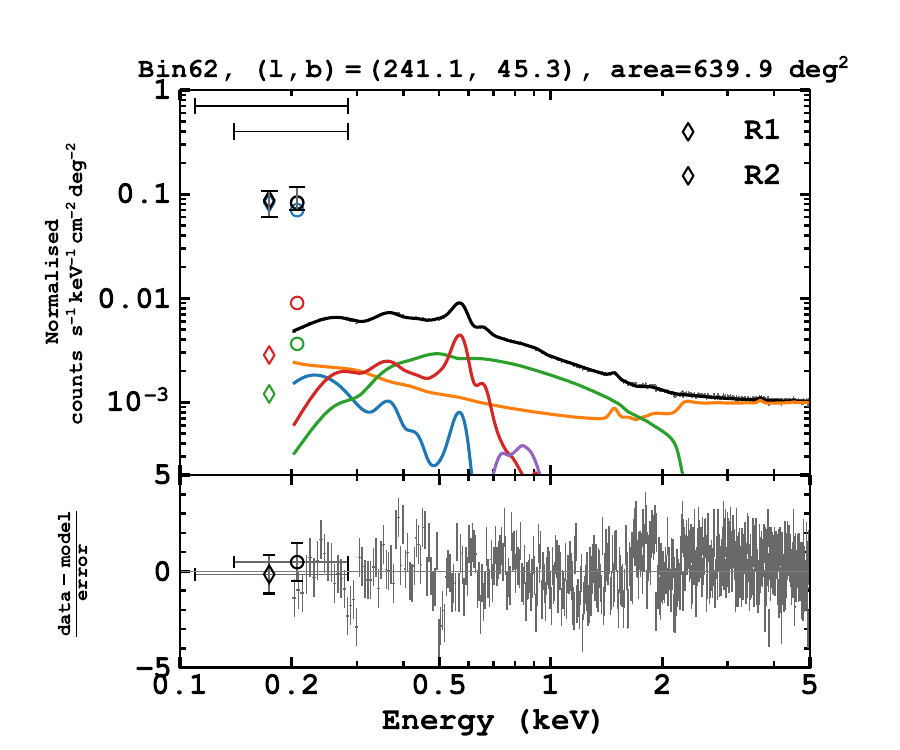}
    \includegraphics[width=0.49\textwidth]{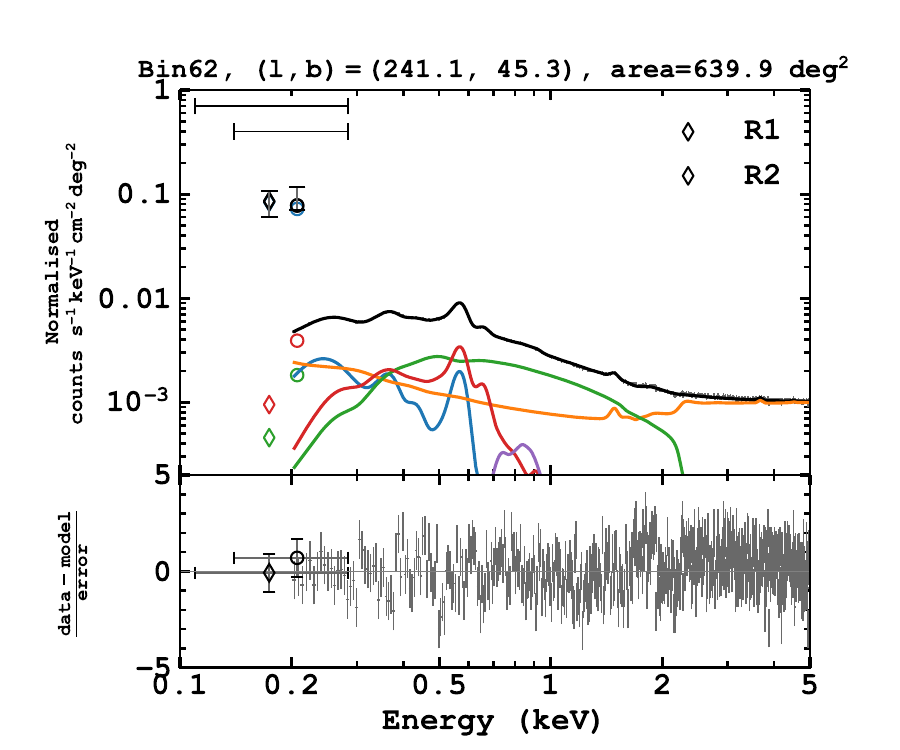}
    \caption{Comparison of spectral fits assuming the LHB in CIE (\textit{left}) and NEI (\textit{right}) of the region showing a potential NEI signature. The residual highlights the  improvement below $0.5\,$keV.}
    \label{fig:cie_nei}
\end{figure*}

We summarise the fitting results in Table~\ref{tab:nei}. Six of the seven regions have the fitted density-weighted ionisation timescale $\tau>10^{12}\,{\rm cm^{-3}\,s}$. At $\tau>10^{12}\,{\rm cm^{-3}\,s}$, \citet{Smith_2010} showed that 90\% of carbon and oxygen ions (which dominate the observed LHB emission) would reach CIE for plasma at $\sim 0.1\,$keV. The one exception is the region centred at ($l$,\,$b$)=($241\degr$,\,$45\degr$), which returns $\tau=(5.6^{+0.8}_{-0.7})\times10^{11}\,{\rm cm^{-3}\,s}$, suggesting a small under-ionisation. Fig.~\ref{fig:cie_nei} compares the CIE and NEI models in this region. From the residual spectra, one can see the improvement below $\sim0.5\,$keV. Nevertheless, because the NEI models of most regions reduce to the CIE case even at a high S/N of 400, we conclude our data can be sufficiently explained by CIE and keep this assumption throughout our analysis in the main text.

\begin{table*}[h]
    \centering
    \begin{tabular}{cc|ccc|cc}
    \hline\hline
    \multicolumn{2}{c|}{} & \multicolumn{3}{c|}{NEI} & \multicolumn{2}{c}{CIE}\\
      $l\,(\degr)$   & $b\,(\degr)$ & $kT_{\rm LHB}\,({\rm keV})$ & $\tau_{\rm LHB}\,(10^{12}\,{\rm cm^{-3}\,s})$ & ${\rm EM_{LHB}\,(10^{-3}\,cm^{-6}\,pc)}$ & $kT_{\rm LHB}\,({\rm keV})$ & ${\rm EM_{LHB}\,(10^{-3}\,cm^{-6}\,pc)}$ \\\hline
      $228$ & $-72$ & $0.120\pm0.002$ & $25^{+16}_{-17}$ & $3.41^{+0.14}_{-0.13}$ & $0.115\pm0.002$ & $3.09^{+0.15}_{-0.14}$\\
      $202$ & $-61$ & $0.114^{+0.003}_{-0.002}$ & $30^{+13}_{-19}$ & $3.11^{+0.12}_{-0.16}$ & $0.112^{+0.002}_{-0.003}$ & $2.84^{+0.13}_{-0.12}$ \\
      $229$ & $-48$ & $0.119\pm0.003$ & $26\pm15$ & $4.34^{+0.27}_{-0.37}$ & $0.121\pm0.002$ & $4.05^{+0.27}_{-0.18}$ \\
      $205$ & $39$ & $0.103\pm0.002$ & $23^{+19}_{-17}$ & $3.48^{+0.14}_{-0.13}$ & $0.098\pm0.002$ & $3.19^{+0.14}_{-0.17}$ \\
      $241$ & $45$ & $0.116\pm0.003$ & $0.55^{+0.08}_{-0.07}$ & $3.12^{+0.12}_{-0.11}$ & $0.097^{+0.003}_{-0.002}$ & $2.96^{+0.18}_{-0.17}$\\
      $208$ & $66$ & $0.104\pm0.002$ & $25\pm16$ & $5.63^{+0.19}_{-0.21}$ & $0.098\pm0.002$ & $5.41^{+0.26}_{-0.25}$\\
      $258$ & $67$ & $0.104^{+0.003}_{-0.002}$ & $17^{+22}_{-16}$ & $4.72^{+0.18}_{-0.17}$ & $0.097\pm0.002$ & $4.43^{+0.21}_{-0.20}$\\\hline
    \end{tabular}
    \caption{Fitted LHB parameters of the seven high-SN spectra under the NEI and CIE assumptions. All but one region show an ionisation timescale of $\tau>10^{12}\,{\rm cm^{-3}\,s}$, essentially reducing to the CIE case. }
    \label{tab:nei}
\end{table*}

\section{Posterior distributions of the dipole and $l_{\rm max}=6$ models of local hot bubble temperature} \label{app:sph_harm}
The best-fit parameters of the dipole model in Fig.~\ref{fig:dipole} are presented in Table\,\ref{tab:dipole}. The $z$-axis points towards the Galactic north pole in our convention, and the $x$-axis points towards the Galactic centre. Two equivalent representations are presented: the complex $a_{lm}$ coefficients and the dipole vector. In the complex representation, each multipole gives $2l+1$ free parameters; Each $a_{l|m|}$ contributes to two, one for the real part and one for the imaginary part, except from $a_{l0}$ coefficients, which do not have a complex part for real-valued functions.

\begin{table}[h]
    \centering
    \caption{Best-fit parameters of the dipole ($l_{\rm max}=1$) model of $kT_{\rm LHB}$. The resulting temperature model is in units of keV.}
    \begin{tabular}{cc}
    \hline\hline
    \multicolumn{2}{c}{Complex}\\\hline
        $a_{00}$ & $0.368\pm0.002$ \\
        $a_{10}$ & $-0.027\pm0.001$ \\
        Re\{$a_{11}$\}\tablefootmark{i} & $0.014\pm0.001$\\
        Im\{$a_{11}$\}\tablefootmark{i} & $-0.036\pm0.002$\\
        \hline
    \multicolumn{2}{c}{Dipole vector}\\\hline
    $A_0$\tablefootmark{ii} & $0.104\pm0.001$ \\
    $A_1$\tablefootmark{ii} & $0.030\pm0.001$\\
    $\phi_{\rm max}$ & $291\fdg1\pm1\fdg8$\\
    $\theta_{\rm max}$& $116\fdg5^{+1\fdg5}_{-1\fdg6}$\\\hline
    \end{tabular}
    \tablefoot{
    \tablefoottext{i}{Re\{$a$\} and Im\{$a$\} refer to the real and imaginary part of the complex coefficient $a$.}\\
    \tablefoottext{ii}{$A_l$ has the unit of keV and can be interpreted as the amplitude of the corresponding multipole.}
    }
    \label{tab:dipole}
\end{table}

For the dipole, it can be shown that $a_{10}$ and $a_{11}$ are related to the dipole vector by the following equations:
\begin{empheq}[left={\empheqlbrace}]{alignat=2}
    &A_1 =  a_{10}\sqrt{\frac{3}{4\pi}}\cos{\theta_{\rm max}} - 2\sqrt{\frac{3}{8\pi}}\sin{\theta_{\rm max}}f,\\
    &\tan{\theta_{\rm max}} = -\frac{\sqrt{2}}{a_{10}}f\\
    &\tan{\phi_{\rm max}} = -\frac{{\rm Im}\{a_{11}\}}{{\rm Re}\{a_{11}\}},
\end{empheq}
where $f = {\rm Re}\{a_{11}\}\cos{\phi_{\rm max}}-{\rm Im}\{a_{11}\}\sin{\phi_{\rm max}}$, $A$ is the amplitude of the dipole and ($\theta_{\rm max}$, $\phi_{\rm max}$; colatitude and azimuth in spherical coordinates) is the direction where this amplitude is obtained.
% In this work, we used the code \texttt{polyMV} \citep{polyMV} to obtain the quadrupole vectors numerically.

\begin{figure}[ht]
    \centering
    \includegraphics[width=0.5\textwidth]{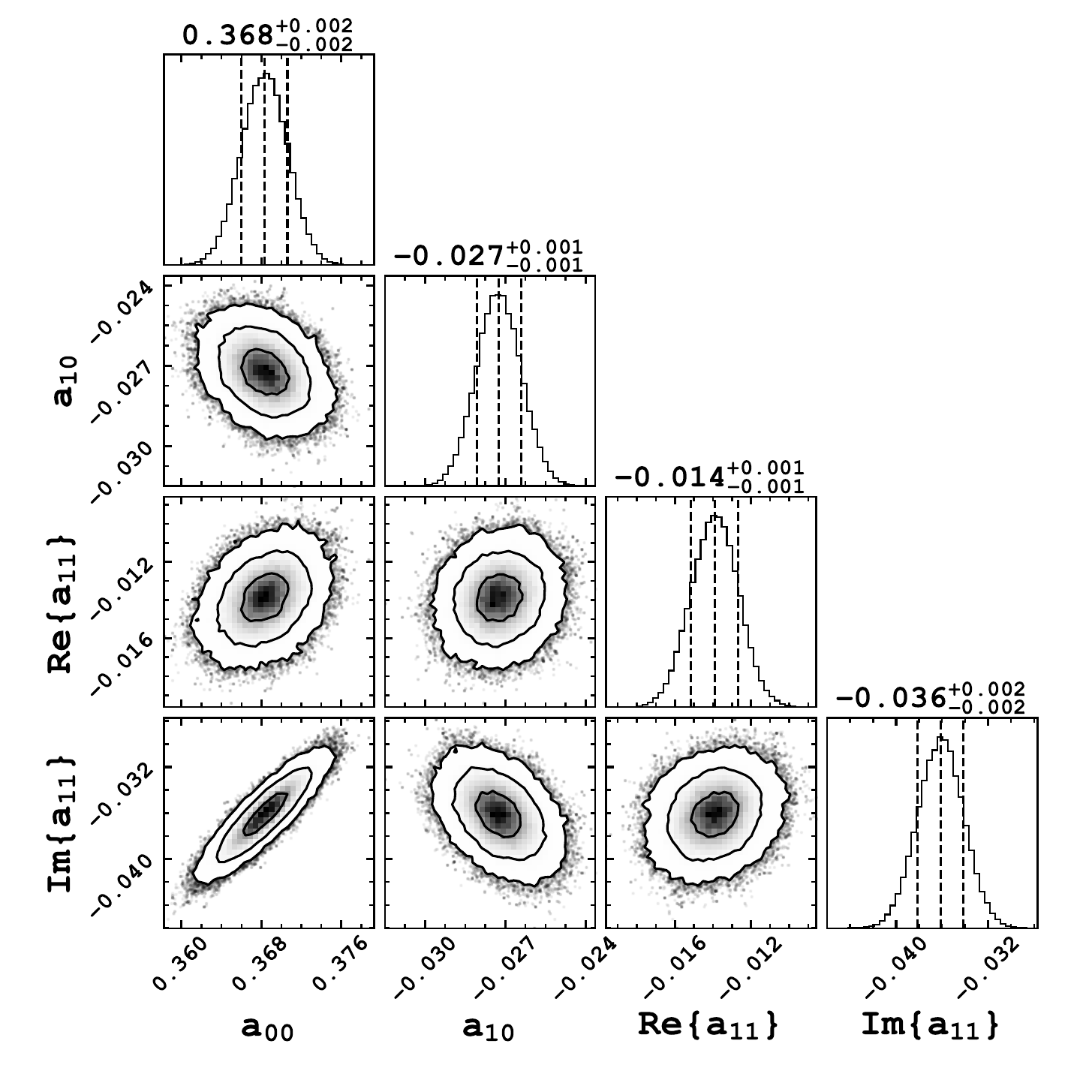}
    \includegraphics[width=0.5\textwidth]{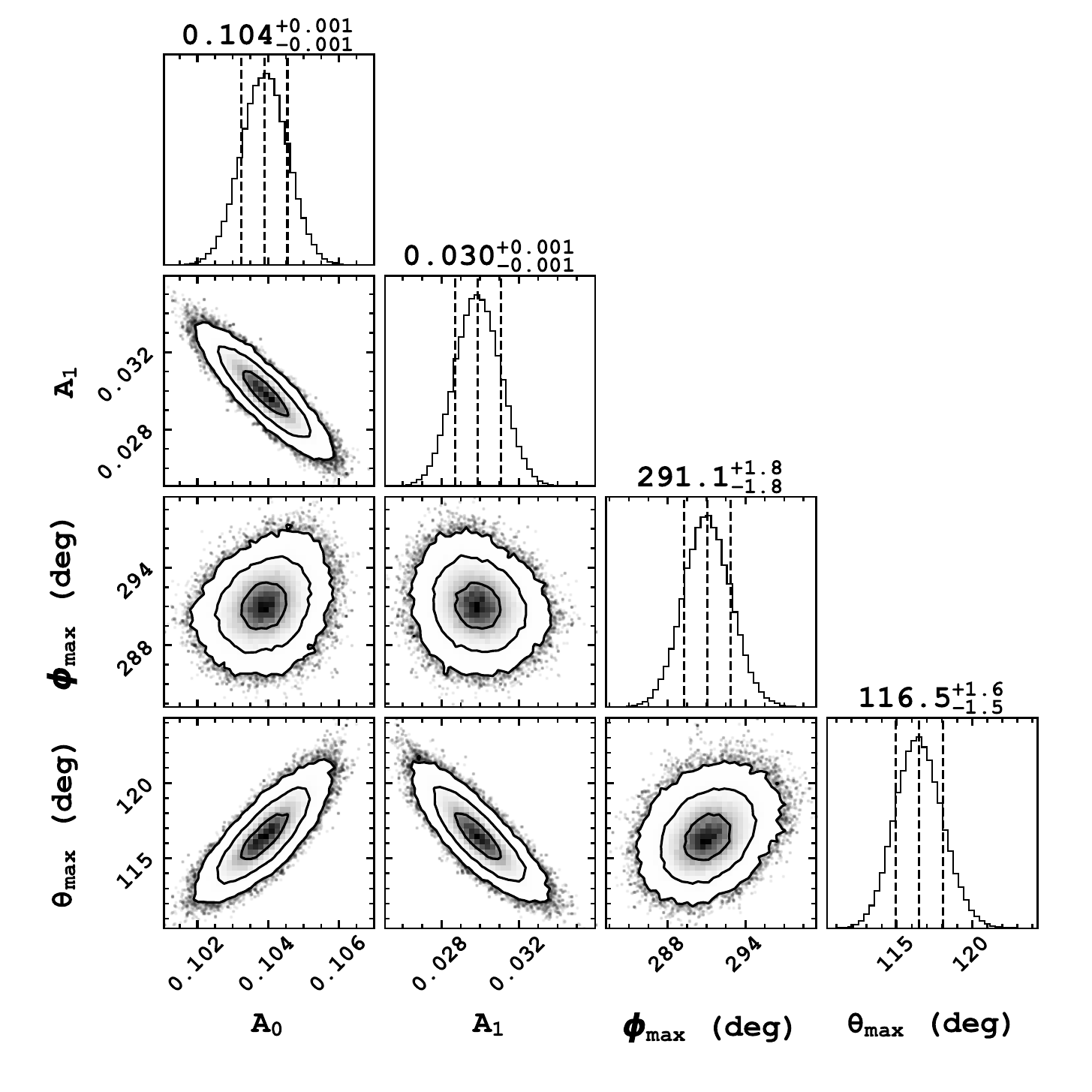}
    \caption{Posterior distributions of the dipole ($l_{\rm max} =1$) model in complex (\textit{top}) and multipole vector (\textit{bottom}) representations. The contours show the 1, 2, and 3\,$\sigma$ uncertainties.}
    \label{fig:dipole_corner}
\end{figure}

The posterior distributions of the dipole are shown in Fig.\,\ref{fig:dipole_corner} both in complex and dipole vector representations.

We list the best-fit parameters of the $l_{\rm max}=6$ spherical harmonics model (Fig.~\ref{fig:lmax6}) in Table~\ref{tab:lmax6}.

\begin{table}[ht]
    \centering
    \caption{Best-fit parameters of the $l_{\rm max}=6$ model of $kT_{\rm LHB}$. The resulting temperature model is in units of keV.}
    \begin{tabular}{cc}
    \hline\hline
$a_{00}$ & $6.165\pm8.591$ \\
$a_{10}$ & $-2.583\pm0.730$ \\
${\rm Re}\{a_{11}\}$ & $-1.425\pm0.983$ \\
${\rm Im}\{a_{11}\}$ & $5.622\pm8.654$ \\
$a_{20}$ & $-3.707\pm5.337$ \\
${\rm Re}\{a_{21}\}$ & $-0.452\pm0.112$ \\
${\rm Im}\{a_{21}\}$ & $-2.798\pm0.785$ \\
${\rm Re}\{a_{22}\}$ & $-3.858\pm6.466$ \\
${\rm Im}\{a_{22}\}$ & $-2.161\pm1.484$ \\
$a_{30}$ & $2.327\pm0.624$ \\
${\rm Re}\{a_{31}\}$ & $0.556\pm0.346$ \\
${\rm Im}\{a_{31}\}$ & $-1.990\pm2.931$ \\
${\rm Re}\{a_{32}\}$ & $2.067\pm0.566$ \\
${\rm Im}\{a_{32}\}$ & $-0.690\pm0.165$ \\
${\rm Re}\{a_{33}\}$ & $1.952\pm1.294$ \\
${\rm Im}\{a_{33}\}$ & $-1.798\pm3.709$ \\
$a_{40}$ & $0.884\pm1.186$ \\
${\rm Re}\{a_{41}\}$ & $0.213\pm0.051$ \\
${\rm Im}\{a_{41}\}$ & $1.300\pm0.332$ \\
${\rm Re}\{a_{42}\}$ & $0.775\pm1.232$ \\
${\rm Im}\{a_{42}\}$ & $0.510\pm0.299$ \\
${\rm Re}\{a_{43}\}$ & $0.564\pm0.130$ \\
${\rm Im}\{a_{43}\}$ & $1.099\pm0.292$ \\
${\rm Re}\{a_{44}\}$ & $0.476\pm1.594$ \\
${\rm Im}\{a_{44}\}$ & $1.177\pm0.752$ \\
$a_{50}$ & $-0.520\pm0.122$ \\
${\rm Re}\{a_{51}\}$ & $-0.091\pm0.044$ \\
${\rm Im}\{a_{51}\}$ & $0.242\pm0.341$ \\
${\rm Re}\{a_{52}\}$ & $-0.480\pm0.116$ \\
${\rm Im}\{a_{52}\}$ & $0.164\pm0.038$ \\
${\rm Re}\{a_{53}\}$ & $-0.254\pm0.135$ \\
${\rm Im}\{a_{53}\}$ & $0.178\pm0.357$ \\
${\rm Re}\{a_{54}\}$ & $-0.389\pm0.103$ \\
${\rm Im}\{a_{54}\}$ & $0.290\pm0.062$ \\
${\rm Re}\{a_{55}\}$ & $-0.475\pm0.283$ \\
${\rm Im}\{a_{55}\}$ & $-0.008\pm0.473$ \\
$a_{60}$ & $-0.049\pm0.054$ \\
${\rm Re}\{a_{61}\}$ & $-0.019\pm0.005$ \\
${\rm Im}\{a_{61}\}$ & $-0.119\pm0.024$ \\
${\rm Re}\{a_{62}\}$ & $-0.032\pm0.054$ \\
${\rm Im}\{a_{62}\}$ & $-0.035\pm0.015$ \\
${\rm Re}\{a_{63}\}$ & $-0.053\pm0.012$ \\
${\rm Im}\{a_{63}\}$ & $-0.104\pm0.022$ \\
${\rm Re}\{a_{64}\}$ & $-0.010\pm0.056$ \\
${\rm Im}\{a_{64}\}$ & $-0.069\pm0.030$ \\
${\rm Re}\{a_{65}\}$ & $-0.079\pm0.015$ \\
${\rm Im}\{a_{65}\}$ & $-0.082\pm0.021$ \\
${\rm Re}\{a_{66}\}$ & $0.040\pm0.078$ \\
${\rm Im}\{a_{66}\}$ & $-0.098\pm0.057$ \\ \hline

    \end{tabular}
    \tablefoot{Re\{$a_{lm}$\} and Im\{$a_{lm}$\} refer to the real and imaginary part of the complex coefficient $a_{lm}$.}
    \label{tab:lmax6}
\end{table}

\section{Latitudinal profiles of local hot bubble temperature} \label{app:lat_profile}
\begin{figure*}
    \centering
    \includegraphics[width=0.99\textwidth]{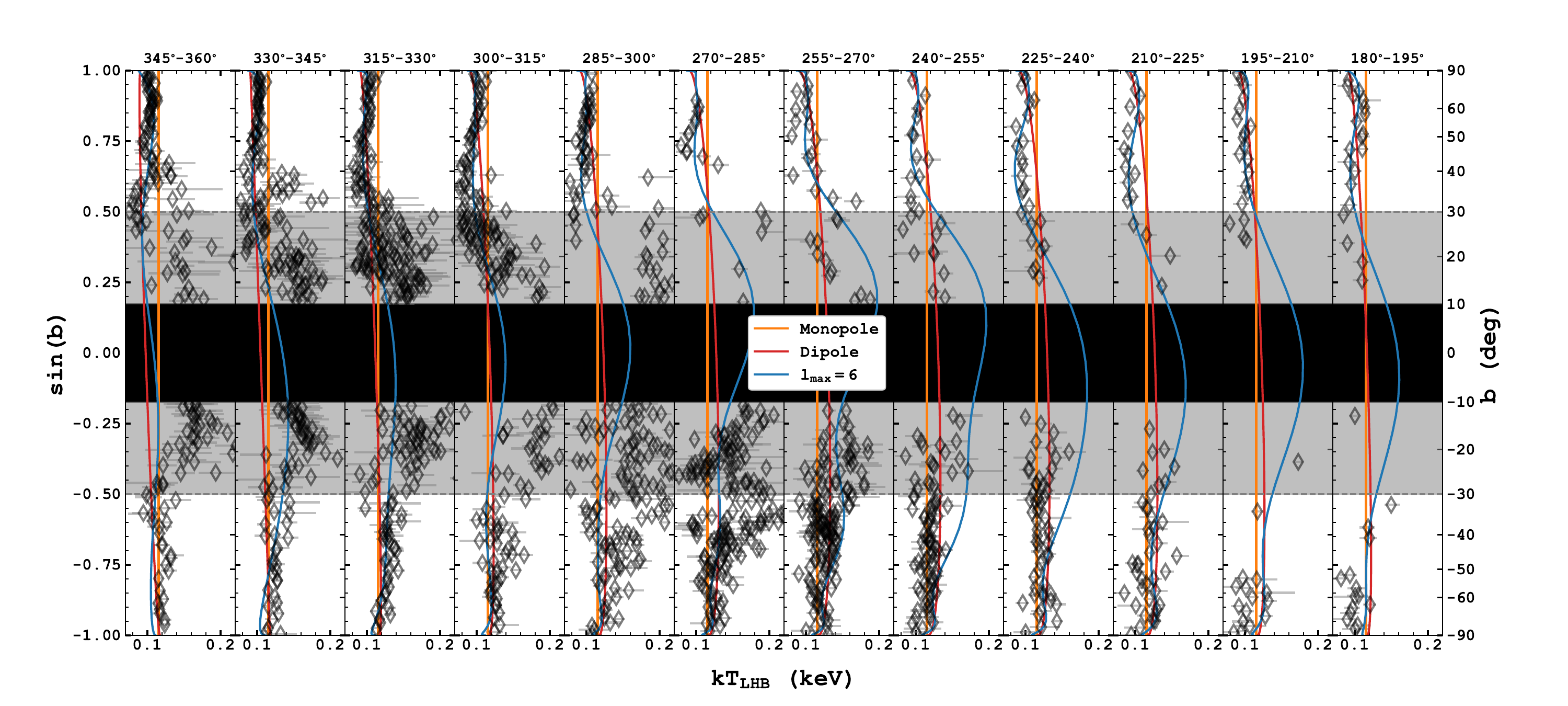}
    \caption{Latitudinal profiles of $kT_{\rm LHB}$ with their $1\,\sigma$ error bars in all longitudinal stripes of $15\degr$ width in the western Galactic hemisphere. The vertical orange, red and blue lines show the corresponding monopole, dipole and $l_{\rm max}=6$ models, respectively. The black region indicates the Galactic plane region, which we masked for the temperature gradient analysis. Regions plotted on the white background indicate where the dipole model was fitted ($\left\vert b\right\vert > 30\degr$). The $l_{\rm max}$ model was fitted with the inclusion of data from the shaded region ($\left\vert b\right\vert > 10\degr$).}
    \label{fig:lat_profile}
\end{figure*}

Fig.~\ref{fig:lat_profile} shows the $kT_{\rm LHB}$ data in another light by dividing the western Galactic hemisphere in $15\degr$-wide longitudinal stripes and shows $kT_{\rm LHB}$ as a function of latitude. This presentation has the advantage of enabling visual inspection of the uncertainty associated with individual contour bins and their comparison with the large-scale gradient.  Lines of the monopole (orange), dipole (red) and $l_{\rm max}=6$ (blue)  models are overlaid in addition. With the help of the best-fit monopole model line, it is apparent that $kT_{\rm LHB}$ is higher in the southern hemisphere than in the north. The dipole model can capture the primary latitudinal gradient, especially in the south, but fails to follow the data closely above $b>30\degr$.
% such as those near the Large Magellanic Cloud ($270\degr \lesssim l \lesssim 315\degr$,$-30\degr\lesssim b \lesssim-15\degr$) and on the west of Antlia supernova remnant ($240\degr \lesssim l \lesssim 270\degr$,$15\degr\lesssim b \lesssim30\degr$).
While the model complexity can increase to accommodate the temperature variation on a smaller scale, we note the presence of a temperature gradient in LHB is clear. The scatter is large in the region ($270\degr \lesssim l \lesssim 300\degr$,$-40\degr\lesssim b \lesssim-10\degr$), which partially covers the Large Magellanic Cloud but cannot be fully attributed to it. The relatively large scatter below $10\degr<b<30\degr$ and $l>300\degr$ is caused by the lower ${\rm EM_{LHB}}$ in this direction, resulting in higher uncertainties in the determination of $kT_{\rm LHB}$.

\section{Local hot bubble temperature dichotomy seen from high S/N spectra} \label{app:highSN_spec}
Sect.~\ref{sec:LHB_kT} demonstrated the north-south dichotomy of the LHB temperature, mainly by the systematic temperature offset between the hemispheres. However, this offset is difficult to appreciate purely by looking at individual spectra of $\text{S/N}\simeq80$ because the temperature difference of $0.02$\,keV is only approximately twice the typical fitting uncertainty, as shown in Fig.~\ref{fig:LHB_kT_hist}.

\begin{figure*}
    \centering
    \includegraphics[width=0.49\textwidth]{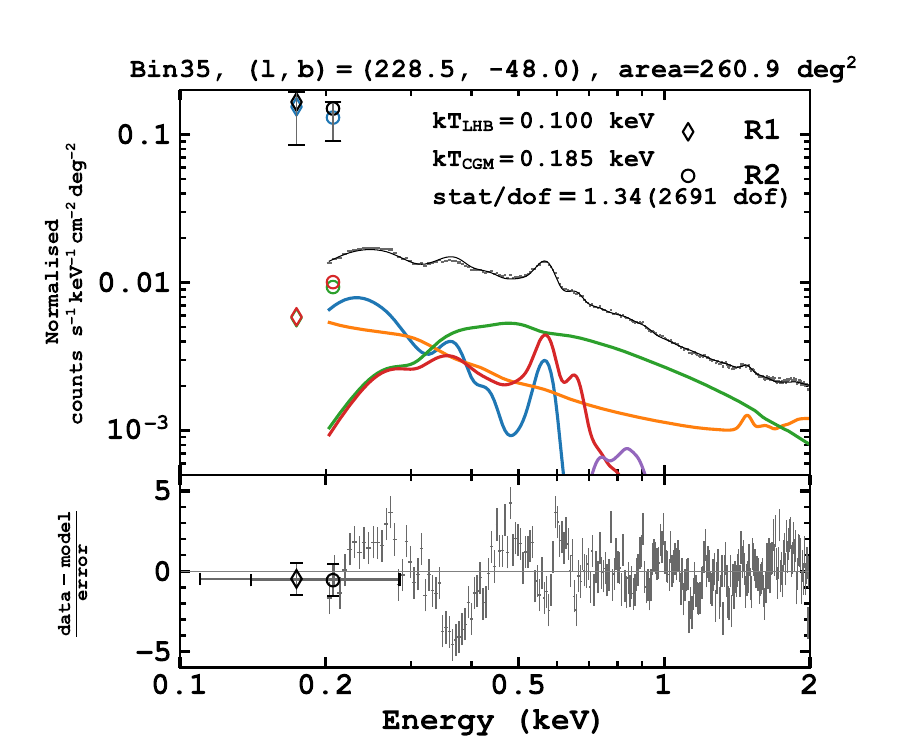}
    \includegraphics[width=0.49\textwidth]{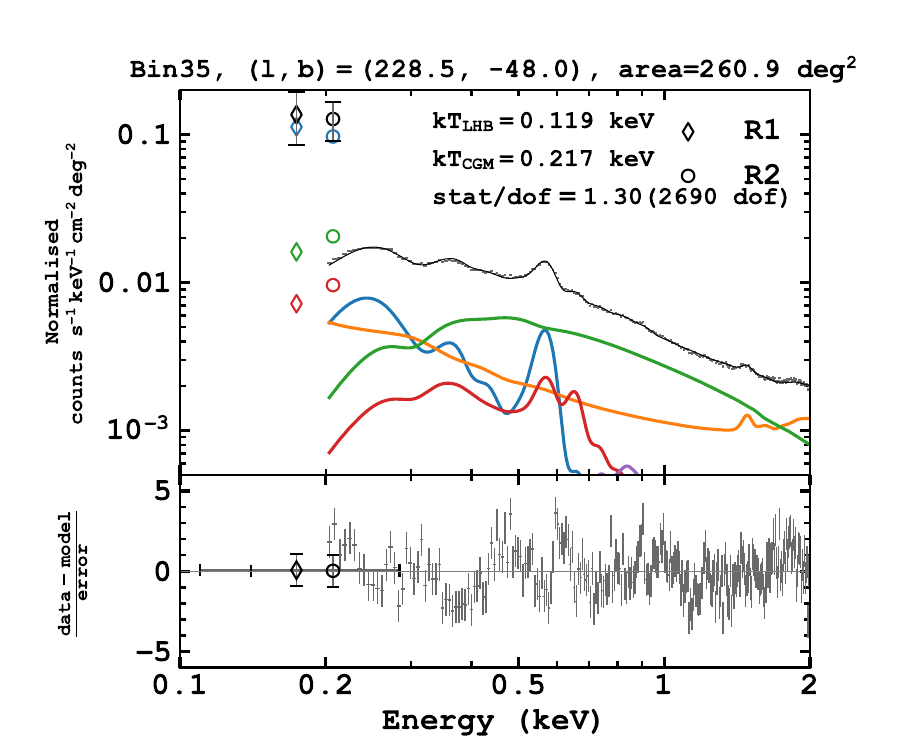}
    \caption{North-south temperature dichotomy shown by a high S/N spectra in the south. \textit{Left: }Spectrum in the southern hemisphere fitted while keeping $kT_{\rm LHB}$ fixed at the northern value ($kT_{\rm LHB}=0.10\,$keV). \textit{Right: }Same as on the left but all the parameters were allowed to vary while fitting.}
    \label{fig:N2S}
\end{figure*}

\begin{figure*}
    \centering
    \includegraphics[width=0.49\textwidth]{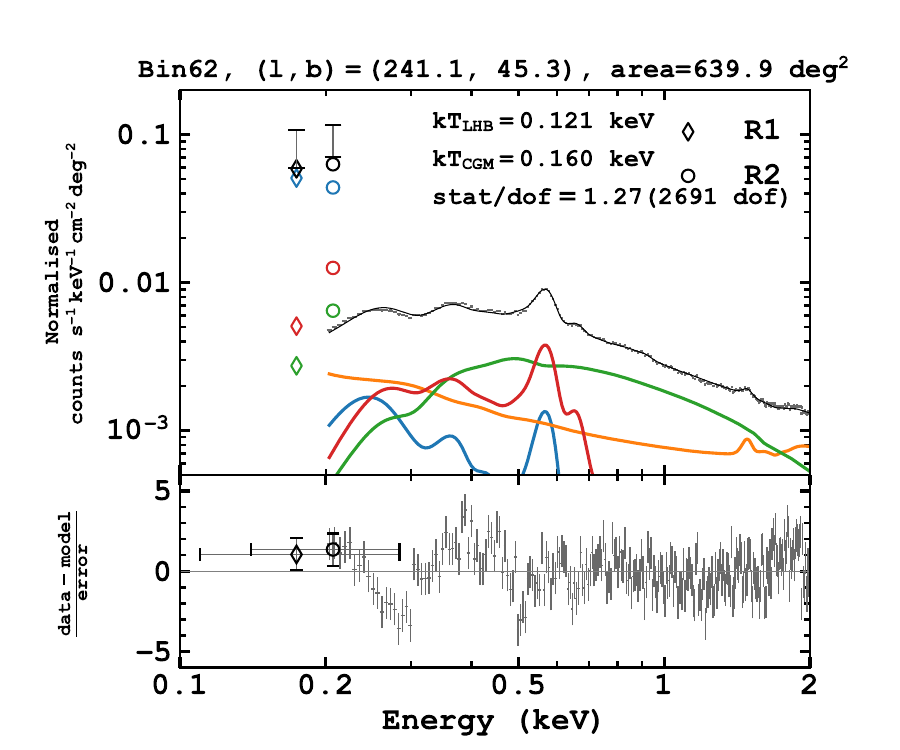}
    \includegraphics[width=0.49\textwidth]{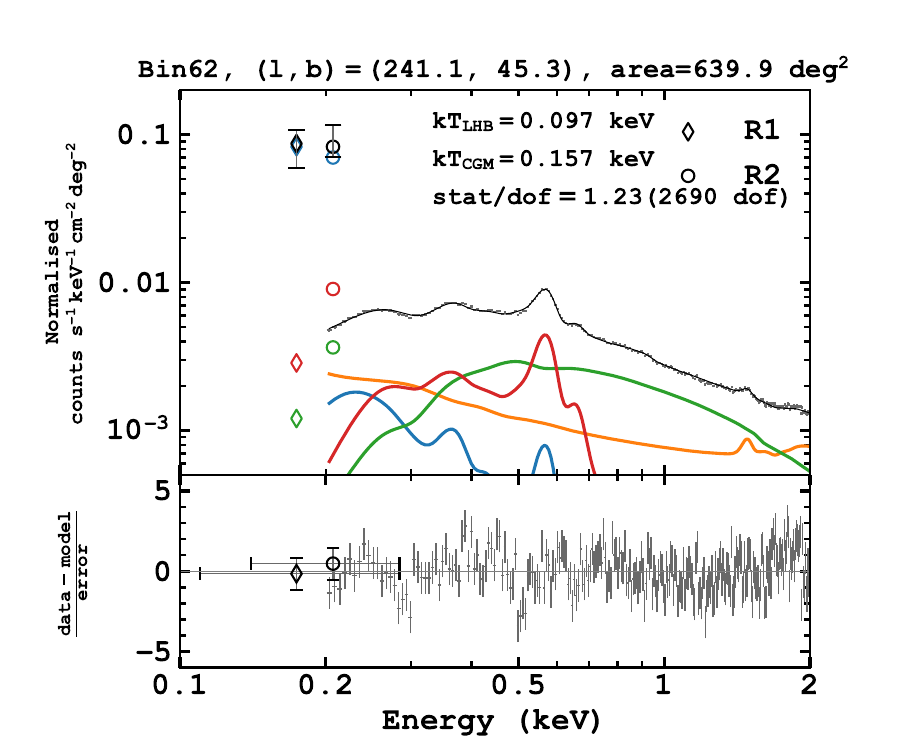}
    \caption{Similar to Fig.~\ref{fig:N2S}, but showing a high S/N spectra in the north. \textit{Left: }Spectrum in the northern hemisphere fitted while keeping $kT_{\rm LHB}$ fixed at the southern value ($kT_{\rm LHB}=0.12\,$keV). \textit{Right: }Same as on the left but all the parameters were allowed to vary while fitting.}
    \label{fig:S2N}
\end{figure*}

To highlight the spectral difference, we took two examples of high-S/N spectra ($\text{S/N}=400$), one from each hemisphere. Fig.~\ref{fig:N2S} shows the spectrum in the south. The difference in the left and right panels lies only with the model: the left panel has the $kT_{\rm LHB}$ kept at the northern value, while all model parameters were allowed to vary in the right panel. Fig.~\ref{fig:S2N} shows the reverse: showing the northern spectrum and keeping the southern temperature of the LHB in the left panel. The comparisons demonstrate the incompatibility of the spectrum from either hemisphere with the LHB temperature measured from the opposite hemisphere. This lends further support for a real LHB temperature dichotomy in the hemispheres.

\end{document}